\documentclass[11pt,a4paper,preprintnumbers]{article}
\pdfoutput=1
\usepackage{jheppub}
\makeatletter
\gdef\@fpheader{} 
\makeatother
\usepackage{amsfonts, amsmath, amssymb, latexsym, siunitx}
\usepackage{graphicx, caption, subcaption, xspace, booktabs}
\usepackage{comment}
\usepackage[normalem]{ulem}
\usepackage[dvipsnames]{xcolor}

\usepackage{hyperref}
\usepackage[acronym,nonumberlist,nopostdot,nogroupskip]{glossaries-extra}
\usepackage{glossary-mcols}
\setabbreviationstyle[acronym]{long-short}
\usepackage[capitalize]{cleveref} 

\binoppenalty=\maxdimen
\relpenalty=\maxdimen
\setglossarystyle{mcolindex}

\makenoidxglossaries

\newacronym{BPL}{BPL}{broken power law}
\newacronym{BSM}{BSM}{beyond the \gls{SM}}
\newacronym{CGB}{CGB}{Compact Galactic Binary}
\newacronym[longplural={confidence intervals (for likelihood analysis) or credible intervals (for Bayesian analysis)}]{CL}{CI}{confidence interval (for likelihood analysis) or credible interval (for Bayesian analysis)}
\newacronym{CMB}{CMB}{cosmic microwave background}
\newacronym{DBPL}{DBPL}{double broken power law}
\newacronym{DM}{DM}{dark matter}
\newacronym{ESA}{ESA}{European Space Agency}
\newacronym{ET}{ET}{Einstein Telescope}
\newacronym{FCC}{FCC}{Future Circular Collider}
\newacronym{GW}{GW}{gravitational wave}
\newacronym{KAGRA}{KAGRA}{Kamioka Gravitational Wave Detector}
\newacronym{LHC}{LHC}{Large Hadron Collider}
\newacronym{LIGO}{LIGO}{Laser Interferometer Gravitational Wave Observatory}
\newacronym{LISA}{LISA}{Laser Interferometer Space Antenna}
\newacronym{LVK}{LVK}{\acrshort{LIGO}/Virgo/\acrshort{KAGRA}}
\newacronym{MHD}{MHD}{magnetohydrodynamic}
\newacronym[longplural={neutron-star binaries}]{NSB}{NSB}{neutron-star binary}
\newacronym{OMS}{OMS}{optical measurement system}
\newacronym{PT}{PT}{phase transition}
\newacronym{PTA}{PTA}{Pulsar Timing Array}
\newacronym{QCD}{QCD}{quantum chromodynamics}
\newacronym{SGWB}{SGWB}{stochastic \gls{GW} background}
\newacronym{SKA}{SKA}{Square Kilometre Array}
\newacronym{SM}{SM}{Standard Model}
\newacronym{SNR}{SNR}{signal-to-noise ratio}
\newacronym[longplural={stellar-origin black hole binaries}]{SOBHB}{SOBHB}{stellar-origin black hole binary}
\newacronym{TDI}{TDI}{time delay interferometry}
\newacronym[longplural={test masses}]{TM}{TM}{test mass}
\newacronym{VBF}{VBF}{vector boson fusion}

\DeclareSIUnit{\yr}{year}

\crefname{section}{Sec.}{Secs.}
\Crefname{section}{Section}{Sections}
\crefname{table}{Tab.}{Tabs.}
\Crefname{table}{Table}{Tables}

\graphicspath{{figures/}}

\newcommand{\OmM}{\Omega_{\rm mag}}
\newcommand{\vA}{\ensuremath{\bar{v}_{\rm A}}\xspace}
\newcommand{\OmK}{\Omega_{\rm kin}}

\newcommand*{\GW}{\ensuremath{\mathrm{\glsentryshort{GW}}}\xspace}
\newcommand*{\MHD}{\ensuremath{\mathrm{\glsentryshort{MHD}}}\xspace}
\newcommand*{\BPL}{\ensuremath{\mathrm{\glsentryshort{BPL}}}\xspace}
\newcommand*{\DBPL}{\ensuremath{\mathrm{\glsentryshort{DBPL}}}\xspace}
\newcommand*{\TM}{\ensuremath{\mathrm{\glsentryshort{TM}}}\xspace}
\newcommand*{\OMS}{\ensuremath{\mathrm{\glsentryshort{OMS}}}\xspace}
\newcommand*{\SNR}{\ensuremath{\mathrm{\glsentryshort{SNR}}}\xspace}

\newcommand{\dd}{\; \textrm{d}}

\newcommand*{\Polychord}{\texttt{Polychord}\xspace}

\title{Gravitational waves from first-order phase transitions in LISA: reconstruction pipeline and physics interpretation}

\author[a,b]{Chiara~Caprini,}
\emailAdd{chiara.caprini@unige.ch}
\affiliation[a]{D\'epartement de Physique Th\'eorique and Center for Astroparticle Physics, Universit\'e de Gen\`eve, Quai E. Ansermet 24, CH-1211 Geneve 4, Switzerland}
\affiliation[b]{Theoretical Physics Department, CERN, 1211 Geneva 23, Switzerland}

\author[c]{Ryusuke~Jinno,}
\emailAdd{ryusuke.jinno@resceu.s.u-tokyo.ac.jp}
\affiliation[c]{Research Center for the Early Universe (RESCEU), University of Tokyo,\\ Hongo 7-3-1, Bunkyo-ku, Tokyo 113-003, Japan}

\author[d,1]{Marek~Lewicki,%
\note{Project coordinator: \href{mailto:marek.lewicki@fuw.edu.pl}{marek.lewicki@fuw.edu.pl}}}
\emailAdd{marek.lewicki@fuw.edu.pl}
\affiliation[d]{Faculty of Physics, University of Warsaw, ul. Pasteura 5, 02-093 Warsaw, Poland}

\author[e,2,*]{Eric~Madge,%
\note{Corresponding author: \href{mailto:eric.madgepimentel@uam.es}{eric.madgepimentel@uam.es}}%
\note[*]{Now at \itshape Instituto de F\'isica Te\'orica UAM-CSIC, Universidad Aut\'onoma de Madrid,\\\hspace*{1.7em}C/ Nicol\'as Cabrera 13--15, Cantoblanco, Madrid 28049, Spain}%
}
\emailAdd{eric.madgepimentel@uam.es}
\affiliation[e]{Department of Particle Physics and Astrophysics, Weizmann Institute of Science,\\ Herzl Street 234, Rehovot, 7610001, Israel}

\author[f,g]{Marco~Merchand,}
\emailAdd{marcomm@kth.se}
\affiliation[f]{Department of Physics, KTH Royal Institute of Technology, SE-10691 Stockholm, Sweden}
\affiliation[g]{The Oskar Klein Centre for Cosmoparticle Physics, AlbaNova University Centre,\\ SE-10691 Stockholm, Sweden}

\author[h,3]{Germano~Nardini,\note{Project coordinator: \href{mailto:germano.nardini@uis.no}{germano.nardini@uis.no}}}
\emailAdd{germano.nardini@uis.no}
\affiliation[h]{Department of Mathematics and Physics, University of Stavanger, NO-4036 Stavanger, Norway}

\author[b,4]{Mauro~Pieroni,\note{Corresponding author: \href{mailto:mauro.pieroni@cern.ch}{mauro.pieroni@cern.ch}}}
\emailAdd{mauro.pieroni@cern.ch}

\author[a]{Alberto~Roper~Pol,}
\emailAdd{alberto.roperpol@unige.ch}

\author[i,j,k]{Ville~Vaskonen}
\emailAdd{ville.vaskonen@pd.infn.it}
\affiliation[i]{Dipartimento di Fisica e Astronomia, Universit\`a degli Studi di Padova,\\ Via Marzolo 8, 35131 Padova, Italy}
\affiliation[j]{Istituto Nazionale di Fisica Nucleare, Sezione di Padova, Via Marzolo 8, 35131 Padova, Italy}
\affiliation[k]{Keemilise ja bioloogilise f\"u\"usika instituut, R\"avala pst. 10, 10143 Tallinn, Estonia}

\author[]{\\ \centering \texttt{(For the LISA Cosmology Working Group)}}

\abstract{
We develop a tool for the analysis of stochastic gravitational wave backgrounds from cosmological first-order phase transitions with LISA: we initiate a template databank for these signals, prototype their searches, and forecast their reconstruction.
The templates encompass the gravitational wave signals sourced by bubble collisions, sound waves and turbulence. Accounting for Galactic and extra-Galactic foregrounds, we forecast the region of the parameter space that LISA will reconstruct with better than $\sim\SI{10}{\%}$ accuracy, if certain experimental and theoretical uncertainties are solved by the time LISA flies.
We illustrate the accuracy with which LISA can reconstruct the parameters on a few benchmark signals, both in terms of the template parameters and the phase transition ones.
To show the impact of the forecasts on physics beyond the Standard Model, we map the reconstructed benchmark measurements into the parameter spaces of the singlet extension of the Standard Model and of the classically conformal invariant $U(1)_{B-L}$ model.
}

\toccontinuoustrue

\begin{document}
\begin{figure}
\begin{flushright}
\href{https://lisa.pages.in2p3.fr/consortium-userguide/wg_cosmo.html}{\includegraphics[width = 0.2 \textwidth]{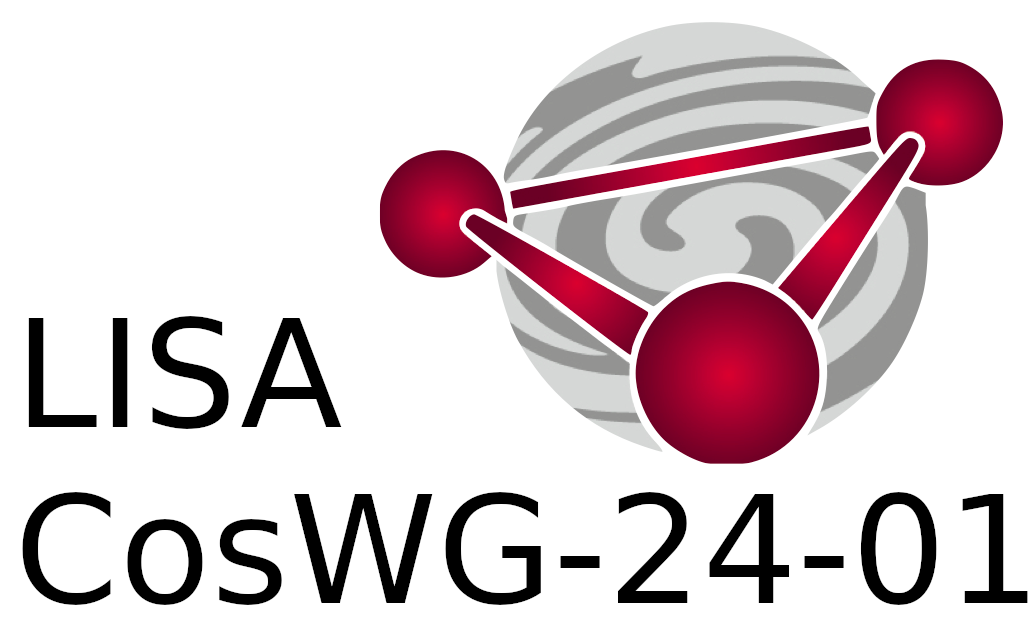}}\\[5mm]
\end{flushright}
\end{figure}

\maketitle

\section{ Introduction}
\label{sec:intro}

Cosmology and astrophysics are currently at the dawn of a new golden age thanks to the discovery of \glspl{GW}, which started with the first 
direct detections by the \gls{LVK} collaboration~\cite{Abbott:2016blz, Abbott:2016nmj,Abbott:2017vtc,Abbott:2017gyy, Abbott:2017oio, LIGOScientific:2018mvr, LIGOScientific:2020ibl, LIGOScientific:2021usb, LIGOScientific:2021djp}
of \glspl{GW} emitted by black hole and neutron star mergers in the late Universe. 
Recently, \gls{PTA} collaborations have presented strong evidence for the presence of a \gls{SGWB} at nano-Hertz frequencies~\cite{NANOGrav:2023gor,EPTA:2023sfo,Reardon:2023gzh,Xu:2023wog}.
The front-runner candidate for the source of this signal is associated with binaries of supermassive black holes, colliding relatively recently in cosmological terms~\cite{NANOGrav:2023hfp,EPTA:2023xxk,Ellis:2023dgf}, and interacting with their environment. 
However, at this stage, it is also possible that this signal is of primordial origin~\cite{NANOGrav:2023hvm,EPTA:2023xxk,Figueroa:2023zhu,Bian:2023dnv,Ellis:2023oxs,Madge:2023dxc}.
The first detection of an \gls{SGWB} gives hope that future \gls{GW} observations will allow us to probe high-energy processes taking place in the early Universe.

Among the many candidates of primordial \gls{GW} sources \cite{Caprini:2018mtu,LISACosmologyWorkingGroup:2022jok}, our analysis focuses on first-order \glspl{PT} in the early Universe. These are expected when a new minimum emerges in the effective potential of the theory
as the Universe cools and thermal corrections to the potential diminish. 
If the \gls{PT} is of first order, the new vacuum and the initial one are separated by a barrier that the scalar field will eventually cross through bubble nucleation. After nucleation, the bubbles grow 
due to the latent heat released from the difference between the true and the false vacua
and eventually collide, filling the whole volume with the new phase.
An \gls{SGWB} is expected to be sourced by the anisotropic stresses due to the collisions of bubbles and highly relativistic fluid shells in very strong transitions \cite{Kosowsky:1991ua, Kosowsky:1992vn, Huber:2008hg, Cutting:2018tjt, Cutting:2020nla, Lewicki:2020jiv, Ellis:2020nnr, Lewicki:2020azd, Lewicki:2022pdb}, as well as the subsequent development of bulk fluid motion \cite{Kamionkowski:1993fg,Caprini:2015zlo, Caprini:2019egz}. The latter initially takes the form of sound waves \cite{Hindmarsh:2013xza,Hindmarsh:2015qta,Hindmarsh:2017gnf,Cutting:2019zws,Hindmarsh:2019phv,Jinno:2020eqg,Jinno:2022mie}, but can also lead to the development of \gls{MHD} turbulence \cite{Kosowsky:2001xp,Dolgov:2002ra,Gogoberidze:2007an,Kahniashvili:2008pe,Caprini:2009yp,Brandenburg:2017neh,Niksa:2018ofa,RoperPol:2019wvy,Kahniashvili:2020jgm,RoperPol:2021xnd,RoperPol:2022iel,Auclair:2022jod,RoperPol:2023bqa}.
Depending on the underlying \gls{PT} model and the consequent bubble and plasma dynamics, 
any of these \gls{GW} sourcing mechanisms may dominate the \gls{SGWB} production, and we therefore include all of them in our analysis.
Turbulence has often been neglected in previous analyses for two reasons: because it was considered subdominant, and because of the absence of a simulation-validated template for the \gls{GW} signal \cite{Caprini:2019egz}. 
However, it is now clear that the non-observation of turbulence in simulations of weakly first-order \glspl{PT} does not imply that it is never formed or always subdominant, in particular, if the lifetime of the sound waves is shorter than one Hubble time \cite{Ellis:2018mja,Caprini:2019egz,Ellis:2020awk,RoperPol:2023dzg}. Furthermore, a simulation-validated template has been developed for the  \gls{SGWB} produced by fully developed turbulence~\cite{RoperPol:2022iel,RoperPol:2023bqa}, which we will use in the following.

Our focus will be on the \gls{LISA} mission~\cite{LISA:2017pwj}, aimed at probing the currently unexplored \gls{GW} frequency band around the milli-Hertz. In terms of cosmology, this band is associated with energies in the early Universe around the electroweak scale~\cite{LISACosmologyWorkingGroup:2022jok}. 
Many extensions of the \gls{SM} predict the electroweak \gls{PT} to be of first order, which is a strong motivation to look for associated \gls{GW} signals in \gls{LISA}~\cite{Caprini:2015zlo, Caprini:2019egz}. 
In this work, we pursue two main goals. First, we select a set of templates for \gls{GW} sourcing processes
associated with first-order \glspl{PT}, in order to initiate a template repository for the pipeline of SGWB searches that we have developed.
We stress that some underlying assumptions and approximations still affect these templates. 
While they cannot be considered as carved into stone, we are confident that they represent the best available choices combining simplicity and accuracy, given the state of the art in the field.
Second, we implement these templates in the \texttt{SGWBinner} code~\cite{Caprini:2019pxz, Flauger:2020qyi}, and use them to produce forecasts of the sensitivity of \gls{LISA} to \glspl{SGWB} from first order \glspl{PT}, studying in particular its  
impact on the reconstruction of the parameters of two examples of particle physics models underlying the transition. 

With respect to previous works focused on the case of \gls{GW} production by sound waves~\cite{Gowling:2021gcy, Giese:2021dnw, Boileau:2022ter, Gowling:2022pzb}, we add parameter estimation also for the \gls{GW} signals from bubble collisions and \gls{MHD} turbulence.
We assess the \gls{LISA} sensitivity in terms of
two different sets of parameters: 
in terms of thermodynamic parameters (cf.~\cref{sec:reconstruction}), such as the kinetic energy fraction, the temperature, and the duration of the transition, as well as in terms of geometric parameters (cf.~\cref{sec:reconstruction_geometric}), such as the frequency breaks and the amplitude of the \gls{SGWB} and provide a mapping between the two sets. 
Compared to previous analyses~\cite{Gowling:2021gcy, Giese:2021dnw, Boileau:2022ter, Gowling:2022pzb}, we go one step further in the process and translate the estimation of the \gls{SGWB} template parameters into constraints on the parameter space of two specific particle physics models (cf.~\cref{sec:interpretation_models}). 
Concerning the \gls{SGWB} from sound waves, we adopt a somewhat simpler template than in Refs.~\cite{Gowling:2021gcy, Giese:2021dnw, Boileau:2022ter, Gowling:2022pzb}, based on the results of Ref.~\cite{Jinno:2022mie}. 
The main features of the \gls{SGWB} spectrum (spectral shape and position of the spectral breaks) are consistent with the results developed in Refs.~\cite{Hindmarsh:2019phv, Gowling:2021gcy, Gowling:2022pzb}. However, the spectral slopes of the \gls{SGWB} are taken from Ref.~\cite{Jinno:2022mie}. These are compatible with the recent results of Refs.~\cite{RoperPol:2023dzg, Sharma:2023mao} (see also Ref.~\cite{RoperPol:2023bqa}).
Note that the spectral shape of the \gls{SGWB} sourced by sound waves in the context of the sound shell model
\cite{Hindmarsh:2016lnk, Hindmarsh:2019phv} is still a subject of active study~\cite{Cai:2023guc, Sharma:2023mao, RoperPol:2023dzg}.

The \gls{LISA} mission has recently been adopted by the \gls{ESA} and its construction will start soon for a launch scheduled in the 2030s. 
Our results serve as a starting point for the main task ahead of the community in preparation of the experiment, which is the development of the pipelines for the data analysis and the physics interpretation. 
We make some simplifying assumptions in our analyses of the forecasts. We neglect the theoretical uncertainties associated with the computation of the transition parameters and the resulting \gls{GW} signal in a given model~\cite{Croon:2020cgk, Schicho:2021gca, Athron:2022jyi, Gould:2023jbz}. The impact of these uncertainties on the parameter reconstruction was recently studied in Ref.~\cite{Lewicki:2024xan}.
However, this is an active research topic~\cite{Gould:2021ccf, Schicho:2022wty, Lofgren:2021ogg, Hirvonen:2021zej, Ekstedt:2022bff, Ekstedt:2022zro, Ekstedt:2023sqc, Gould:2023ovu} and we assume significant progress will be made towards reducing this issue by the time the data becomes available.
Another kind of theoretical uncertainty that we neglect are general relativity effects, important especially when the average bubble size is comparable in size to the Hubble scale~\cite{Giombi:2023jqq}. Their exact effect on the \gls{SGWB} spectra is still uncertain. Furthermore, we adopt a simple two-parameter noise model to predict the performance of the mission. While it is simplistic to assume that one can directly apply a pre-flight validated instrument noise model in a realistic search for the \gls{SGWB} in real data, the development of methodologies to incorporate noise uncertainties in \gls{SGWB} searches is a subject of ongoing study \cite{Baghi:2023qnq, Muratore:2023gxh, Pozzoli:2023lgz}. 
We plan to update our predictions as more realistic noise estimates, as well as methods of coincident noise and \gls{SGWB} estimation, are produced by the \gls{LISA} Consortium. 
We use simplified, template-based predictions for the astrophysical foregrounds. The impact of astrophysical foregrounds on the reach of \gls{LISA} will very likely be significant~\cite{Boileau:2020rpg, Pieroni:2020rob, Boileau:2021sni, Lewicki:2021kmu, Racco:2022bwj}.
Also in this aspect, we plan to improve our results 
in the future, when the methods for disentangling the primordial \gls{SGWB} from the astrophysical foregrounds improve further. 
We work on simulated data including only the \gls{SGWB}, the foregrounds, and the instrument noise: therefore, we assume that the data stream can be perfectly cleaned from residuals due to the imperfect subtraction of the signals of other \gls{GW} sources. 
Moreover, the data are also simulated adopting the very same templates for the noise, foregrounds and \gls{SGWB} which we adopt in the reconstruction: hence, we also implicitly assume that any theoretical and experimental uncertainties are absent, and the models reproduce faithfully the reality of the signals and the noise.
Last but not least, we limit our discussion to parameter reconstruction only, and leave the question of model discrimination, taking into account also other \gls{SGWB} sources not related to \glspl{PT}, for future work.

This paper is organised as follows.
In \cref{sec:models}, we introduce the \gls{SGWB} templates in terms of the ``geometric parameters'' such as the amplitude, characteristic frequencies, and spectral slopes, and specify how these depend on the ``thermodynamic \gls{PT}  parameters'' such as kinetic energy fraction in the plasma, the transition temperature and inverse timescale of the transition. \Cref{sec:SGWBinner} describes the \texttt{SGWBinner} code and the data analysis strategy.
We then proceed with evaluating the parameter reconstruction for cosmological \glspl{PT} in terms of different sets of parameters. In \cref{sec:reconstruction_geometric}, we perform the \gls{SGWB} spectrum reconstruction in terms of the geometric parameters. In \cref{sec:reconstruction}, we address the reconstruction of the thermodynamic parameters related to the dynamics of the \gls{PT}.
Finally, \cref{sec:interpretation_models} connects the \gls{PT} parameters to two particular underlying particle physics models, reconstructing the ``fundamental model parameters'' such as masses and couplings. 
We conclude in \cref{sec:conclusions}.

\section{SGWB templates for first-order phase transitions}
\label{sec:models}

Cosmological first-order \glspl{PT} generate \glspl{GW} predominantly via three mechanisms: collisions of the vacuum bubble walls, sound waves in the primordial plasma, and turbulent motion.
The corresponding \gls{SGWB} spectra of the former can be modeled as a \gls{BPL}, whereas the latter two can be described in terms of \glspl{DBPL}. The respective amplitudes and frequency breaks depend on several parameters, which under some assumptions (to be clarified later for each source) can be related to the strength of the \gls{PT} $\alpha$, its inverse duration normalised to the Hubble rate $\beta/H_*$, the temperature $T_*$ after the end of the transition, and the bubble wall velocity $\xi_w$ \cite{Espinosa:2010hh}. Furthermore, the turbulence \gls{SGWB} will also depend on the fraction of energy $\varepsilon$ converted into turbulent motion in the \gls{PT}.

The present-day frequencies and amplitudes
are related to the quantities at the time of \gls{GW} production (with subscript~``*'') via the relations
\begin{align}
    f &= f_* \frac{a_*}{a_0} = H_{*,0} \frac{f_*}{H_*}  \,,
    &
    h^2\Omega &= h^2 F_{\GW,0} \, \Omega_*\,,
\end{align}
where $H_{*,0}$ is the Hubble rate at the time of \gls{GW} production $H_*$ redshifted to today, 
\begin{equation}
    H_{*,0} = \frac{a_*}{a_0} H_* \simeq \SI{1.65e-5}{\Hz}\,\left(\frac{g_*}{100}\right)^{\frac{1}{6}} \left(\frac{T_*}{\SI{100}{\GeV}}\right) , 
    \label{redsh_H}
\end{equation}
and $F_{\GW,0}$ is the red-shift factor for the fractional energy density,
\begin{equation}
    h^2 \, F_{\GW,0} = h^2 \left(\frac{a_*}{a_0}\right)^{4}
    \left(\frac{H_{*}}{H_{0}}\right)^2 \simeq \num{1.64e-5} \left(\frac{100}{g_{*}}\right)^{1/3}\,,
    \label{FGW0}
\end{equation}
with $H_0 = 100\, h$ km/s/Mpc. Here, $g_*$ is the effective number of degrees of freedom at the end of the \gls{PT}, which we assume to approximately be the same as the number of entropy degrees of freedom $g_* \approx g_{*s}$, as appropriate for $T_*>\SI{0.1}{\MeV}$~\cite{Kolb:1990vq}. We will now proceed to describe the \gls{SGWB} templates through both sets of geometric and thermodynamic parameters corresponding to the underlying sources, and refer to them as $\vec\theta_{\rm Cosmo}$ in each case.

\subsection{Template I: broken power law for bubble collisions and highly relativistic fluid shells}
\label{sec:bubble_template}

Template I is a \glsfirst{BPL} that can be expressed as
\begin{equation}
	\label{eq:BPL}
	\Omega_\GW^\BPL(f, \vec \theta_{\rm Cosmo}) = \Omega_b \left(\frac{f}{f_b}\right)^{\! n_1} \left[ \frac{1}{2}+\frac{1}{2} \left(\frac{f}{f_b}\right)^{\! a_1} \right]^{\frac{n_2-n_1}{a_1}} ,
\end{equation} 
where $\vec\theta_{\rm Cosmo}=\{\Omega_b, f_b\}$ is the set of free geometric parameters of the template, while $n_1, n_2$, and $a_1$ are geometric parameters fixed by the underlying modelling of the source, as described below.
The amplitude is normalised such that $\Omega(f_b) = \Omega_b$.
At $f\ll f_b$ the spectrum grows as $f^{n_1}$, and at $f\gg f_b$ it decreases (assuming $n_2<0$) as $f^{n_2}$. The parameter $a_1$ describes how broad is the transition around the spectral peak.
Note that, in general, $f_b$ does not coincide with the peak of the spectrum, located at $f_p = f_b\, (-n_1/n_2)^{{1/ a_1}}$, and $\Omega_b$ does not correspond to the value
of the spectrum at the peak (unless $n_1 = - n_2$), which is
\begin{equation}\label{peak_single_bpl}
\Omega_p = \Omega_b \left[\frac{1}{2} \left(- \frac{n_2}{n_1}\right)^{\frac{n_1}{n_1 - n_2}} + \frac{1}{2} \left(-\frac{n_1}{n_2}\right)^{-\frac{n_2}{n_1 - n_2}} \right]^{-\frac{n_1 - n_2}{a_1}} \ .
\end{equation}
Then, the \gls{SGWB} spectrum can alternatively be expressed in terms of the
spectral peak $f_p$ and its value at the peak $\Omega_p$,
\begin{equation}
	\Omega_\GW^\BPL(f, \vec \theta_{\rm Cosmo}) = \Omega_p \frac{\left(n_1 - n_2\right)^{\frac{n_1 - n_2}{a_1}}}{\left[-n_2 \left(\frac{f}{f_p}\right)^{-\frac{n_1 a_1}{n_1 - n_2}} + n_1 \left(\frac{f}{f_p}\right)^{-\frac{n_2 a_1}{n_1 - n_2}}\right]^{\frac{n_1 - n_2}{a_1}}} \ .
 \label{eq:BPLalt}
\end{equation}
While the functional form \cref{eq:BPLalt} is more commonly used in the literature~\cite{Huber:2008hg, Konstandin:2017sat, Lewicki:2020azd, Lewicki:2022pdb}, 
in the following we will work with \cref{eq:BPL}, which 
directly resembles the \gls{DBPL} given in \cref{eq:DBPL}.

We use the \gls{BPL} template, \cref{eq:BPL}, for \gls{GW} production in very strong \glspl{PT}.
These transitions are defined by the vacuum energy of the scalar field dominating over the radiation background, i.e.~$\alpha \gg 1$.  
One possibility for the time evolution of the bubbles is that the walls interact sufficiently weakly with the surrounding plasma and thus most of the released vacuum energy goes to the bubble wall gradient and kinetic energies~\cite{Bodeker:2009qy}.
In such a transition, plasma motion can be neglected and the dominant source of \glspl{GW} comes from the scalar field.
The production of \glspl{GW} in such a system was modelled with the envelope approximation~\cite{Kosowsky:1992vn,Kosowsky:1992rz}, in which the collided parts of the bubble walls are neglected in the thin wall limit, and further developed in e.g.~Refs.~\cite{Huber:2008hg,Weir:2016tov,Jinno:2016vai}.

However, the collided part of the bubble walls has recently been found to be essential for a correct estimation of the \gls{SGWB} spectrum, since in reality, the shells cannot lose their momentum instantaneously.
Such a possibility has been pointed out in numerical simulations~\cite{Child:2012qg,Cutting:2018tjt,Jinno:2019bxw,Cutting:2020nla} and from analogies with particle collisions~\cite{Jinno:2017fby}.
While for maximally long-lived shells a modeling called bulk flow was proposed and has been studied numerically and analytically~\cite{Konstandin:2017sat,Jinno:2017fby}, it is not obvious how the collided part decays in specific particle physics models.
In this regard, we adopt a hybrid approach developed in Refs.~\cite{Lewicki:2020jiv,Lewicki:2020azd} for the time evolution of the collided walls and the resulting \gls{GW} production.
In these studies, time evolution of the momentum decay in the radial direction is embedded in multi-bubble nucleation and collision histories, allowing for the calculation of the \gls{SGWB} spectrum in microphysics potentials that can realistically cause strongly supercooled \glspl{PT}.

While the dominant energy carrier in the weak coupling limit is the scalar field, supercooled \glspl{PT} can also proceed with relativistic detonations, in which the vacuum energy is dominantly transferred to highly relativistic fluid shells.
In this energy transfer, particle splitting plays a key role~\cite{Bodeker:2017cim, Hoche:2020ysm,Azatov:2021irb,Gouttenoire:2021kjv,Azatov:2023xem}.
As a result, the wall reaches a terminal velocity and the main energy carriers are the highly relativistic fluid shells surrounding the bubbles.
The production of \glspl{GW} for this case is expected to be qualitatively different from that of sound waves because, in the former case, the fluid shells become very thin and highly concentrated around the walls,
and thus the macroscopic distribution of the energy-momentum tensor is expected to be rather similar to that of scalar walls \cite{Kamionkowski:1993fg,Jinno:2019jhi,Lewicki:2022pdb}.
In Ref.~\cite{Lewicki:2022pdb}, the \gls{SGWB} spectrum has been estimated by incorporating the radial time evolution of highly relativistic fluid shells. It was shown that the resulting \gls{SGWB} spectrum is closer to the one found in the case of bubble wall collisions in models featuring the breaking of a gauge symmetry discussed in Ref.~\cite{Lewicki:2020azd}, which allows us to use the same \gls{SGWB} template in both cases.

\begin{figure}[p]
    \vspace*{-2\baselineskip}
    \centering
    \begin{subfigure}{\textwidth}
        \centering
        \includegraphics[width=.63\textwidth]{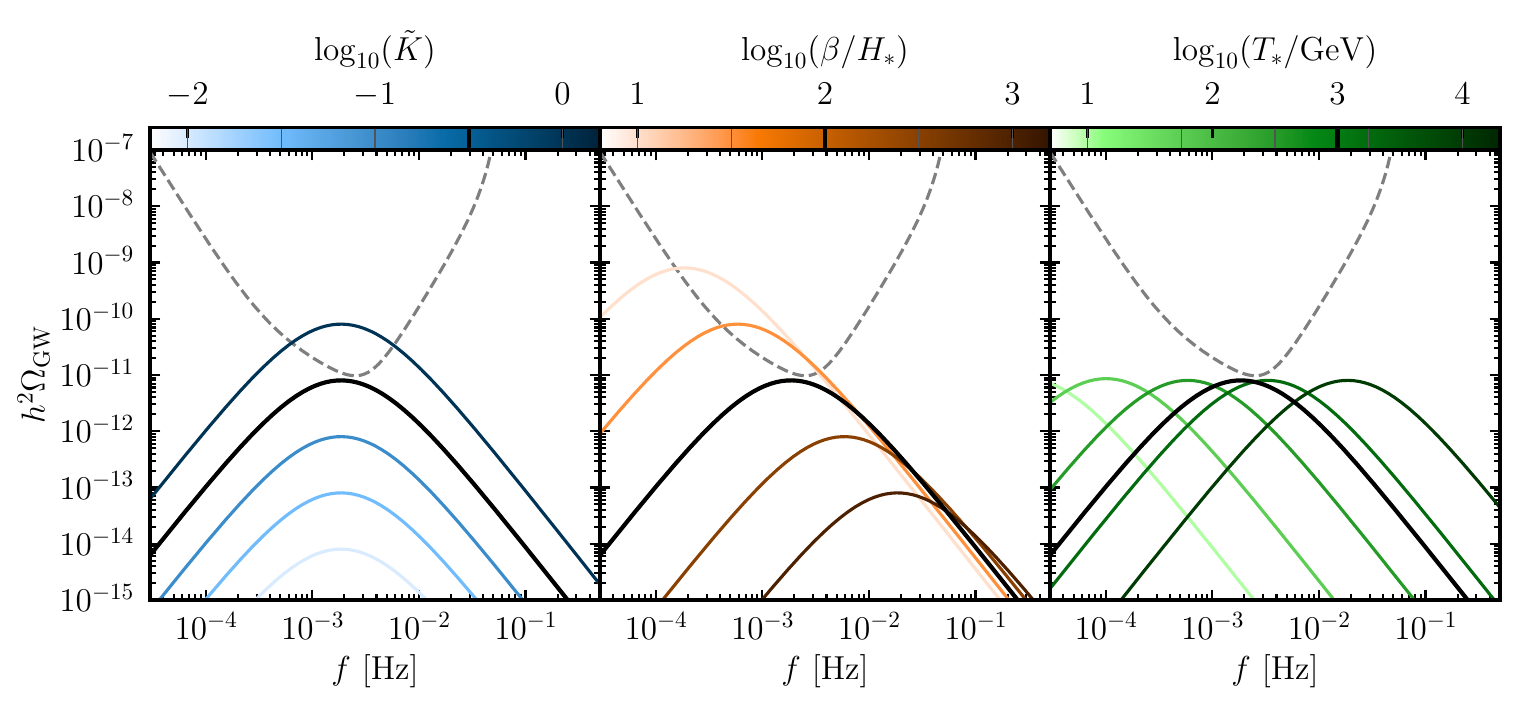}
        \caption{bubble collisions and highly relativistic fluid shells
        (black: $\tilde{K} = 0.32$, $\beta/H_* = 
 100$, $T_* = \SI{1}{\TeV}$)}\label{fig:pars_PT_coll}
    \end{subfigure}
    \begin{subfigure}{\textwidth}
        \centering
        \includegraphics[width=.81\textwidth]{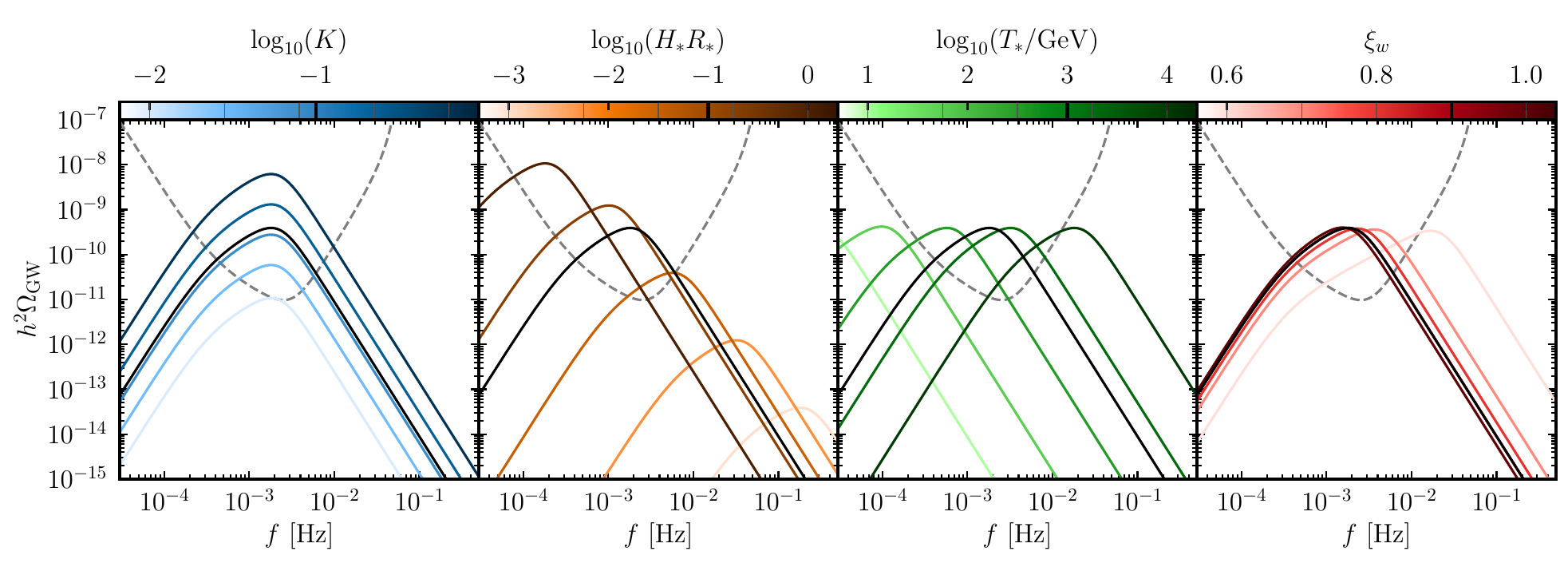}
        \caption{sound waves (black: $K=0.1$, $H_* R_* = 0.1$, $\xi_w = 0.9$, $T_* = \SI{1}{\TeV}$)}\label{fig:pars_PT_sw}
    \end{subfigure}
    \begin{subfigure}{\textwidth}
        \centering
        \includegraphics[width=.63\textwidth]{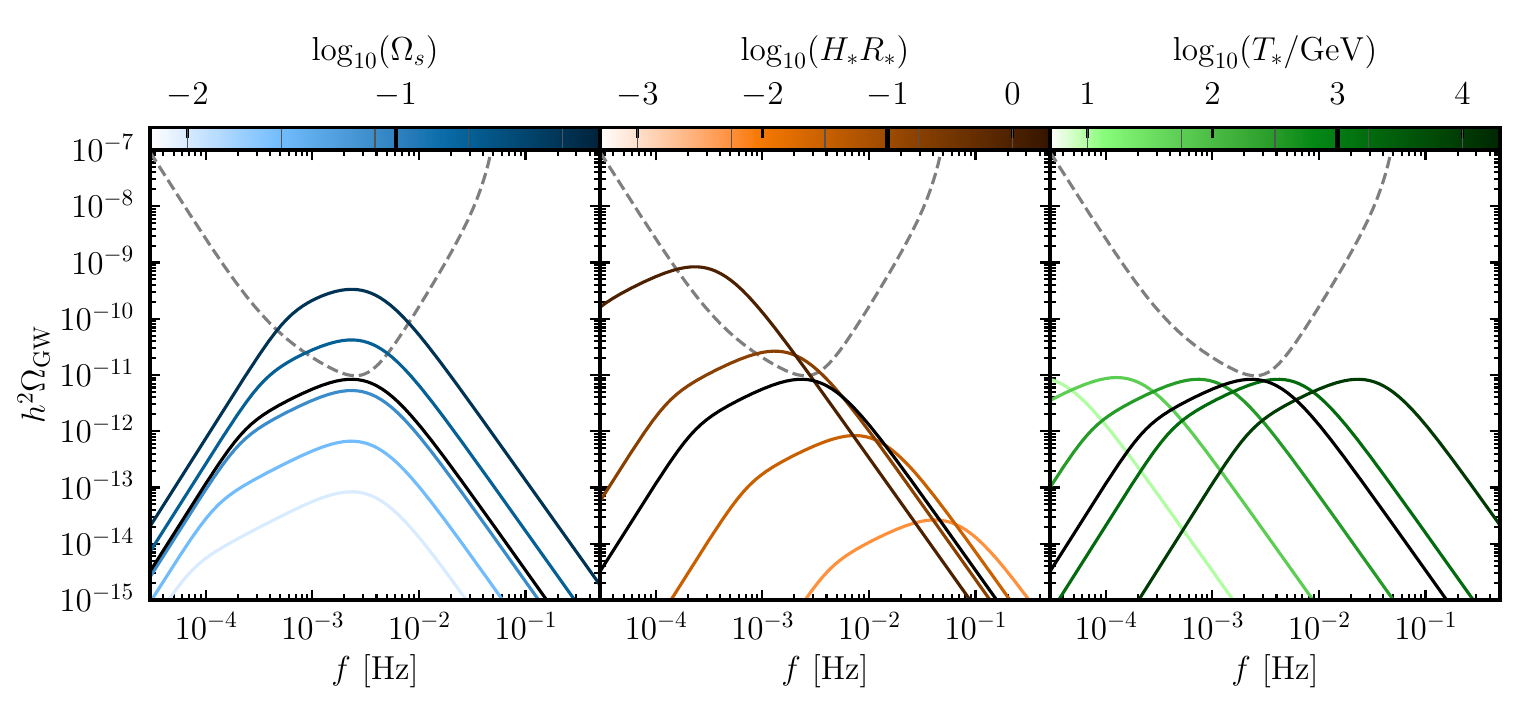}
        \caption{turbulence (black: $\Omega_s = 0.1$, $H_* R_* = 0.1$, $T_* = \SI{1}{\TeV}$)}\label{fig:pars_PT_turb}
    \end{subfigure}
    \begin{subfigure}{\textwidth}
        \includegraphics[width=\textwidth]{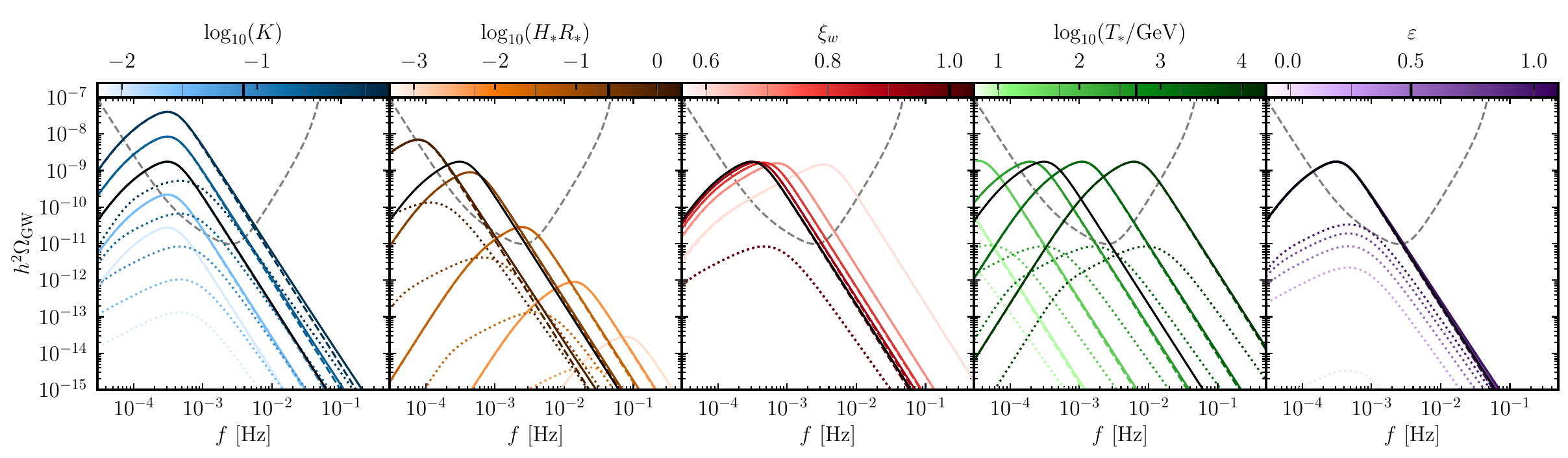}
        \caption{sound waves + turbulence (black: $K=0.08$, $H_*R_* = 0.25$, $\xi_w = 1$, $T_* = \SI{500}{\GeV}$, $\varepsilon=0.5$)}
        \label{fig:pars_PT_sw_turb}
    \end{subfigure}
    \caption{%
        The coloured lines show the parameter dependence of the \gls{SGWB} templates for \glspl{PT} in terms of the thermodynamic parameters, varying one parameter at a time. The remaining parameters are fixed to the benchmark values indicated in the subcaptions, corresponding to the black spectrum in each plot. 
        From top to bottom we show bubble collisions, sound waves, turbulence, and sound waves + turbulence.
        For the latter, the spectra of the sound waves and turbulence contributions are shown as dashed and dotted lines, respectively.
        The dashed gray line is the \gls{LISA} noise curve.
    }
    \label{fig:pars_pts}
\end{figure}

In this context, the peak amplitude and frequency at the present time are related to the parameters $\alpha$, $\beta/H_*$, and $T_*$ via
\begin{equation}\label{eq:amp_and_freq_strongPT}
\begin{aligned}
h^2 \Omega_p  &= h^2 F_{\GW,0} \,A_\text{str} \,\tilde{K}^2\,\left(\frac{H_*}{\beta}\right)^{\!2} \,, 
\quad
f_p &\simeq 0.11\,H_{*,0}\, \frac{\beta}{H_*}\,,
\end{aligned}
\end{equation}
where $\tilde{K} \equiv \alpha/(1 + \alpha)$
is the fractional energy density of the \gls{GW} source. The spectrum was computed assuming a very strong transition with $\alpha\gg 1$, corresponding to $\tilde{K}\approx 1$. However, we leave $\tilde{K}$ as a free parameter in the reconstruction (cf.~\cref{sec:reconstruction}) to take into account non-standard scenarios such as a \gls{PT} taking part in a dark sector whose energy density is only a fraction of the total~\cite{Breitbach:2018ddu, Fairbairn:2019xog}. 
We will focus on the most generic case of either gauged scalar field bubble collisions or highly relativistic fluid shells sourcing the \gls{SGWB}.
In this case, the numerical simulations of Ref.~\cite{Lewicki:2022pdb} indicate $A_\text{str}\,\simeq\,0.05$, and the 0.11 factor in $f_p$ of \cref{eq:amp_and_freq_strongPT}.
The corresponding parameters in \cref{eq:BPL} are found to be $n_1 = - n_2 \simeq 2.4$ and $a_1 \simeq 1.2$ \cite{Lewicki:2022pdb}. In this case, $f_b=f_p$ and $\Omega_b=\Omega_p$.
Note that at scales larger than the horizon scale at the time of the transition, the \gls{SGWB} spectrum would be expected to scale as $f^3$ because the source is uncorrelated \cite{Caprini:2009fx}. We do not directly include this effect because the spectra we use were computed with simulations that do not take into account the expansion and the exact impact is uncertain~\cite{Lewicki:2022pdb}.
For convenience, \cref{tab:slopes} summarizes the spectral slopes used in this work for the respective contribution to the \gls{SGWB} spectrum. The geometric free parameters $\{f_b, \Omega_b\}$ can be expressed through the thermodynamic free parameters $\vec{\theta}_{\rm Cosmo}=\{ \tilde{K}, \beta/H_*, T_*\}$ via \cref{eq:amp_and_freq_strongPT}.
The dependence of the spectrum on $\tilde{K}$, $\beta/H_*$, and $T_*$ is illustrated in \cref{fig:pars_PT_coll}.

\begin{table}
    \centering
    \begin{tabular}{l|c|ccccc}
        \toprule
         \gls{GW} source & power law template & $n_1$ & $n_2$ & $n_3$ & $a_1$ & $a_2$ \\ \midrule
         strong \gls{PT}, \cref{sec:bubble_template} & \gls{BPL}, \cref{eq:BPL} & $2.4$ & $-2.4$ & --- & $1.2$ & --- \\
         sound waves, \cref{sec:sw_template} & \gls{DBPL}, \cref{eq:DBPL} & $3$ & $1$ & $-3$ & $2$ & $4$ \\
         turbulence, \cref{sec:mhd_template} & \gls{DBPL}, \cref{eq:DBPL} & $3$ & $1$ & $-8/3$ & $4$ & $2.15$ \\
        \bottomrule
    \end{tabular}
    \caption{%
        Spectral slopes of the \gls{BPL} and \gls{DBPL} templates used for the respective \gls{GW} sources from \glspl{PT}.
    }
    \label{tab:slopes}
\end{table}

\subsection{Template II: double broken power law}
\label{sec:dbpl}

Template II is a \glsfirst{DBPL} which takes the form
\begin{align}
\Omega^{\DBPL}_{\GW}(f, \vec \theta_{\rm Cosmo}
) &=
\Omega_\text{int} \times S(f) = \Omega_2 \times S_2(f), \label{eq:DBPL}
\\
S(f) &= N \left( \frac{f}{f_1} \right)^{n_1}
\left[
1 + \left( \frac{f}{f_1} \right)^{a_1}
\right]^{\frac{- n_1 + n_2}{a_1}}
\left[
1 + \left( \frac{f}{f_2} \right)^{a_2}
\right]^{\frac{- n_2 + n_3}{a_2}}, \nonumber
\end{align}
where the \gls{SGWB} spectrum is decomposed into the amplitude $\Omega_\text{\text{int}}$, corresponding to the value of the spectrum logarithmically integrated over the frequency, and the shape function $S(f)$.
The spectrum behaves as $\Omega^{\DBPL}_{\GW} \propto f^{n_1}$ ($f < f_1$), $f^{n_2}$ ($f_1 < f < f_2$), and $f^{n_3}$ ($f_2 < f$) from low to high frequencies. 
The parameters $a_1$ and $a_2$ describe how smoothly the power law behavior changes at $f \sim f_1$ and $f \sim f_2$, respectively.
In principle, the normalization factor $N$ is determined from $\int_{- \infty}^\infty \!\dd \ln f \, S (f) = 1$.
However, as we cannot write an analytical expression for $N$ for arbitrary slopes and frequency breaks, we normalise the spectrum so that $S_2(f_2) = 1$, i.e.\ $S_2(f) = S(f)/S(f_2)$, and use the amplitude $\Omega_2$ at the second frequency break instead of the integrated amplitude $\Omega_\text{int}$.
Thus, $\vec\theta_{\rm Cosmo}=\{\Omega_2, f_1, f_2\}$ is the set of free geometric parameters of this template, while $n_1$, $n_2$, $n_3$, $a_1$, and $a_2$ are parameters fixed by the underlying modelling of the source.

\subsubsection{Sound waves}
\label{sec:sw_template}

If the coupling between the bubble walls and plasma particles is large enough, the latent heat is transferred also to the bulk motion of the fluid. 
While the fluid motion in general has both compression and shear components, the former dominates the latter at the initial stage of the transition because of the initial condition set by expanding spherical bubbles.
After bubbles collide, compression waves propagate acting as a source of \glspl{GW}.
This behavior is found in numerical simulations~\cite{Hindmarsh:2013xza,Hindmarsh:2015qta,Hindmarsh:2017gnf,Cutting:2019zws}, and the sound shell model~\cite{Hindmarsh:2016lnk,Hindmarsh:2019phv} is proposed to explain it. The results show that both the sound shell thickness and the bubble size at the collision time are imprinted on the \gls{SGWB} spectrum. At later times, vortical motion of the fluid can be produced
by, for example, the non-linearities of the fluid or the presence
of primordial magnetic fields.
In this case, \gls{MHD} turbulence would develop in the primordial plasma.
This additional source of \glspl{GW} is presented in the next subsection.

The aforementioned simulations solve the system of the fluid variables and the scalar field on a lattice, coupling both sectors via a phenomenological friction parameter. On the other hand, taking into account the large hierarchy between the scale of bubbles of cosmological size (with a sizable fraction of the Hubble radius) and that of the Higgs field, which varies within the bubble thickness (of the electroweak scale or higher), the effects of the Higgs field can be considered local to the bubble.
Therefore, one can integrate out the dynamics of the Higgs field and treat the expansion of the true vacuum as a space- and time-dependent background, thus removing numerical difficulties associated with resolving the scalar field in the fluid dynamics, which can then be solved separately.
Such a ``Higgsless'' approach has been pursued in Refs.~\cite{Jinno:2020eqg,Jinno:2022mie}.
In these studies, shock waves in the plasma are also resolved with an appropriate numerical scheme~\cite{Kurganov:2000ovy} and are accurately incorporated into the simulation.

The \gls{SGWB} spectrum from sound waves is well described by the \gls{DBPL} of \cref{eq:DBPL} that encodes the typical fluid shell thickness and the bubble size.
In particular, the frequency breaks are found to be~\cite{Jinno:2022mie,Caprini:2024gyk}
\begin{align}
f_1 & \simeq 0.2 \, H_{*,0} \, (H_* R_*)^{-1}\,, &
f_2 & \simeq 0.5 \, H_{*,0} \, \Delta_w^{-1} \,
(H_* R_*)^{-1} \,,
\label{eq:sw_shape}
\end{align}
where $\Delta_w = \xi_{\rm shell}/\max(\xi_w, c_s)$, with $\xi_{\rm shell}$ being the dimensionless sound shell thickness and $c_s$ the speed of sound.
$H_{*,0}$ is given in \cref{redsh_H}.
For the integrated amplitude $\Omega_{\rm int}$, the parametric dependence is \cite{Jinno:2022mie,Caprini:2024gyk}
\begin{align}
h^2 \Omega_\mathrm{int}
&=
h^2 F_{\GW,0} \, A_\text{sw}\, K^2 \left( H_* \tau_{\rm sw} \right) \left( H_*R_* \right),
\label{eq:sw_amplitude}
\end{align}
where $F_{\GW, 0}$ is given in \cref{FGW0}.
Here, $\tau_{\rm sw}$ is the duration of the sound wave source and $R_*$ the average bubbles size, estimated as~\cite{Caprini:2019egz}
\begin{equation}
H_* R_* = (8 \pi)^{1/3} \max(\xi_w, c_s) \,\frac{H_*}{\beta}\, .
\label{eq:Rstar}
\end{equation}
The constant $A_\text{sw} \simeq 0.11$ in \cref{eq:sw_amplitude}
is computed 
from the average over the simulation data points with different values of $\alpha$ and $\xi_w$ \cite{Jinno:2022mie,Caprini:2024gyk}. The kinetic energy fraction is $K \simeq 0.6 \, \kappa \,\alpha / (1 + \alpha)$, where $\kappa$ denotes the kinetic energy fraction of a single expanding bubble, derived in Ref.~\cite{Espinosa:2010hh} for a perfect fluid with an ultra-relativistic equation of state, i.e.\ with speed of sound~$c_s^2 = 1/3$ (see Refs.~\cite{Giese:2020rtr,Giese:2020znk} for a generalization). 
The factor 0.6 accounts for the efficiency in producing kinetic energy in the bulk fluid motion with respect to the single bubble case, as found in Ref.~\cite{Jinno:2022mie} (see discussion in Ref.~\cite{Caprini:2024gyk}).
It constitutes a conservative estimate with respect to what was assumed, e.g.~in Ref.~\cite{Caprini:2019egz}.

The dependence of the other parameters entering the sound waves \gls{SGWB} spectrum in \cref{eq:sw_shape,eq:sw_amplitude} is as follows.
The dimensionless shell thickness is calculated from the 
profile of an expanding bubble as $\xi_{\rm shell} = \xi_{\rm front} - \xi_{\rm rear} = |\xi_w-c_s|$, where the last equality holds for subsonic deflagrations and detonations, the latter being the cases we focus on in this work, as we set $\xi_w$ close to unity.
The relative duration of the \gls{GW} source,
$H_* \tau_{\rm sw} = {\rm min} \left[ H_* \tau_{\rm sh}, 1 \right]$, corresponds either to
the estimated decay time into turbulence, $H_* \tau_{\rm sh} = H_* R_*/\sqrt{\bar v_f^2}$, 
if the latter is shorter than the Hubble time \cite{Caprini:2019egz}; or to the Hubble time itself. 
The first case is expected for relatively strong transitions corresponding to large bulk fluid velocities
\cite{Ellis:2020awk}. Note that we will nevertheless adopt the average fluid velocity expression $\bar{v}_f^2 =\Gamma^{-1} \, K$ valid in
the subrelativistic bulk velocity limit, where $\Gamma = \bar w/\bar e$ is the mean adiabatic index, which is $\Gamma = 4/3$ for a radiation fluid~\cite{Caprini:2019egz}.
In the second case, i.e.~when the sound source duration is comparable to the Hubble time,
the scaling of the \gls{SGWB} spectral amplitude becomes $\Omega_\mathrm{int} \propto (H_* R_*) \propto (H_*/\beta)$, 
in contrast to $\Omega_{p} \propto (H_*/\beta)^2$ for bubble collisions, indicating that the sourcing lasts longer for sound waves.

For the exponents we use the result of Ref.~\cite{Jinno:2022mie} and take $n_1 = 3$, $n_2 = 1$, $n_3 = -3$, $a_1 = 2$, and $a_2 = 4$, cf.\ \cref{tab:slopes}.
The amplitude at $f_2$ is then related to the integrated amplitude, \cref{eq:sw_amplitude}, via
\begin{equation}
    \Omega_2 = \frac{1}{\pi}\left( \sqrt{2} + \frac{2\,f_2/f_1}{1+f_2^2/f_1^2}\right) \Omega_{\rm int}
    \approx 0.55\,\Omega_{\rm int}\,,
    \label{eq:Om2_sound}
\end{equation}
where the approximate sign holds for $\xi_w\approx 1$. The \gls{SGWB} spectrum is then completely fixed once the thermodynamic free parameters $\vec{\theta}_{\rm Cosmo}=\{ K, H_*R_*, \xi_w, T_*\}$ are set. 
The dependence of the \gls{SGWB} spectrum on the input parameters $K$, $H_* R_*$, $\xi_w$, and $T_*$ is illustrated in \cref{fig:pars_PT_sw}.

\subsubsection{Magnetohydrodynamic turbulence}
\label{sec:mhd_template}

The source of \glspl{GW} from \gls{MHD} turbulence is a
combination of the anisotropic stresses produced by velocity and magnetic fields. 
Following Ref.~\cite{Caprini:2019egz},
we insert an unknown parameter $\varepsilon$ to represent the fraction of overall kinetic energy 
in bulk motion that is converted to \gls{MHD} turbulence. 
We define the parameter
\begin{equation}
    \Omega_s =\OmK + \OmM = \varepsilon K \,,
    \label{Oms_turb}
\end{equation}
representing the total energy density fraction in the \gls{GW} source
from turbulence, where $\OmM$ and $\OmK$ are the corresponding magnetic (mag) and kinetic (kin) energy density fractions.
Furthermore, we assume that the turbulence is
fully developed at the beginning
of the sourcing process and that
equipartition between the turbulent kinetic and magnetic energies is reached \cite{Banerjee:2004df}, $\OmM \approx \OmK$, so that
\begin{equation}
    \bar{v}_f^2 \approx \vA^2 \approx \Gamma^{-1} \varepsilon\, K
    = \frac{3}{4}\,\Omega_s\,,
\end{equation}
where $\bar v_f$ is the characteristic fluid velocity and \vA is the Alfv\'en speed, both approximated
in the subrelativistic limit.
The turbulence \gls{SGWB} spectral peak is located at the energy injection scale, which we assume to be the bubbles separation at the end of the \gls{PT}, $R_*$ \cite{Caprini:2009yp}.
The characteristic length scale of the \gls{MHD} turbulence relative to the Hubble scale at the moment of the \gls{GW} sourcing determines the frequency at which the \gls{SGWB} spectrum peaks, $f_p \simeq
1.4\,H_{*,0} \, (H_* R_*)^{-1}$
\cite{RoperPol:2022iel,RoperPol:2023bqa}, where $H_{*,0}$ and $H_* R_*$ are given by~\cref{redsh_H,eq:Rstar}, respectively.
Another important parameter in the model is the duration of the \gls{GW} 
source, which can be inferred from \gls{MHD} simulations and is of
the order of a few eddy turnover times, ${\cal N} \tau_{\rm e}=\, {\cal N}  R_*/[2 \pi \max(\bar{v}_f, \vA)] \simeq {\cal N}  R_*/(2 \pi \vA)$, with $\mathcal{N}\simeq 2$ \cite{RoperPol:2022iel,RoperPol:2023bqa} (note that the definition of $\tau_{\rm e}$ differs from the one in Ref.~\cite{RoperPol:2023bqa} by a factor $1/2\pi$).

\begin{figure}
    \centering
    \includegraphics[width=.5\textwidth]{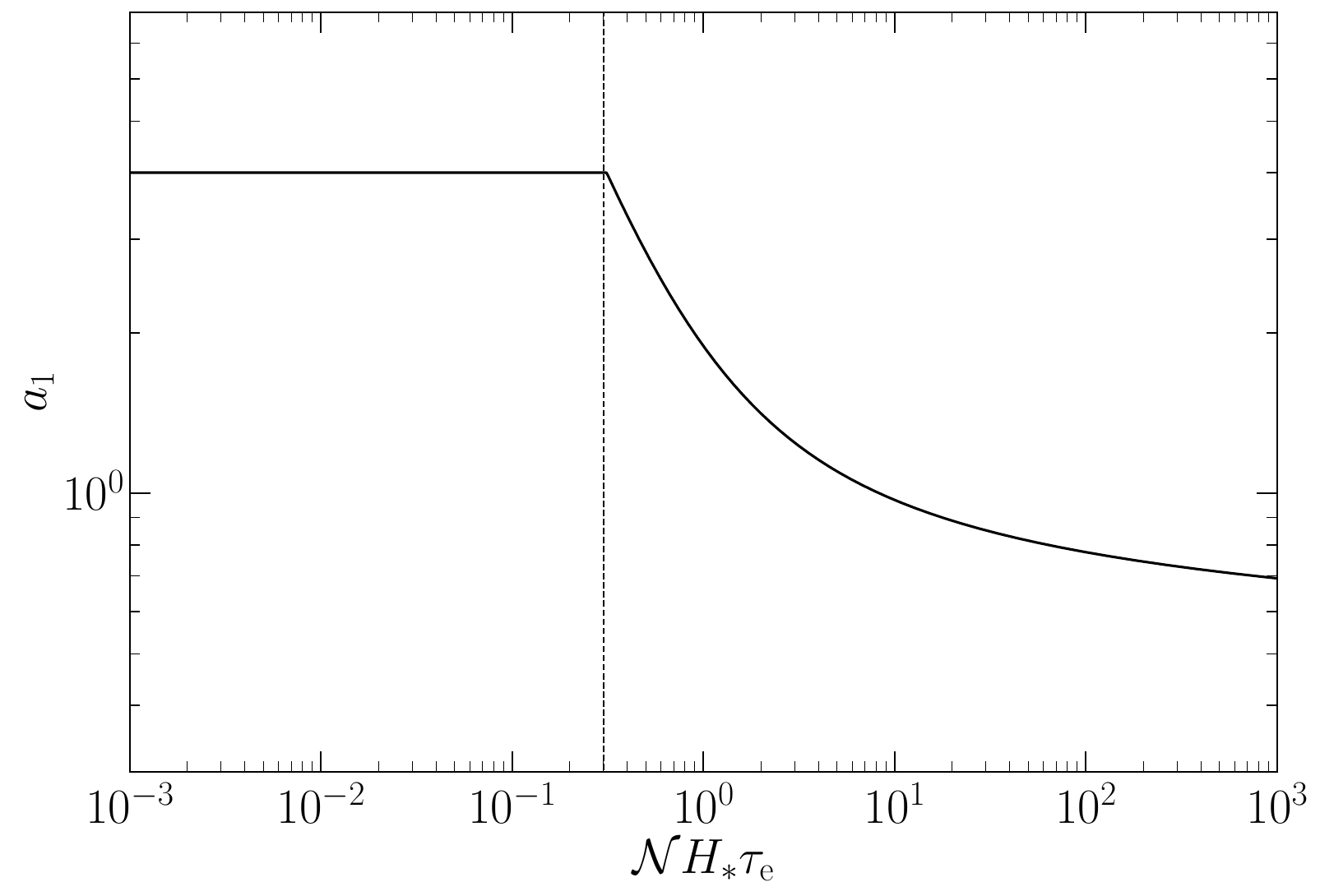}
    \caption{Smoothness parameter $a_1$ of the \gls{MHD} turbulence \gls{SGWB} as a function of the normalised eddy turnover time. It is set to 4 for values below the dashed vertical line corresponding to ${\cal N} H_* \tau_{\rm e} = 0.3$ as larger values of $a_1$ do not significantly influence the \gls{DBPL}
    template.}.
    \label{fig:alpha_tau}
\end{figure}

The \gls{SGWB} from \gls{MHD} turbulence can be
written as \cite{RoperPol:2022iel,RoperPol:2023bqa}
\begin{equation}
\Omega_\GW^{\MHD}(f, \vec \theta_{\rm Cosmo}) = 3 \, {\cal A} \, \Omega_s^2
\, F_{\GW}\,   (H_*R_*)^3\, (f/H_{*,0})^3\, p_{\tilde\Pi}(f,H_*R_*)\,\mathcal{T} (f,\mathcal{N}{H_*}\tau_{\rm e}),
\end{equation}
where the amplitude is\footnote{For simplicity and to take into account the average
of the \gls{SGWB} spectrum over oscillations in time, the parameter ${\cal A}$ here corresponds to ${\cal C}/(2 {\cal A}^2)$ in Ref.~\cite{RoperPol:2022iel}, similar to Ref.~\cite{RoperPol:2023bqa}.} ${\cal A} \simeq 0.085$, the spectrum of the anisotropic stresses can be approximated as \cite{RoperPol:2023bqa}
\begin{equation}\label{f2_turbulence}
    p_{\tilde \Pi} 
    (f,H_*R_*)= \biggl[1 + \biggl(\frac{f}{f_2}
    \biggr)^{a_2} \biggr]^{-\frac{11}{3 a_2}},
    \quad \text{\ with \ }
    f_2 \simeq 2.2 \, \frac{H_{*,0}}{H_*R_*}
    \text{\ \, and \ }a_2 \simeq 2.15,
\end{equation}
and the function ${\cal T}$ is \cite{RoperPol:2022iel,RoperPol:2023bqa}
\begin{eqnarray}
   \mathcal{T} (f,{\cal N} H_*\tau_{\rm e})= 
   \begin{cases}
      \ln^2 \bigl[1+\mathcal{N}{H}_*\tau_{\rm e}\bigr]\,, & \text{if } 
      \  2 \pi f/H_{*,0} \leq ({\cal N} H_*\tau_{\rm e})^{-1}, \\
      \ln^2 \bigl[1+H_{*,0}/(2 \pi f)\bigr] \,, &   \text{if } 
      \  2 \pi f/H_{*,0} > ({\cal N} H_*\tau_{\rm e})^{-1}\,.
   \end{cases}
\end{eqnarray}
The function ${\cal T}$ can be approximated to a \gls{BPL} as
\begin{equation}
    {\cal T} (f, {\cal N} H_*\tau_{\rm e}) = 
    \biggl(\frac{H_{*,0}}{2 \pi f_1}
    \biggr)^2 \biggl[1 + \biggl(\frac{f}{f_1}\biggr)^{a_1
    } \biggr]^{-\frac{2}{a_1}},
\end{equation}
where the position of the break~$f_1$
and the smoothness parameter~$a_1$, shown
in \cref{fig:alpha_tau},
depend on the eddy turnover time,
\begin{equation}
\label{eq:f1_turbu}
    f_1 ({\cal N} \tau_{\rm e}) = \frac{H_{*,0}}{2\pi} \ln^{-1} \bigl(
    1 + \, {\cal N} H_* \tau_{\rm e} \bigr)
    \approx 
    \frac{H_{*,0}}{2 \pi {\cal N} H_*\tau_{\rm e}} = \frac{\vA}{{\cal N}} \, \frac{H_{*,0}}{H_* R_*}\,.
    \quad 
\end{equation}
In the above equation, we have
assumed that ${\cal N} H_* \tau_{\rm e}
= {\cal N} {H}_* R_*/(2 \pi \vA) \ll 1$: this holds in general whenever
the \gls{MHD} turbulence \gls{SGWB} is large enough to
be observable, 
i.e.~in the regime where $\vA \sim \sqrt{\Omega_s}$ is large, meaning that the \gls{PT} is strongly first order (large $\alpha$). In this limit, the parameter $a_1$, can be approximated as a
constant, $a_1 = 4$, for ${\cal N} H_* \tau_{\rm e} \lesssim 0.3$ 
(or, equivalently, for $H_* R_* \lesssim \vA$ since ${\cal N} \simeq 2$).

The resulting \gls{MHD} turbulence \gls{SGWB} can be expressed 
in terms of the \gls{DBPL} model of \cref{eq:DBPL} setting
$n_1 = 3$, $n_2 = 1$, $n_3 = -8/3$,
$a_1 = 4$, and $a_2 \simeq 2.15$,
\begin{align}\begin{aligned} \label{Omega_GW_MHD}
    \Omega_\GW^{\MHD} (f, \vec \theta_{\rm Cosmo}) =  \frac{3 \, {\cal A} \, \vA \,
    \Omega_s^2 \, ({H}_* R_*)^2}{4 \pi^2 {\cal N}} &\,
    F_{\GW, 0} \\ \times
    \biggl(\frac{f}{f_1}\biggr)^3  &\, \biggl[1 + 
    \biggl(\frac{f}{f_1}\biggr)^{a_1}\biggr]^{-\frac{2}{a_1}} \biggl[1 + \biggl(\frac{f}{f_2}\biggr)^{a_2}\biggr]^{-\frac{11}{3 a_2}}.
\end{aligned}\end{align}
Note that expressed in this form, the amplitude of the \gls{SGWB} seems to be proportional to
$\vA \Omega_s^2 \sim \Omega_s^\frac{5}{2}$.
This is actually not the case, because
$f_1$ is a function of $\vA$.
We find from \cref{f2_turbulence,eq:f1_turbu} that the ratio between the frequency breaks is 
\begin{equation}
\label{eq:f1f2_turb}
    \frac{f_2}{f_1} = \frac{2.2 \, {\cal N}}{\vA}
    \simeq \frac{2.5 \, {\cal N}}{\sqrt{\Omega_s}} \gg 1.
\end{equation}
This ratio is in general large since $\vA < 1$, so that $f_2/f_1 \gtrsim 5$ with ${\cal N} \simeq 2$.
This property allows us to find the amplitude $\Omega_2$ used in the \gls{DBPL} template in \cref{eq:DBPL}.
At frequencies $f/f_1 \gg 1$, the \gls{SGWB} spectrum can be approximated 
by a single \gls{BPL},
\begin{equation}
    \Omega_\GW^{\MHD} (f \gg f_1, \vec \theta_{\rm Cosmo}) = \frac{3 \, {\cal A} \, \vA \, \Omega_s^2 \, (H_* R_*)^2}{4 \pi^2 {\cal N}} F_{\GW, 0} \biggl(\frac{f}{f_1}\biggr) \biggl[1 + \biggl(\frac{f}{f_2}\biggr)^{a_2} \biggr]^{-\frac{11}{3 a_2}}\,,
\end{equation}
from which one can derive
\begin{equation}\label{OmGW0_MHD}
    h^2 \Omega_2 = h^2 F_{\GW, 0} \, A_{\MHD} \, \Omega_s^2 \, 
    (H_* R_*)^2\,,
\end{equation}
with $A_{\MHD} = 3 \times 2.2 \, {\cal A}/(4 \pi^2) \times 2^{-11/(3a_2)} \simeq \num{4.37e-3}$.
Here the 
correct physical scaling $\Omega_2 \sim \Omega_s^2$ is recovered.
Finally, using the value of $\Omega_2$ in \cref{OmGW0_MHD} and the expression for the \gls{SGWB} in \cref{Omega_GW_MHD}, we can reproduce the \gls{DBPL} template in \cref{eq:DBPL} with
 \begin{equation}
     S_2 (f) = 2^\frac{11}{3 a_2} \biggl(\frac{f_1}{f_2}\biggr)
     \biggl(\frac{f}{f_1}\biggr)^3 \biggl[1 + \biggl(\frac{f}{f_1}\biggr)^{a_1}\biggr]^{-\frac{2}{a_1}} \biggl[1 + \biggl(\frac{f}{f_2} \biggr)^{a_2} \biggr]^{-\frac{11}{3 a_2}}\,.
 \end{equation}

The peak of the \gls{SGWB} spectrum in the template of \cref{eq:DBPL} is 
not located exactly at $f_2$.
Under the assumption $f_p/f_1 \sim f_2/f_1 \gg 1$, we can find the position of the
spectral peak $f_p
    = f_2 (-n_2/n_3)^{1/a_2} \simeq 0.6 f_2 \simeq 1.4 \, H_{*,0} \,
    (H_* R_*)^{-1}$ \cite{RoperPol:2022iel,RoperPol:2023bqa}, and the \gls{SGWB} amplitude at the peak using \cref{peak_single_bpl}, $\Omega_p \simeq 1.2 \, \Omega_2$.

To wrap up, the parameters that characterise the \gls{SGWB}
spectrum from \gls{MHD} turbulence are $\Omega_s$, $H_*R_*$,
and $T_*$,
from which we can derive the free geometric parameters of
the \gls{DBPL},
\begin{align}\begin{gathered}
    f_1 = \frac{\sqrt{3 \, \Omega_s}}{2\, {\cal N}} \, H_{*,0} \, (H_* R_*)^{-1}, 
    \quad 
    f_2 \simeq 2.2 \, H_{*,0} \, (H_* R_*)^{-1},
    \\ h^2 \Omega_2 = h^2 F_{\GW, 0} \, A_{\MHD}
    \, \Omega_s^2 \, (H_* R_*)^2 \,,
\label{geom_pars_turb}
\end{gathered}\end{align}
with the exponents of the \gls{DBPL} summarised in \cref{tab:slopes}.
Note that $\Omega_s$ and $H_* R_*$ can be related to the \gls{PT} parameters using \cref{Oms_turb,eq:Rstar}, and the  relation $K \simeq 0.6 \, \kappa \,\alpha / (1 + \alpha)$ given in \cref{sec:sw_template}.

The parameter dependence on $\Omega_s$, $R_*H_*$, and $T_*$ of the spectrum is depicted in \cref{fig:pars_PT_turb}. 
In realistic scenarios of first-order \glspl{PT}, the full \gls{SGWB} spectrum consists of the sum of both contributions due to the fluid motion, sound waves and turbulence, in which case it can be parameterised in terms of $\vec{\theta}_{\rm Cosmo}=\{ K, R_*H_*, \xi_w, T_*, \varepsilon\}$, as illustrated in \cref{fig:pars_PT_sw_turb}.

\section{Parameter reach and reconstruction}
\label{sec:SGWBinner}

The \texttt{SGWBinner} code has been originally designed to carry out agnostic searches and reconstructions of the primordial \gls{SGWB} signal at \gls{LISA}~\cite{Caprini:2019pxz, Flauger:2020qyi}, and used to quantify the instrument capabilities~\cite{Seoane:2021kkk, Colpi:2024xhw}. In such agnostic searches, the code 
reconstructs the primordial signal as power laws in multiple bins of optimal widths, while using templates only for the astrophysical foregrounds. The agnostic search is flexible but sub-optimal when a precise template of the primordial signal is available.
For this work (and the companion papers~\cite{Blanco-Pillado:2024aca, Braglia:2024kpo}), we have developed a new functionality of the \texttt{SGWBinner}: a template-based analysis of the primordial signal. 
We present it below, starting with a review of the foundations of the \gls{LISA} measurement. We also initiate a template bank within the code, including all the signals discussed in the previous section,  preparing the tools to facilitate SGWB searches of cosmological first-order \glspl{PT} in \gls{LISA}.

\subsection{LISA measurement channels}

\Gls{LISA} consists of three satellites, $i=1,2,3$, with relative distances $L_{ij}$. At every time~$t$, the satellite~$i$ emits and receives a laser beam to and from the satellite~$j\neq i$, and measures their phases. Due to the laser frequency instabilities, the direct comparison of the phase measurements is too noisy to perform interferometry sensitive to the expected \gls{GW} signals. However, this noise contribution is largely suppressed by post-processing the data and performing \gls{TDI}~\cite{Tinto:2020fcc}, i.e.~a comparison of the phase measurements at opportune different times.  
Different \gls{TDI} approaches exist~\cite{Tinto:1999yr, Prince:2002hp, Shaddock:2003bc, Shaddock:2003dj, Tinto:2003vj, Vallisneri:2005ji, Muratore:2020mdf, Muratore:2021uqj,Hartwig:2021mzw}. Here we consider the ``first generation'' one, which works under the assumption that the \gls{LISA} satellites are in an equilateral configuration (i.e.~$L_{ij}=L$ for any $i,j$) and have identical noise levels.\footnote{The robustness of the results obtained in this simplified approach against the relaxation of both these assumptions has been investigated in Ref.~\cite{Hartwig:2023pft}. Such work demonstrated that, for a power law signal, a non-equilateral \gls{LISA} configuration and the inclusion of up to twelve noise parameters (one per each \acrshort{TM}/laser noise component) do not affect the reconstruction of the signal parameters significantly. Further studies considering different signal templates, e.g.~the ones considered in the present work, are required to generalise this statement.} In the equilateral approximation, the laser phase emitted from the satellite $j$ at time $t-L/c$ is recorded at time $t$ in satellite $i$, with~$c$ being the light speed. 
We denote the phases $\eta_{ij}(t)$. Three measurements (suppressing the frequency instabilities) can be recast from these phases~\cite{Prince:2002hp},
\begin{equation}
    {\rm A} = \frac{{\rm Z} - {\rm X}}{\sqrt{2}}\;, \qquad  
    {\rm E} = \frac{{\rm X} - 2 {\rm Y} + {\rm Z}}{\sqrt{6}} \;, \qquad  
    {\rm T}=\frac{{\rm X} + {\rm Y} + {\rm Z}}{\sqrt{3}}  \; ,
    \label{eq:basis_rotation}
\end{equation}
where 
\begin{equation}\label{eq:tdi-definition}
	{\rm X}  = (1 - D_{13}D_{31})(\eta_{12} + D_{12} \eta_{21}) + (D_{12}D_{21} - 1)(\eta_{13} + D_{13} \eta_{31}) \; ,
\end{equation}
and the Y and Z variables are obtained through cyclic permutations of the indices. Here, $D_{ij} \eta_{kl}(t) = \eta_{kl}(t - L_{ij}/c)$ is the one-arm length delay operator. 
Remarkably, in the above approximations, the channels A, E, and T constitute three virtual, orthogonal \gls{GW} interferometers, with the former two being identical and the latter being a (quasi-)null channel, i.e.~a channel whose sensitivity to the signal is largely suppressed. The \texttt{SGWBinner} simulates and analyses the data in these three channels, and leverages the T channel property to reconstruct the satellite noise sources.

\subsection{Data in each channel}
\label{sec:data_in_each_channel}

The data collected via the A, E, and T channels will contain all the transient and non-transient signals.
This includes individually resolvable \gls{GW} emitting events
and a stochastic component, given by
the sum of all unresolved astrophysical sources, of the stochastic instrumental noise components, and of possible \gls{SGWB} signals. 
The \texttt{SGWBinner} simulates data streams of the stochastic component and forecasts 
posteriors for the parameters of its aforementioned
contributions, under the simplifying assumption that their correlations with the transient signals and transient noises are negligible. Thus, for our purpose, the data $\tilde d_i$ in the channel $i={\rm A,E,T}$ reduce to
\begin{equation}
 \tilde{d}_i(t) = \sum_\nu \tilde{n}^\nu_i(t) + \sum_\sigma \tilde{s}^\sigma_i(t) \; ,
 \label{eq:data_time_domain}
\end{equation}
where the first sum includes the stochastic noise components, $\tilde{n}^{\nu}_i$, and the second sum runs over the signal contributions, $\tilde{s}_i^\sigma$. 
As we will clarify in \cref{sec:noise_and_signals}, the noise components can be summarised in two contributions (from the test masses and the optical measurement systems), while we will focus on three signal contributions (two foregrounds and one primordial \gls{SGWB}).

For simplicity, we assume all dominant noise sources to be practically stationary and Gaussian, with zero mean.  Moreover, we expect the noise produced by a source common to different channels to create negligible correlations after \gls{TDI} operations. 
In this way, the components $\tilde{n}^{\nu}_i$
effectively  are independent, random, and Gaussian variables. 
In particular, $\tilde{n}^{\nu}_A$ and $\tilde{n}^{\nu}_E$ just differ by 
their statistical realization, while $\tilde{n}^\nu_A(t)$  (or $\tilde{n}^\nu_E(t)$) also differs from $\tilde{n}^\nu_T(t)$ due to their response function. 
Indeed, for identical satellites in an equilateral configuration,  every satellite has statistically equivalent noise sources; furthermore, A and E have the same response functions, but A (or E) and T do not.

We also treat $\tilde{s}^\sigma_i(t)$ as independent, (zero-mean) Gaussian, and stationary variables. While this choice simplifies and speeds up the data simulation and analysis, it is suboptimal. Indeed, among our stochastic signals, the Galactic foreground is anisotropic
in the \gls{CMB} frame, yielding 
a non-stationary signal in the data due to the yearly rotation of the \gls{LISA} satellites' plane.  Averaging this non-stationary signal over several years effectively converts it to a broader Gaussian contribution that can be treated as stationary. This should not bias the whole parameter reconstruction but, at least in principle, should be less precise than an analysis including the yearly variability (cf.~\cref{sec:noise_and_signals} for details on the averaged Galactic foreground).

Within these assumptions, we Fourier transform the data in \cref{eq:data_time_domain} from time to frequency domain,
\begin{equation}
	d_i(f) = \int_{-{\tau}/2}^{\tau/2} \tilde d_i(t) \; \textrm{e}^{- 2 \pi i f t }  \dd t\; ,
\label{eq:FT}
\end{equation}
with statistical properties
\begin{equation}
    \label{eq:data_statistics}
	\langle  d_i(f) \rangle  = 0  \; ,  \qquad 
    \langle d_i(f) d^{*}_j(f^{\prime})  \rangle  = \delta_{ij} \frac{\delta(f -f')}{2} \left[  P_{N,ii}(f)  +  P_{S,ii}(f) \right]   \; , 
\end{equation}
where the brackets denote the ensemble average and $\tau$ is the time over which the Fourier transform is performed (as will be clarified in \cref{sec:data_gen_sgwbinner}).\footnote{In \cref{eq:data_statistics}, the Dirac delta only arises in the limit of infinite observation time. It is however a good approximation in the frequency regime $f \ll 1/\tau$ we are interested in. The Kronecker delta instead accounts for the orthogonality of the A, E, and T channel basis emerging in the case of three identical satellites in perfectly equilateral configuration.}
Moreover, $P_{N,ii}$ 
and $P_{S,ii}$ are the noise and signal power spectra. They are respectively the sum of the power spectra of each noise source $\nu$ and each
signal $\sigma$ recorded in the $i$-$i$ correlator. $P_{N,ii}$ can be expressed in terms of the $i$-channel response function  $T^\nu_{ii}$ relative to the strain $S^{\nu}_{\rm N}$ of the noise sources $\nu$, reading as 
\begin{equation}
    P_{N,ii}(f)=\sum_\nu T^{\nu}_{ii}(f) S_N^{\nu}(f)       \;.
\end{equation}
The same can be done for $P_{S,ii}$, with the peculiarity that all
signal strain sources share the same response function:
\begin{equation}
    P_{S,ii}
    = 
    R_{ii}(f) \sum_\sigma S_S^{\sigma}(f) 
    =
    \frac{3H_0^2}{4\pi^2 f^3} R_{ii}(f) \sum_\sigma h^2 \Omega_\GW^\sigma (f)\,,
    \label{eq:P_Sii}
\end{equation}
with $R_{ii}$ denoting the response function to the signals and $S_{S}^\sigma$ the strain of the \glspl{SGWB}.

For forecast studies and our level of approximations, it is reasonable to neglect the uncertainties in the response functions. For $T_{ii}^\nu$ and $R_{ii}$ we thus adopt the response functions already implemented in the \texttt{SGWBinner} code, where all parameters have known values so that there is no need to fit for them. The expressions of $T_{ii}^\nu$ and $R_{ii}$ can be found in \cref{app:response}.\footnote{For studies on the response functions and their uncertainties see, e.g.~Refs.~\cite{Flauger:2020qyi, Hartwig:2021mzw, Nam:2022rqg}.}  

\subsection{Injected (and reconstructed) noises and signals}
\label{sec:noise_and_signals}

We now describe the noise and signal source strains we implement in \texttt{SGWBinner} for the present work. Regarding the signals, we only consider the astrophysical foregrounds whose presence in the \gls{LISA} data is guaranteed from other observations, namely the Galactic foreground and the one from \glspl{SOBHB}. At the same time, we omit any potential cosmological \gls{SGWB} besides the \gls{PT} one.

\begin{description}
    \item[Noise:] \Gls{LISA} performs its measurements by monitoring the relative distances of the \glspl{TM} inside the satellites. 
    The statistical errors in the \gls{GW} detection are then primarily due to perturbations of the \gls{TM} free fall regime, and uncertainties in the measurement of the \gls{TM} relative positions. This qualitatively explains a well-investigated result: the dominant noise strains $S_N^\nu$ are given by the \gls{TM} and the \gls{OMS} strains \cite{LISA:2017pwj,Babak:2021mhe}
        \begin{align}
        \label{eq:Acc_noise_def}
        S^\TM_{ii}(f) & = A^2 \;  \left(1 + \left(\frac{\SI{0.4}{\milli\Hz}}{f}\right)^2\right)\left(1 + \left(\frac{f}{\SI{8}{\milli\Hz}}\right)^4\right) \left(\frac{1}{2 \pi f c}\right)^2 \;  \left( \frac{\si{\femto\m^2} }{ \si{\s^3} } \right) \;,  \\ 
        \label{eq:OMS_noise_def}
        S^\OMS_{ii}(f) & = P^2 \; \left(1 + \left(\frac{\SI{2e-3}{\milli\Hz}}{f}\right)^4 \right) 
        \left(\frac{2 \pi f}{c}\right)^2 \; 
        \left(\frac{ \si{\pico\m}^2 }{ \si{\Hz} }\right) \;,
    \end{align}
    where $S^{\TM}_{ii}$ models the noise relative to the non-free fall effects, and $S^{\OMS}_{ii}$ models the noise due to mismeasuring the \gls{TM} distances.\footnote{Such an effective description must be accompanied by opportune manipulation of the response functions; see \cref{app:response}
}

    In our simulation, we inject the noise with the nominal values $A=3$ and $P=15$. In the parameter estimation of the noise and signals we adopt Gaussian priors centered on these nominal values and with \SI{20}{\%} width~\cite{LISA:2017pwj}.
    
    \item[Extra-Galactic foreground:] It results from the superposition of the \glspl{SOBHB} and\linebreak \glspl{NSB} that are outside the Milky Way and that \gls{LISA} does not individually resolve. \Gls{LISA} is sensitive to these sources when they are in the inspiral phase. Because of the statistical properties of the \gls{SOBHB} and \gls{NSB} populations as well as the \gls{LISA} angular resolution, the extra-Galactic foreground in \gls{LISA} can be approximated as an isotropic signal with a power law frequency shape~\cite{Phinney:2001di, Regimbau:2011rp, Perigois:2020ymr, Babak:2023lro, Lehoucq:2023zlt}
    \begin{equation}
    \label{eq:Ext_template}
    h^2 \Omega^{\textrm{Ext}}_\GW (f)  = h^2 \Omega_{\rm Ext}  \left(\frac{f}{\SI{1}{\milli\Hz}}\right)^{2/3} \; . 
    \end{equation}

    The fiducial value and prior for $\Omega_{\rm Ext}$ can be obtained from the population inference analyses at terrestrial interferometers~\cite{KAGRA:2021duu}. By approximating one of the recent estimates~\cite{Babak:2023lro}, we use $\log_{10} (h^2 \Omega_{\rm Ext}) = -12.38 $ for the injections and a Gaussian prior around it, with $\sigma = 0.17$, for the parameter estimation.

    \item[Galactic foreground:]
    The \gls{CGB} population produces a loud signal in \gls{LISA} coming from the Galactic disk~\cite{Nissanke:2012eh}. After removing the \glspl{CGB} that \gls{LISA} can individually resolve, there remains a strong stochastic component which, in the \gls{LISA} satellite's frame, originates from a sky patch that moves with time and has a cycle of one year (i.e.~the yearly periodicity of the \gls{LISA} orbit around the Sun). Thus, due to the angular dependence of the response functions, the unresolved \gls{CGB} signal integrated over the whole sky dome exhibits a strength with a yearly modulation~\cite{Nissanke:2012eh,Boileau:2021sni}. To forecast the sensitivity to cosmological \glspl{PT}, one would ideally leverage the yearly modulation to better separate the astrophysical signal from the cosmological one, as done, e.g.~in Refs.~\cite{Boileau:2021sni,Boileau:2022ter}. This however requires data analysis techniques which are not yet implemented in the \texttt{SGWBinner}. In the following, we therefore only
    consider the whole sky-dome signal integrated over the total observing time of the mission, $T_{\rm obs}$. This assumption is a conservative one.
    For such a signal we adopt 
    the following template~\cite{Karnesis:2021tsh}:
    \begin{equation}
    \label{eq:SGWB_gal}
    h^2\Omega^{\textrm{Gal}}_\GW(f)=
     \frac{f^3}{2}\left(\frac{f}{\SI{1}{\Hz}}\right)^{-\frac{7}{3}}   
    \left[1+\tanh \left({\frac{f_{\textrm{knee}}-f}{f_2}} \right) \right] e^{-(f/f_1)^\upsilon} 
    h^2 \Omega_{\rm Gal} \;,
    \end{equation}
    where
    \begin{eqnarray}
    \log_{10} (f_1/\si{\Hz}) &=& a_1 \log_{10}(T_\mathrm{obs}/\si{\yr}) + b_1\,,\nonumber\\
    \log_{10} (f_{\textrm{knee}}/\si{\Hz}) &=& a_k \log_{10}(T_\mathrm{obs}/\si{\yr}) + b_k\,.
    \end{eqnarray}
    For the injection, we use the fiducial values $a_1 = -0.15$, $b_1=-2.72$, $a_k = -0.37$, $b_k=-2.49$, $\upsilon = 1.56$, $f_2 = \SI{6.7e-4}{\Hz}$ and $\log_{10} (h^2 \Omega_{\rm Gal})=-7.84$~\cite{Karnesis:2021tsh}. We assume all of them but $h^2 \Omega_{\rm Gal}$ to be known in the parameter estimation and adopt a Gaussian prior on $\log_{10} (h^2 \Omega_{\rm Gal})$ centered at $-7.84$ with width $\sigma = 0.21$.
    
    \item[\gls{PT} \gls{SGWB}:] We implement Template~I (i.e.~the \gls{BPL} representing the \gls{SGWB} from bubble collisions and highly relativistic fluid shells) and Template~II  (i.e.~the \gls{DBPL} representing the \gls{SGWB} from sound waves and \gls{MHD} turbulence), given respectively in \cref{sec:bubble_template,sec:dbpl}. 
    We implement the templates both in terms of the geometric parameters that characterise the spectral templates and of the thermodynamic parameters that describe the \gls{PT} and the subsequent fluid motion. The injection values and priors of the geometric and thermodynamic
    parameters are described in \cref{sec:reconstruction_geometric,sec:reconstruction}, respectively.
\end{description}

\subsection{Data generation and analysis}
\label{sec:data_gen_sgwbinner}
 
\Gls{LISA} is planned to operate for at least 4.5 years up to a maximum of 10 years. However, due to antenna repointing and similar operations, some periodic data gaps are scheduled, yielding a duty cycle of about \SI{82}{\%}~\cite{Colpi:2024xhw}. This would for instance correspond to interrupting the data acquisition for two days every two weeks.
We consider an intermediate scenario and set the \texttt{SGWBinner} code to work with $N_d=126$ data segments of duration $\tau=11.4$ days each (see \cref{eq:FT}), summing up to $T_{\rm obs} =4$ effective years of data.

Furthermore, the \texttt{SGWBinner} works in frequency domain.
We denote as $d^s_i(f_\textrm{k})$ the Fourier-transformed datum of the segment 
$s=1, ..., N_{d}$ in the bin of frequency $f_{\rm k}$ and spacing $\Delta f = 1/\tau\simeq \SI{e-6}{\Hz}$. 
To simulate the \gls{LISA} data, for every signal and noise component described in \cref{sec:noise_and_signals}, the \texttt{SGWBinner} code generates $N_d$ Gaussian realizations at each frequency bin $f_\textrm{k}$. These bins cover the whole \gls{LISA} band, which in this work we assume to be $[\num{3e-5},0.5]\,\si{\Hz}$, with the required resolution $\Delta f$.

The \texttt{SGWBinner} code then performs the analysis of the mock data.  
In our stationary approximation, one can analyse the average value at each frequency $f_\mathrm{k}$ after averaging over the $N_d$ segments. In practice, we compute ``segment-averaged'' data defined as $\mathcal{D}^\textrm{k}_{ii} \equiv \sum_{s = 1}^{N_d} d^s_i(f_\textrm{k}) d^s_i(f_\textrm{k}) / N_d$. In addition, the resolution $\Delta f$ is beyond the characteristic scale of the typical primordial \gls{SGWB} frequency shape, making the data file heavy and slow to manipulate. For this reason, the \texttt{SGWBinner} code computes a new data set $D^\mathrm{k}_{ii}$, 
defined on a coarser set of frequencies $\bar{f}^\mathrm{k}_{ii}$ statistically equivalent to the original data~\cite{Caprini:2019pxz, Flauger:2020qyi}. In practice, $\bar{f}^\mathrm{k}_{ii}$ and $D^\mathrm{k}_{ii}$ are computed through inverse variance weighting, and each of the new data points is associated with a weight $w^\mathrm{k}_{ii}$ that accounts for the number of high-resolution points (and their variance) falling into the coarse-grained bin.

For the parameter reconstruction, the code then probes the following likelihood:
\begin{equation}
\label{eq:likelihood}
\ln \mathcal{L} (\vec{\theta}\, ) = \frac{1}{3} \ln \mathcal{L}_{\rm G} (\vec{\theta} | D^\mathrm{k}_{ii}) +  \frac{2}{3} \ln \mathcal{L}_{\mathrm{LN}} (\vec{\theta} | D^\mathrm{k}_{ii}) \; ,
\end{equation}
with
\begin{equation}
\ln \mathcal{L}_{\rm G} (\vec{\theta} | D^\mathrm{k}_{ii}) = -\frac{N_d}{2} \sum_{\mathrm{k}} \sum_{i} w^\mathrm{k}_{ii} \left[ 1 - D^\mathrm{k}_{ii} / D^{\rm Th}_{ii} (\bar{f}^\mathrm{k}_{ii}, \vec{\theta}) \right]^2  \; ,
\end{equation}
\begin{equation}
\ln \mathcal{L}_{\rm LN} (\vec{\theta} | D^\mathrm{k}_{ii}) = -\frac{N_d}{2} \sum_{\mathrm{k}} \sum_{i} w^\mathrm{k}_{ii} \ln^2 \left[ D^{\rm Th}_{ii} (\bar{f}^\mathrm{k}_{ii}, \vec{\theta}) / D^\mathrm{k}_{ii}   \right] \; .
\end{equation}
Here $D^{\rm Th}_{ii} (\bar{f}^\mathrm{k}_{ii}, \vec{\theta}) $ denotes the data predicted in our theoretical model (i.e.~our overall template of the injected noises and signals), while $\vec{\theta} = \{\vec{\theta}_S, \vec{\theta}_N\} $ is the vector of the free parameters involved in the injected noises and signals' templates, 
namely  
$$\vec{\theta}_N=\{A,P\} \;, \qquad 
\vec{\theta}_S=\{
\log_{10} (h^2 \Omega_{\rm Gal}),
\log_{10} (h^2 \Omega_{\rm Ext}),
\vec{\theta}_{\rm Cosmo}
\}\;,$$
where   
$\vec{\theta}_{\rm Cosmo}$ denotes the thermodynamic or geometric parameters of the \gls{PT} templates.
Notice that $\ln \mathcal{L}_{\rm LN}$ is introduced to take into account the mild non-Gaussianity that the coarse graining introduces~\cite{Bond:1998qg, Sievers:2002tq, WMAP:2003pyh, Hamimeche:2008ai} and the Gaussian likelihood $\ln \mathcal{L}_{\rm G}$ does not properly treat. 
Moreover, the priors for the noises and \gls{SGWB} signals should be added to~\cref{eq:likelihood} to obtain the posterior distributions, which the \texttt{SGWBinner} code determines and visualizes by calling \texttt{Polychord}~\cite{Handley:2015vkr, Handley:2015fda}, \texttt{Cobaya}~\cite{Torrado:2020dgo}, and \texttt{GetDist}~\cite{Lewis:2019xzd}.

We finally remark that several forecast analyses use the \gls{SNR} as a rule of thumb to assess whether a primordial \gls{SGWB} can be detected. To help the comparison with these analyses, we will quote the \glspl{SNR} of the injected \gls{PT} signals. We compute it as~\cite{Romano:2016dpx}
\begin{equation}
\label{eq:SNR_def}
\SNR = \sqrt{T_\mathrm{obs} \;\sum_{i} \int \left( \frac{S_\GW^{\rm Cosmo}}{S_{i, \rm N}} \right)^2 \dd f }  \; ,
\end{equation}
where the integral is over the \gls{LISA} frequency range and $S_\GW^{\rm Cosmo}$ is related to $h^2 \Omega_\GW^{\rm Cosmo}$ as in \cref{eq:P_Sii} and  $\Omega_\GW^{\rm Cosmo}$ represents  $\Omega_\GW^{\BPL}$ and $\Omega_\GW^{\DBPL}$ depending on the particular case. For $S_{i, \rm N}$, include experimental noises defined in~\cref{eq:Acc_noise_def,eq:OMS_noise_def}.\footnote{It is also possible to include the astrophysical foregrounds, most importantly the galactic noise from \cref{eq:SGWB_gal}, as noise sources which would bring our SNR curves closer to reconstruction including foregrounds.}

We also perform a Fisher analysis, which corresponds to approximating the likelihood function ${\cal L}$ to be Gaussian.
The relation between $\Delta \chi^2$ and the Fisher information matrix is given by 
\begin{align}
\Delta \chi^2 & = \sum_{i,j}
F_{ij}
\Delta x_i \Delta x_j,
\label{eq:Fisher_geometrical}
\end{align}
where $\Delta x_i \equiv x_{i,{\rm postulated}} - x_{i,{\rm fiducial}}$ and $i,j$ run over the set of parameters. 
The Fisher information matrix~$F_{ij}$ is given by 
\begin{align}
    F_{ij}(\vec{x}_\mathrm{fiducial}) = \!\!\!\sum\limits_{\alpha=A,E,T}\!\!\! T_\mathrm{obs} \!\int \!\!\dd f\, \left.\frac{(\partial_{x_i} h^2\Omega_{\alpha}) (\partial_{x_j} h^2\Omega_{\alpha})}{(h^2\Omega_{\alpha})^2}\right|_{\vec{x}=\vec{x}_\mathrm{fiducial}} \,,
\end{align}
where $h^2 \Omega_{\alpha} = h^2 \Omega_\GW^{\rm Cosmo} + h^2 \Omega_\GW^\mathrm{Ext}  + h^2 \Omega_\GW^\mathrm{Gal} + h^2 \Omega_{N,\alpha} $
is the sum of the \glspl{SGWB}, the foregrounds, and the fractional energy density noise $h^2 \Omega_{N, \alpha} = \sum_\nu 4 \pi^2 f^3 S_{\alpha}^{\nu}/(3 H_0^2)$
with $\nu = \text{\gls{TM}}, \text{\gls{OMS}}$.
The sum runs over the \gls{TDI} AET channels. 
The projected relative uncertainty on the reconstruction of the parameters is then obtained from the diagonal values of the inverse of the Fisher information matrix,
$\Delta x_i / x_i = \Delta \ln x_i = \ln 10 \times \Delta \log_{10} x_i = \\ \ln 10 \times \sqrt{(F^{-1})_{ii}}$. Such errors represent the uncertainty on a given parameter after marginalizing over all the others.

\section{Reconstruction of the geometric parameters}
\label{sec:reconstruction_geometric}

In this section, we discuss the prospect for the parameter reconstruction by \gls{LISA} in terms of the geometric parameters, for the three templates presented in \cref{sec:models}. 
The geometric-parameter templates have the advantage of speeding up the likelihood exploration as their parametrisation does not feature degeneracies. 
On the other hand, their non-trivial connection to \gls{PT} parameters might lead to exploration of unphysical parts of the parameter space.

\subsection{Broken power law}

We start with very strong transitions, discussed in \cref{sec:bubble_template}, leading to a \gls{SGWB} from bubble collisions or highly relativistic fluid shells described by the \gls{BPL} spectrum in \cref{eq:BPL}. 
We fix the exponents to the values in \cref{tab:slopes}, so that the geometric parameters of the model are $\vec{\theta}_{\rm Cosmo} =(\log_{10} h^2\Omega_{b},\,\log_{10} f_b/\si{\Hz})$. In this case, we report the \gls{LISA} sensitivity to the geometric parameters using both Fisher and Bayesian analyses.

\begin{figure}
\includegraphics[width=\textwidth]{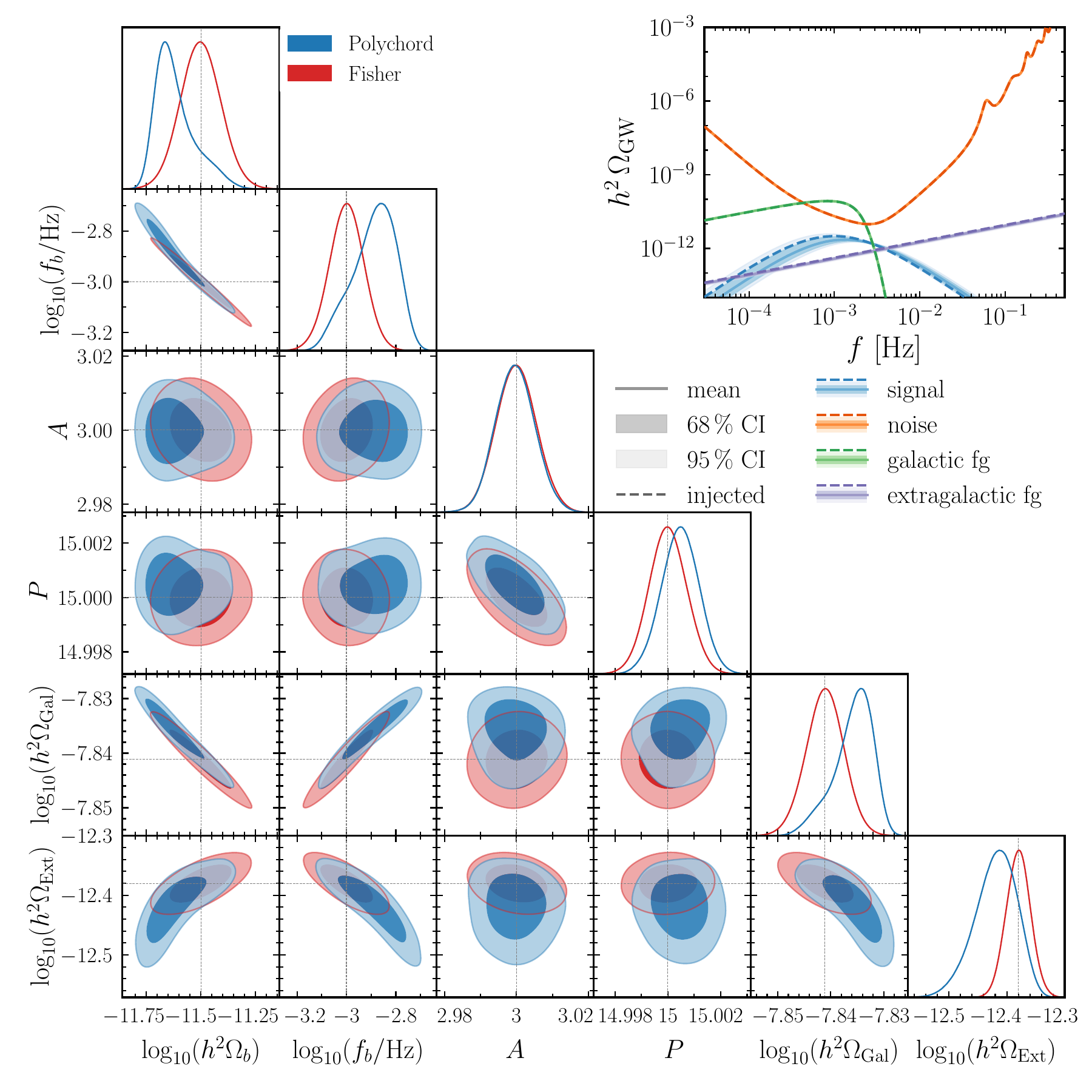}
    
\caption{
Triangle plot comparing the results of the Fisher analysis~(red) and \Polychord~(blue) for a \gls{BPL} spectrum corresponding to bubble collisions including noise and foregrounds. We fix the spectral exponents to the slopes for strong \glspl{PT} in \cref{tab:slopes} and choose an injected signal with break frequency  $f_b = 1$\,mHz and amplitude  $\log_{10}(h^2\Omega_b) = -11.5$.  The plot in the top right corner visualizes the injected signals and noises as well as their reconstructed \SI{68}{\%} (dark areas) and \SI{95}{\%} (light areas) \glspl{CL} obtained with \Polychord.}
\label{fig:BPL-fisher-polychord2}
\end{figure}

As an example of the results of these analyses, \cref{fig:BPL-fisher-polychord2} depicts the expected \SI{68}{\%} and \SI{95}{\%} \glspl{CL} of a \gls{BPL} signal from bubble collisions with a fiducial amplitude $h^2\Omega_b = \num{3e-12}$ and frequency $f_b = \SI{1}{\milli\Hz}$.
We have chosen these values as they provide a signal in the ballpark of the maximum sensitivity of \gls{LISA}.
The priors used for the noise and foreground parameters are listed in \cref{tab:priors_noise_fg}, while those for the \gls{SGWB} parameters are listed in \cref{tab:priors_spectral}, left column.
In addition to the signal reconstruction, \cref{fig:BPL-fisher-polychord2} also shows the posteriors of the noise parameters~$A$ and~$P$, as well as those  of the amplitudes of the Galactic and extra-Galactic binary foregrounds,
provided in \cref{sec:noise_and_signals}.
The outcome of the Fisher analysis is shown in red, whereas the blue contours depict the reconstruction using \Polychord.
The inlay in the upper right shows the corresponding \gls{SGWB} and foreground spectra, as well as the instrument sensitivity.
The dashed curves correspond to the injected spectrum, whereas the solid lines show the spectrum of the mean reconstructed parameters.
The shaded areas depict the \SI{68}{\%} and \SI{95}{\%} \glspl{CL} inferred from the spectra calculated from the parameter points in the posterior samples.
The parameters of this example of signal are reconstructed with an accuracy within a few percent.

\begin{table}
    \centering
    \begin{tabular}{cc|cc}
        \toprule
        parameter & $\mu\pm\sigma$ & 
        parameter & $\mu\pm\sigma$ \\ \midrule
        $A$ & $3 \pm 0.6$ &
        $\log_{10}(h^2\Omega_\text{Gal})$ & $-7.84 \pm 0.21$ \\
        $P$ & $15 \pm 3$ &
        $\log_{10}(h^2\Omega_\text{Ext})$ & $-12.38 \pm 0.17$ \\
        \bottomrule
    \end{tabular}
    \caption{%
        Mean~$\mu$ and standard deviation~$\sigma$ of the Gaussian priors used in \Polychord for the noise parameters~$A$ and~$P$ as well as the amplitudes of the Galactic~($\log_{10}(h^2\Omega_\text{Gal})$) and extra-Galactic~($\log_{10}(h^2\Omega_\text{Ext})$) foregrounds. The injected values correspond to the mean.
    }
    \label{tab:priors_noise_fg}
    \end{table}

\begin{table}
    \centering
    \begin{tabular}{lc|lc}
        \toprule
        parameter & prior range & parameter & prior range \\ \midrule
        $\log_{10}(h^2\Omega_b)$ & $(-30, -5)$ &
            $\log_{10}(h^2\Omega_2)$ & $(-30, -5)$ \\
        $\log_{10}(f_b/\si{\Hz})$ & $(-7, 1)$ &
            $\log_{10}(f_2/\si{\Hz})$ & $(-7, 1)$  \\
        & & $\log_{10}(f_2/f_1)$ & $(0, 2)$ \\
        \bottomrule
    \end{tabular}
    \caption{%
        Ranges of the uniform priors used in \Polychord for the parameters of the \gls{BPL}~(left) and \gls{DBPL}~(right).
        The spectral slopes are fixed to the values in \cref{tab:slopes} corresponding to the source under consideration.}
    \label{tab:priors_spectral}
\end{table}

\begin{figure}
\centering
\includegraphics[width=.7365\textwidth]{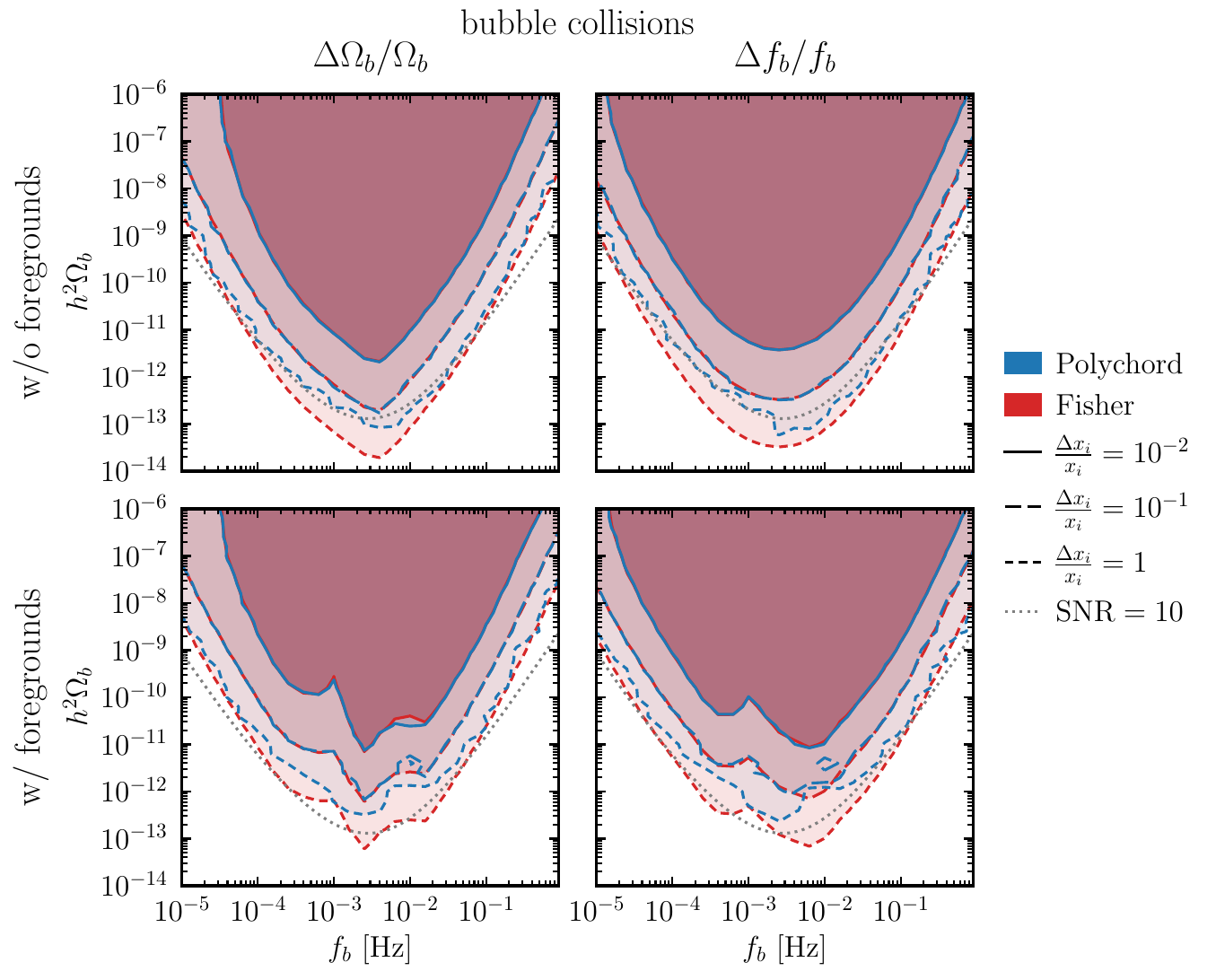}
\caption{
\Gls{LISA} sensitivity to the break frequency and amplitude of the \gls{BPL} template using Fisher analysis~(red) and \Polychord~(blue). The lower panel includes the foregrounds from Galactic and extra-Galactic binaries, whereas the upper panel is without foregrounds. 
The solid, long-dashed, and short-dashed lines correspond to a relative uncertainty of \SI{1}{\%}, \SI{10}{\%}, and \SI{100}{\%} in the reconstruction of the amplitude~$\Omega_b$~(left) and the frequency~$f_b$~(right), respectively.
The gray dotted line indicates an \gls{SNR} of 10 (neglecting foregrounds) in \gls{LISA}.
}
\label{fig:BPL-fisher-polychord}
\end{figure}

Let us now vary the values of the parameters $f_b$ and $\Omega_b$, and consider the corresponding reconstruction. 
The result is shown in \cref{fig:BPL-fisher-polychord}.
We display the relative uncertainty on the reconstructed values of the break amplitude $h^2\Omega_b$~(left) and frequency $f_b$~(right) as a function of the fiducial values. 
The upper panel shows the result of the analysis when disregarding the foreground contributions, whereas the lower panel includes the expected foregrounds from unresolved Galactic and extra-Galactic binaries.
The dark-red regions, delimited by solid lines, are reconstructed with an uncertainty of \SI{1}{\%}, the intermediate regions delimited by long-dashed lines have a reconstruction uncertainty of \SI{10}{\%}, whereas the light-red regions delimited by
short-dashed lines are reconstructed within an order-one factor.
For comparison, we again show the parameter estimation both from the Fisher analysis (in red), and with nested sampling, using the \Polychord sampler (in blue).
In the latter case, the relative uncertainty is calculated from the values of the diagonal of the covariance matrix of the \Polychord samples, approximating the posterior as a normal distribution.
The priors for the amplitude and break frequency are assumed to be uniform, with the corresponding ranges listed in \cref{tab:priors_spectral}.
The parameters yielding an \gls{SNR} of 10 are further indicated by the dotted gray line.
We find that the uncertainty in the Fisher analysis agrees well with \Polychord reconstruction in the region where the parameters are reconstructed within $\sim\SI{10}{\%}$ or better. 
Slight deviations can be seen in the region where the reconstruction works only at $\mathcal{O}(1)$.
The highest sensitivity is achieved for break frequencies around $f_b \sim \SIrange{e-3}{e-2}{\Hz}$.

\subsection{Double broken power law}
\label{sec:geometricMCMC}

\begin{figure}
\centering
\includegraphics[width=\textwidth]{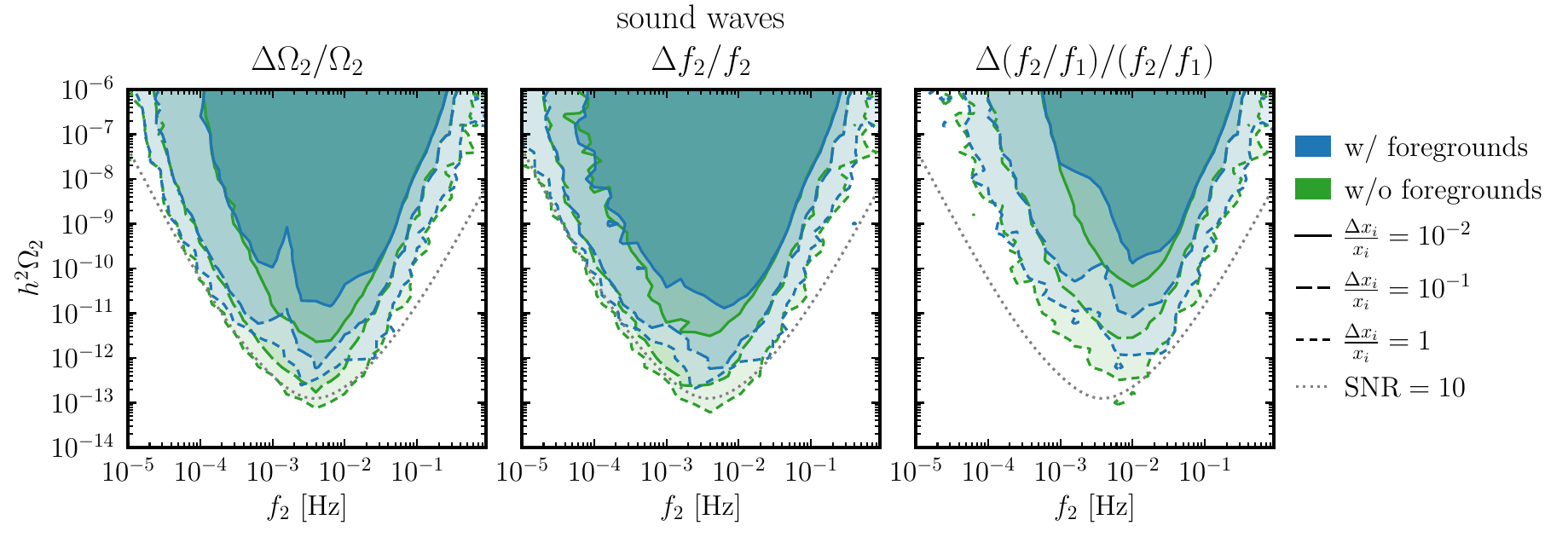}
\caption{%
\Gls{LISA} sensitivity to the \gls{DBPL} template parameters from \Polychord.
The blue regions correspond to the sensitivity including foregrounds, whereas the green curves only take into account the instrument noise.
The fiducial point is taken from the sound wave template with $f_2/f_1 \simeq 2.5(\xi_w/\xi_{\rm shell}) \approx 5.9$ corresponding to a detonation with $\xi_w\approx 1$ and $c_s=1/\sqrt{3}$, see \cref{eq:sw_shape}.
}
\label{fig:Polychord_sw}
\end{figure}

\begin{figure}
\centering
\includegraphics[width=\textwidth]{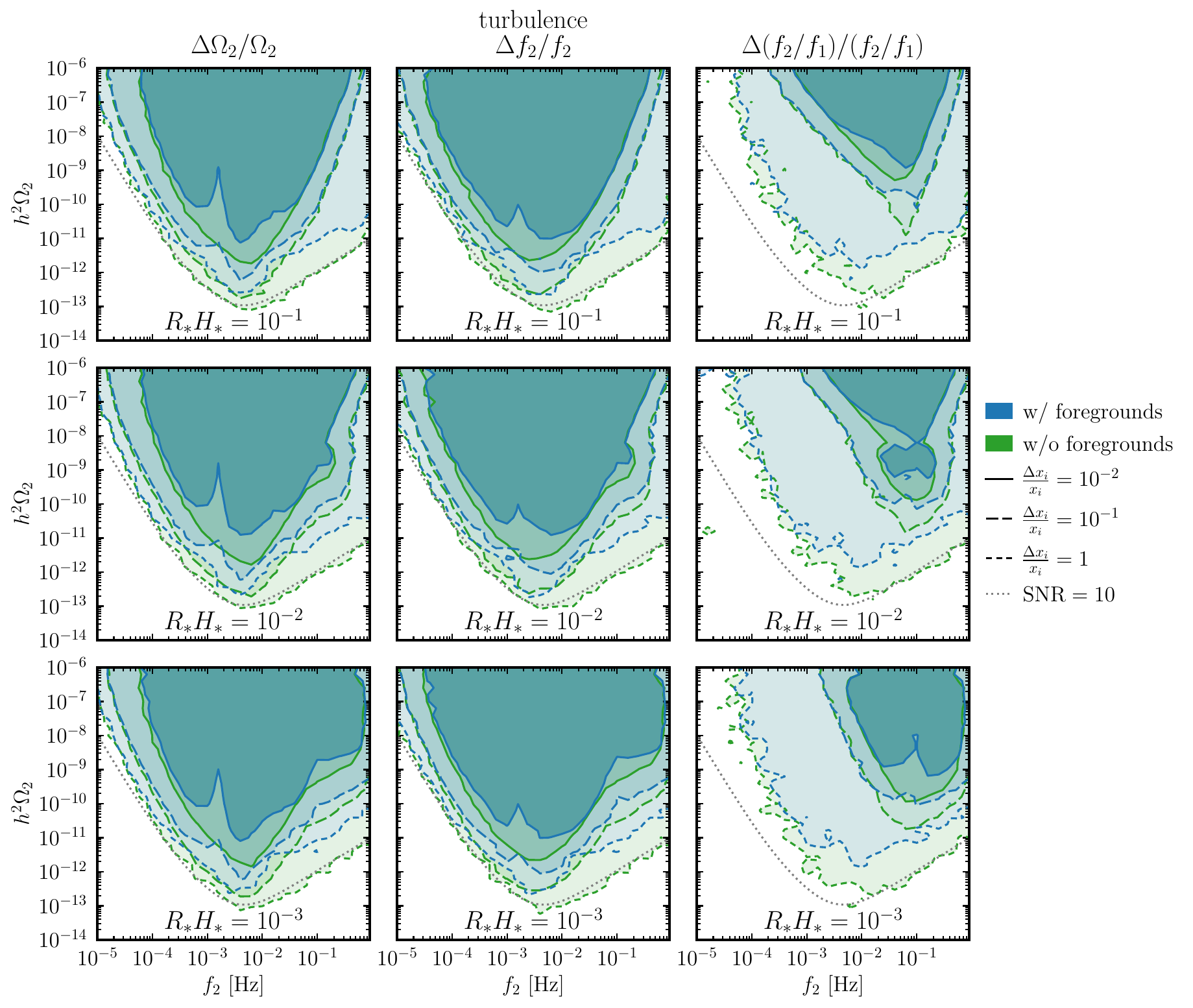}
\caption{
Same as \cref{fig:Polychord_sw}.
The fiducial point is taken from the turbulence template with $f_2/f_1 \simeq 80 \, (R_* H_*)^{1/2} (h^2 \Omega_2/10^{-12})^{-1/4}$, see \cref{geom_pars_turb}. The rows correspond to different values of $R_*H_*=\num{e-1}$, \num{e-2}, and \num{e-3}.
}
\label{fig:Polychord_turb}
\end{figure}

Here we analyse the case of an \gls{SGWB} generated by bulk fluid motion, discussed in \cref{sec:dbpl}, which can be described by the \gls{DBPL} spectrum of \cref{eq:DBPL}. 
In this case, we report the \gls{LISA} sensitivity to the geometric parameters using the Bayesian approach only.

The results for the \gls{DBPL} are shown in \cref{fig:Polychord_sw,fig:Polychord_turb}.
We reconstruct the geometric parameters $\vec{\theta}_{\rm Cosmo}= \{\log_{10} (h^2\Omega_{2}),  \log_{10} (f_2/\si{\Hz}), \log_{10} (f_2 / f_1)\}$ and calculate the uncertainty of the reconstructed values from the corresponding values of the covariance matrix.
In \cref{fig:Polychord_sw}, we adapt the \gls{DBPL} spectrum to the sound waves case, setting the break frequency ratio to
$f_2/f_1 \simeq 2.5(\xi_w/\xi_{\rm shell}) \approx 5.9$, corresponding to a detonation with $\xi_w\approx 1$ and $c_s=1/\sqrt{3}$, see \cref{eq:sw_shape}. In \cref{fig:Polychord_turb}, we adapt the \gls{DBPL} spectrum to the \gls{MHD} turbulence case, and by making use of \cref{geom_pars_turb} we set the frequency ratio to $f_2/f_1 \simeq 2.5 \,\mathcal{N} \sqrt{R_*H_*} \,(F_{\GW, 0}\, A_{\MHD}/\Omega_2)^{1/4}\,\simeq 80 \, (R_* H_*)^{1/2} \, (h^2 \Omega_2/10^{-12})^{-1/4}$ for ${\cal N} = 2$.
Note that, by expressing the amplitude of the \gls{SGWB} in terms of $\Omega_2$ in the case of \gls{MHD} turbulence one introduces a spurious dependence of the ratio $f_2/f_1$ on $R_* H_*$, while physically it only depends on the fractional energy $\Omega_s$. 
Therefore, in \cref{fig:Polychord_turb} we present the reconstruction for a few fixed values of $R_*H_*$.
The priors used for the amplitude and break frequencies are listed in \cref{tab:priors_spectral}.

\subsection{Synergy with other gravitational wave observatories}

\begin{figure}
\centering
\includegraphics[width=.708\textwidth]{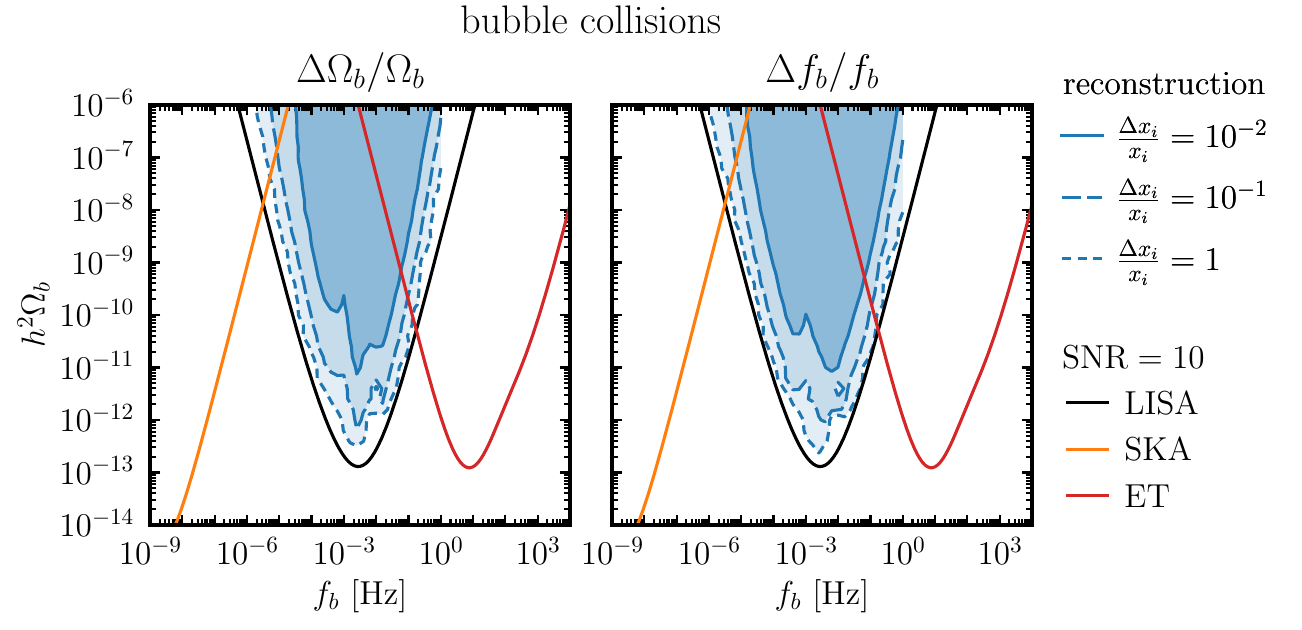}

\vspace{.5\baselineskip}
\includegraphics[width=.9592\textwidth]{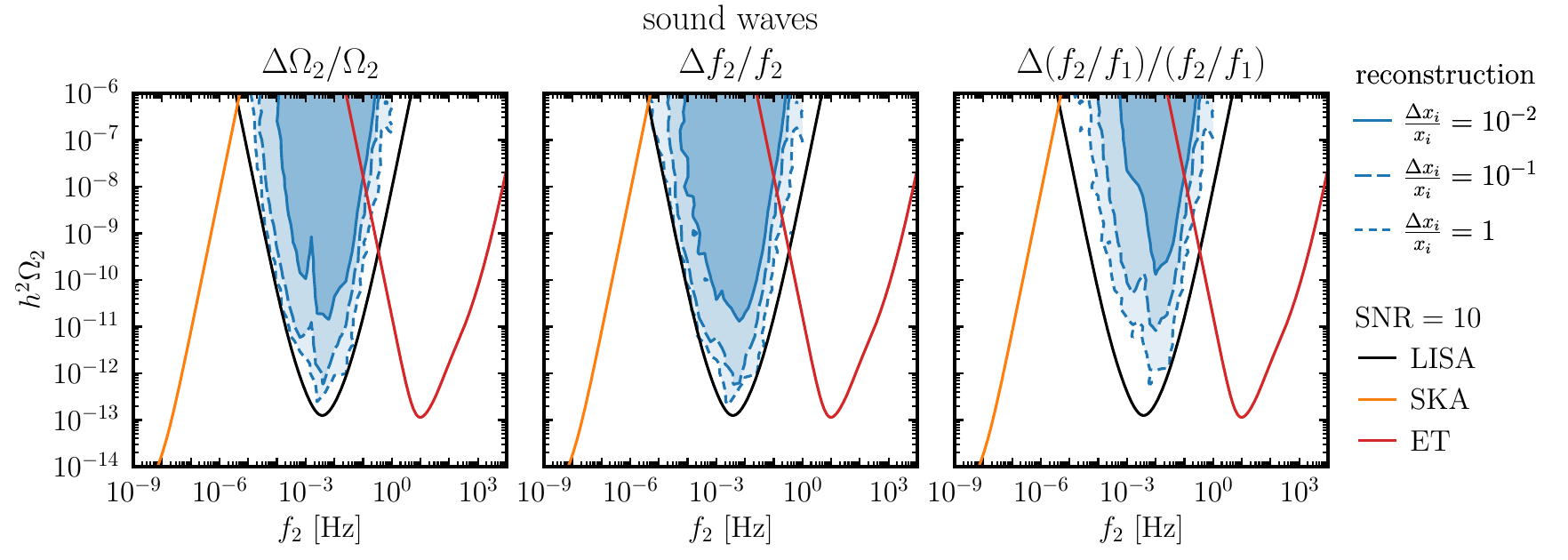}
\caption{Relative uncertainty for the reconstruction of amplitude (left), frequency (top right or bottom center), and frequency ratio (bottom right, only for 
sound waves) in the cases of bubble collisions~(top) and sound waves~(bottom). The black~(\gls{LISA}), red~(\gls{ET}),
and orange~(\gls{SKA}) lines correspond to an \gls{SNR} of 10 for both templates.}
\label{fig:LISA_PTA_ET}
\end{figure}

In \cref{fig:LISA_PTA_ET}, we compare the reconstruction reach of \gls{LISA} to the one of next-generation ground-based \gls{GW} interferometers, in particular the \gls{ET}~\cite{Sathyaprakash:2012jk}, and to the one of future \glspl{PTA}, 
in particular the \gls{SKA}~\cite{Janssen:2014dka}. 
The black line depicts the parameters that produce an \gls{SNR} of 10 in \gls{LISA} after an effective observation time of four years.
The red line indicates the parameters leading to an \gls{SNR} of 10 in \gls{ET}, using the noise curve from Ref.~\cite{Hild:2010id}, whereas the orange line corresponds to the \gls{SKA}, assuming observation of \num{1000} pulsars over 20~years with a timing cadence of two weeks and timing uncertainty of \SI{0.1}{\micro\s}~\cite{Weltman:2018zrl}.
While synergies with \gls{ET} are possible, simultaneous observation of a \gls{SGWB} from a \gls{PT} in the \gls{LISA} and \gls{PTA} bands requires unrealistically high amplitudes, due to the large separation between the frequency bands of the two experiments.

Another problem is the existence of foregrounds, which are not included in the \gls{SNR} curves. The blue regions correspond to the respective reconstruction uncertainty estimated using \Polychord including foregrounds in the \gls{LISA} band (cf.\ \cref{fig:Polychord_sw}), and a similar analysis would be required for the other experiments, which would further diminish the prospects for synergy between the experiments.

\section{Reconstruction of the thermodynamic parameters}
\label{sec:reconstruction}

As shown in \cref{sec:models}, the \gls{SGWB} signal from a first-order \gls{PT} can be expressed in terms of a set of thermodynamic parameters, depending on the sourcing process. 
The \gls{SGWB} from bubble collisions or highly relativistic fluid shells in strong first-order \glspl{PT} is determined by the energy fraction $\tilde K\approx 1$, the relative inverse \gls{PT} duration $\beta/H_*$, and the temperature at the end of the \gls{PT}, $T_*$. The bubble wall speed does not enter, since in this case the bubble walls move close to the speed of light, $\xi_w\simeq 1$.  
In the case of a \gls{SGWB} generated by both sound waves and turbulence, the parameters entering the spectrum are rather connected to the bulk fluid motion: the kinetic energy fraction $K$, the relative size of the bulk fluid motion $H_*R_*$, and the \gls{PT} temperature $T_*$. For sound waves, the fluid shell thickness, which depends on the bubble walls speed $\xi_w$ and on the sound speed $c_s$, also enters the \gls{SGWB} spectral shape; while for turbulence, there is an extra parameter, $\varepsilon$, representing the fraction of kinetic energy which is vortical rather than compressional. 
Note that the gradient and/or kinetic energy fractions $\tilde K$ and $K$ can be related to the \gls{PT} strength $\alpha$, while the typical scale of the bulk fluid motion can be connected to the \gls{PT} inverse duration $\beta/H_*$ and the bubbles wall speed $\xi_w$ (see \cref{eq:Rstar}). 

While the \gls{SGWB} spectra can be described in terms of three to five thermodynamic parameters, depending on the source, 
only two or three geometric parameters are present: the amplitude, and one or two break frequencies, depending on whether the power law is doubly or singly broken, i.e.\ whether the signal stems from strong transitions or is generated via interactions with the plasma.
We are hence confronted with degeneracies when intending to infer the \gls{PT} parameters from the signal. We will demonstrate that these degeneracies can be lifted if more than one contribution is reconstructed.

In this section, we therefore develop two methods to reconstruct the thermodynamic parameters entering the \gls{SGWB} signals, both based on parameter estimation with \Polychord implemented in the \texttt{SGWBinner}. The first one consists in estimating the thermodynamic parameters directly. 
While this is the straightforward way to perform the parameter reconstruction, it is associated with additional computational costs to resolve the degeneracies discussed above.
We hence also employ a second method which consists in estimating the geometric parameters first, and then converting the reconstructed geometric parameters to the corresponding thermodynamic ones. 
As it does not suffer from degeneracies, running the geometric-parameter reconstruction as a first step is computationally more efficient.
It further simplifies the reinterpretation of the results in the case that new simulations indicate a change in the interdependence of the geometric and thermodynamic parameters.

We apply the parameter transformation from geometric to thermodynamic parameters directly to each point in the samples generated with \Polychord, discarding unphysical points that, for instance, yield $\xi_w > 1$ or $\tilde{K}>1$.
This automatically takes into account the Jacobian factor from the change of variables, as it is encoded in the density of parameter points in the samples.
The resulting posterior, however, in principle differs from the posterior obtained via direct sampling, as different priors are used.
These differences can be removed by using the induced prior~\cite{Gowling:2022pzb}, i.e.\ the prior obtained from transforming the prior of the thermodynamic parameters to geometric parameters, when sampling in terms of the geometric parameters.
Alternatively, each parameter point can be assigned a weight given by the ratio of the prior on the thermodynamic parameters to the induced prior.
In the following, though, we refrain from doing so. 
Comparing the two approaches hence corresponds to studying the thermodynamic parameter reconstruction for two different prior assumptions.
As our priors are flat and relations between (the logarithms of) the parameters are mostly linear, the differences in priors are expected to have no significant effect.

\paragraph{Bubble collisions and highly relativistic fluid shells}
In this case we have three independent thermodynamic parameters ($\tilde{K}$, $\beta/H_*$, $T_*$) but only two geometric parameters: the break position and its amplitude $(f_b, h^2\Omega_b)$.
The reconstruction of the thermodynamic parameters hence inevitably suffers from degeneracies.
Fixing either of the three parameters, the other two can be determined uniquely. 
In particular, as strong transitions in sectors that are in equilibrium with the \gls{SM} plasma have $\alpha \gg 1$, fixing $\tilde{K} \sim \frac{\alpha}{1+\alpha} \approx 1$ allows for the reconstruction of $\beta/H_*$ and $T_*$.
Alternatively, as shown by \cref{eq:amp_and_freq_strongPT}, we can obtain constraints on the combinations 
$T_* \beta/H_* \approx \SI{54}{\TeV} f_b/\si{\mHz}$ and $\tilde{K} H_*/\beta \approx \sqrt{\num{e6} h^2\Omega_b}$.\footnote{Note that we fix $g_* = 106.75$ in these expressions.}

\begin{figure}
    \includegraphics[width=\textwidth]{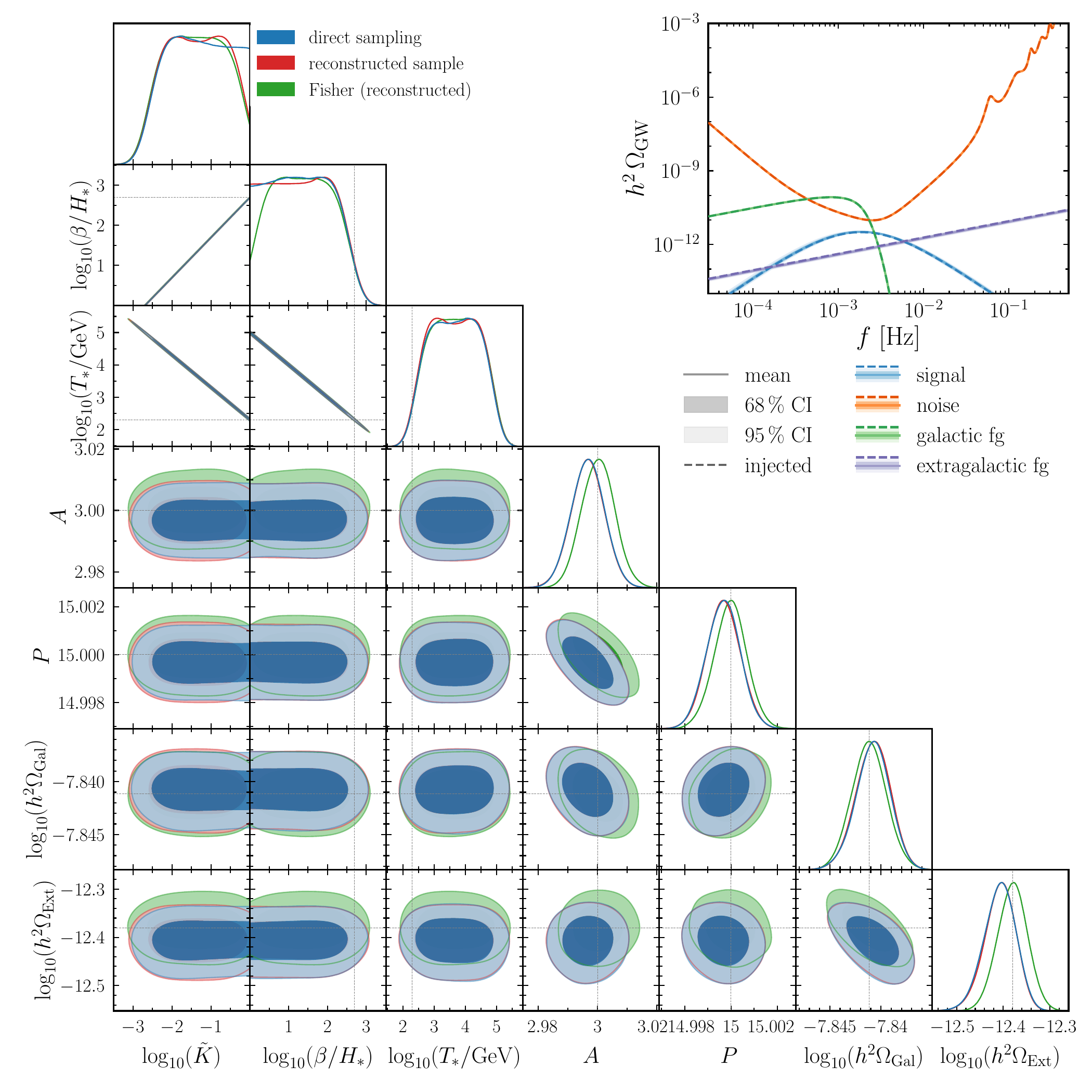}
    \caption{
        Reconstruction of the thermodynamic parameters of the \gls{GW} spectrum from bubble collisions in a strong \gls{PT}.
        The blue contours are sampled directly, whereas the red contour is reconstructed from a sample in terms of the geometric parameters of the \gls{BPL}. 
        The green contours are obtained from the Fisher analysis for the geometric parameters. The plot in the top right corner visualizes the injected signals and noises as well as the \SI{68}{\%} (dark areas) and \SI{95}{\%} (light areas) \glspl{CL} of their reconstruction obtained from direct sampling.
    }
    \label{fig:reco_bubble}
\end{figure}

\Cref{fig:reco_bubble} shows the reconstruction of the thermodynamic parameters of the \gls{SGWB} spectrum from bubble collisions with $\tilde{K}=1$, $\beta/H_*=500$, and $T_*=\SI{200}{\GeV}$. As discussed in \cref{sec:bubble_template}, even though the spectrum we adopt in this work for bubble collisions was computed assuming $\tilde{K}\approx 1$, here we keep $\tilde{K}$ as a free parameter.
The blue contours in \cref{fig:reco_bubble}  depict the reconstruction directly in terms of the thermodynamic parameters. The degeneracies between the parameters are clearly visible.
In addition, the red contours show the reconstruction using a sample in terms of the break frequency $f_b$ and amplitude $\log_{10}(h^2\Omega_b)$ of the corresponding \gls{BPL}.
The respective priors used in \Polychord are listed in \cref{tab:priors_coll}.
In order to obtain the three thermodynamic parameters from the two geometric parameters, for each point in the sample, a random value for the transition temperature $T_*$ was drawn according to the prior in \cref{tab:priors_coll}, and the remaining parameters $\tilde{K}$ and $\beta/H_*$ have been calculated assuming this temperature.\footnote{%
    This corresponds to adding $\log_{10}(T_*/\si{GeV})$ as an additional parameter to the set of geometric parameters, i.e.\ $\left\{\log_{10}(h^2\Omega_b),\, \log_{10}(f_b/\si{Hz})\right\} \to \left\{\log_{10}(h^2\Omega_b),\, \log_{10}(f_b/\si{Hz}),\, \log_{10}(T_*/\si{GeV})\right\}$. 
    The likelihood in this parametrisation remains independent of $T_*$, so that the new posterior is simply given by the original posterior multiplied by the prior on the additional parameter.
}
The green contours indicate the Fisher analysis estimate in terms of the geometric parameters, converted to thermodynamic parameters using the same method.

\begin{table}
    \centering
    \begin{tabular}{lrc|lrc}
        \toprule
        parameter & input & prior range & parameter & input & prior range \\ \midrule
        $\log_{10}(\tilde{K})$ & $0.0$\ & $(-5,0)$ & 
            $\log_{10}(h^2\Omega_b)$ & $-11.5$\ & $(-30,-5)$\\
        $\log_{10}(\beta/H_*)$ & $2.7$\ & $(0,5)$ & 
            $\log_{10}(f_b/\si{\Hz})$ & $-2.7$\ & $(-7,1)$\\
        $\log_{10}(T_*/\si{\GeV})$ & $2.3$\ & $(0,8)$ &
            & & \\
        \bottomrule
    \end{tabular}
    \caption{%
        Input values and flat prior ranges for the injection and reconstruction of our bubble collisions benchmark spectrum in terms of the thermodynamic parameters~(left) and the geometric parameters~(right).
        The spectral slopes are set to $(n_1,n_2,a_1) = (2.4,-2.4,1.2)$, see \cref{tab:slopes}.
    }
    \label{tab:priors_coll}
\end{table}

The three approaches yield consistent results for the parameter estimation. 
Note that in the particular case of the bubble collision spectrum, the priors in the two sampling methods are actually equivalent, as both methods use flat priors for $\log_{10}(T_*/\si{\GeV})$, so that there are no differences induced by the temperature dependence of the effective degrees of freedom, and boundary effects from samples with parameters corresponding to unphysical values of $\log_{10}(\tilde{K})$ are avoided by discarding the corresponding parameter points.

\paragraph{Sound waves}
The \gls{SGWB} spectrum from sound waves is given in terms of three geometric parameters,
the two frequency breaks and the amplitude at the second break, but four thermodynamic parameters, $K$, $H_* R_*$, $T_*$, and $\xi_w$.
The relation between the geometric parameters $f_1$, $f_2$, and $\Omega_{2}$ in the sound waves template and the thermodynamic parameters are given in \cref{eq:sw_shape,eq:sw_amplitude,eq:Om2_sound}.
In this case, the frequencies depend on $H_* R_*$, $\xi_{w}$, and $T_*$, and the integrated amplitude $\Omega_{\mathrm{int}}$ on $K$ and $H_* R_*$, while $\Omega_2$ also involves $\xi_w$ through the ratio $f_2/f_1$ (both amplitudes $\Omega_2$ and $\Omega_{\rm int}$ also depend subdominantly on $T_*$ via the degrees of freedom). This allows us to reconstruct the thermodynamic parameters in the following way: from the ratio of the frequency breaks, $f_2/f_1 = 2.5\,\Delta_w^{-1} (\xi_w)$ (see \cref{eq:sw_shape}), we can derive the compatible values of $\xi_w$ ($\xi_w$ might not be unique since $\Delta_w (\xi_w)$ is in general not invertible, so there can be one or two possible values of $\xi_w$ for each $\Delta_w$), from $f_1$
we can derive constraints (one for each value of $\xi_w$) on the ratio $T_*/(H_* R_*)$, and from the amplitude $\Omega_2$, combined with $f_2/f_1$ and $f_1$, we can derive a constraint on a combination of $K$ and $T_*$.
These constraints are
\begin{align}
    \frac{T_*}{H_* R_*} \simeq \SI{30}{\TeV}\,\frac{f_1}{\si{\mHz}}
    \ \ \text{and}\ \
     \min\biggl[K^2 \,H_*R_*,\,  
     \frac{2}{\sqrt{3}}
     K^\frac{3}{2}(H_*R_*)^2\biggr]\simeq \frac{\Omega_\mathrm{int}}{\num{1.8e-6}}\,,
    \label{constraint_sw}
\end{align}
where $\Omega_{\rm int}$ is computed from $f_2/f_1$ and $\Omega_2$ using \cref{eq:Om2_sound}. 
Hence, from the geometric parameters, we can only uniquely determine $\xi_w$ (one or two values).
The parameters $H_* R_*$ and $T_*$ are degenerate since only the ratio $T_*/(H_*R_*)$ 
is reconstructed (one constraint for each value of $\xi_w$) using \cref{constraint_sw}.
Furthermore, $H_*R_*$ and $K$ are also degenerate via \cref{constraint_sw}.
There are two possible degeneracies, due to the $\min$ function that appears in the computation of the sound wave time $\tau_{\rm sw}$ (cf.~discussion in \cref{sec:sw_template}).
Hence, either the product $K^2 H_*R_*$ or $K^{3/2} (H_*R_*)^2$ is constrained.
Therefore, in general, the parameters $K$, $H_* R_*$, and $T_*$ cannot be uniquely reconstructed from the geometric parameters of the sound waves template. However, the method described in the previous section and displayed in \cref{fig:reco_bubble} can also be applied in this case.

\paragraph{Turbulence}

The geometric parameters $f_1$, $f_2$, and $\Omega_2$ in the turbulence template are given in \cref{geom_pars_turb}, and depend bijectively on $\Omega_s$ (from the ratio of the break frequencies), $H_*R_*$ (from the amplitude), and $T_*$ (via $H_{*,0}$ in $f_1$). 
This gives
\begin{subequations}
\begin{align}
    \label{constraint_turb_2}
    \varepsilon K 
    &= \Omega_s \simeq \frac{6.5\, {\cal N}^2}{(f_2/f_1)^2} \,,\\\label{constraint_turb_1}
    H_* R_* &\,\simeq\,  \frac{\num{6e-4}}{\mathcal{N}^2}  \left(\frac{f_2}{f_1}\right)^2 \!\sqrt{\frac{h^2\Omega_2}{\num{e-12}}}\,,
    \\
    \label{constraint_turb_3}
    T_* &\simeq \frac{\SI{1.6}{\GeV}}{\mathcal{N}^2}\frac{f_2}{\si{\mHz}} \left(\frac{f_2}{f_1}\right)^2 \sqrt{\frac{h^2\Omega_2}{\num{e-12}}}\,.
\end{align}
\end{subequations}
Hence, from the geometric parameters we can determine uniquely $H_* R_*$ and $T_*$, whereas $K$ and $\varepsilon$ are degenerate and only their product $\Omega_s = \varepsilon K$ is constrained. 

\paragraph{Soundwaves + turbulence}

If we combine the \glspl{SGWB} from sound waves and turbulence, $\Omega_\GW = \Omega_\GW^{\rm sw} + \Omega_\GW^{\rm turb}$, and use the fit as a \gls{DBPL} for each of the components, we can break the degeneracies present separately for the sound waves and turbulence and
reconstruct all the thermodynamic parameters $\{K, H_*R_*, \xi_w, T_*, \varepsilon\}$
when the geometric parameters $\{f_1^{\rm sw}, f_2^{\rm sw}, f_1^{\rm turb},\Omega_2^{\rm turb}, \Omega_2^{\rm sw}\}$
can be inferred from the signal parameter estimation.
Note that $f_2^{\rm turb} = 2.2\,H_{*,0}/(H_* R_*) \simeq 11 f_1^{\rm sw}$,
such that $f_2^{\rm turb}$ does not add information as long as we know its exact relation with $f_1^{\rm sw}$.
Indeed, one can see that given this relation, the first constraint in \cref{constraint_sw} and \cref{constraint_turb_1} are equivalent if we use $T_*$ from \cref{constraint_turb_3}.

From the sound waves geometric parameters we are able to reconstruct the possible value/s (one or two) of $\xi_w$ from $f_2^{\rm sw}/f_1^{\rm sw}$, while we can reconstruct $T_*$ from $\Omega_2^{\rm turb}$, $f_1^{\rm turb}$, and $f_1^{\rm sw}/f_1^{\rm turb}$ (see \cref{constraint_turb_3}).
We can solve for $K$ for both cases in \cref{constraint_sw} and then check which solution also satisfies the corresponding condition on the sound wave lifetime.
This could lead to one or two solutions for $K$.
Then, for each solution of $K$, we can reconstruct $\varepsilon = \Omega_s/K$.
In general, the reconstruction of the geometric parameters of a combined sound waves and turbulence template allows us to reconstruct all the thermodynamic parameters in a unique way.

\begin{table}
    \centering
    \begin{tabular}{lrc|lrc}
        \toprule
        parameter & input & prior range & parameter & input & prior range \\ \midrule
        $\log_{10}(K)$ & $-1.1$\ & $(-4,-0.2)$ & 
            $\log_{10}(h^2\Omega_2^\text{sw})$ & $-8.8$\ & $(-30,-5)$\\
        $\log_{10}(R_*H_*)$ & $-0.6$\ & $(-3,0)$ & 
            $\log_{10}(\Omega_2^\text{turb}/\Omega_2^\text{sw})$ & $-1.7$\ & $(-5,0)$\\
        $\xi_w$ & $1.0$\ & $(0.6,1)$ &
            $\log_{10}(f_2^\text{sw}/\si{\Hz})$ & $-3.4$\ & $(-5,-1)$ \\
        $\log_{10}(T_*/\si{\GeV})$ & $2.7$\ & $(0,6)$ &
            $\log_{10}(f_2^\text{sw}/f_1^\text{sw})$ & $0.77$\ & $(0,2)$\\
        $\varepsilon$ & $1.0$\ & $(0.01,1.2)$ &
            $\log_{10}(f_1^\text{turb}/f_1^\text{sw})$ & $-0.21$\ & $(-2,2)$\\
        \bottomrule
    \end{tabular}
    \caption{%
        Input values and flat prior ranges for the injection and reconstruction of the turbulence plus sound waves benchmark spectrum in terms of the thermodynamic parameters~(left) and of the geometric parameters~(right).
        The ratio $f_2^{\rm turb}/f_1^{\rm sw}$ is fixed to $11$ (cf.~\cref{eq:sw_shape,geom_pars_turb}), whereas the spectral slopes are set to $(n_1,n_2,n_3,a_1,a_2) = (3,1,-3,2,4)$ and $(n_1,n_2,n_3,a_1,a_2) = (3,1,-8/3,4,2.15)$ for the sound waves and turbulence \gls{SGWB} spectrum, respectively (cf.~\cref{tab:slopes}).
    }
    \label{tab:priors_sw+turb}
\end{table}

\begin{figure}
    \includegraphics[width=\textwidth]{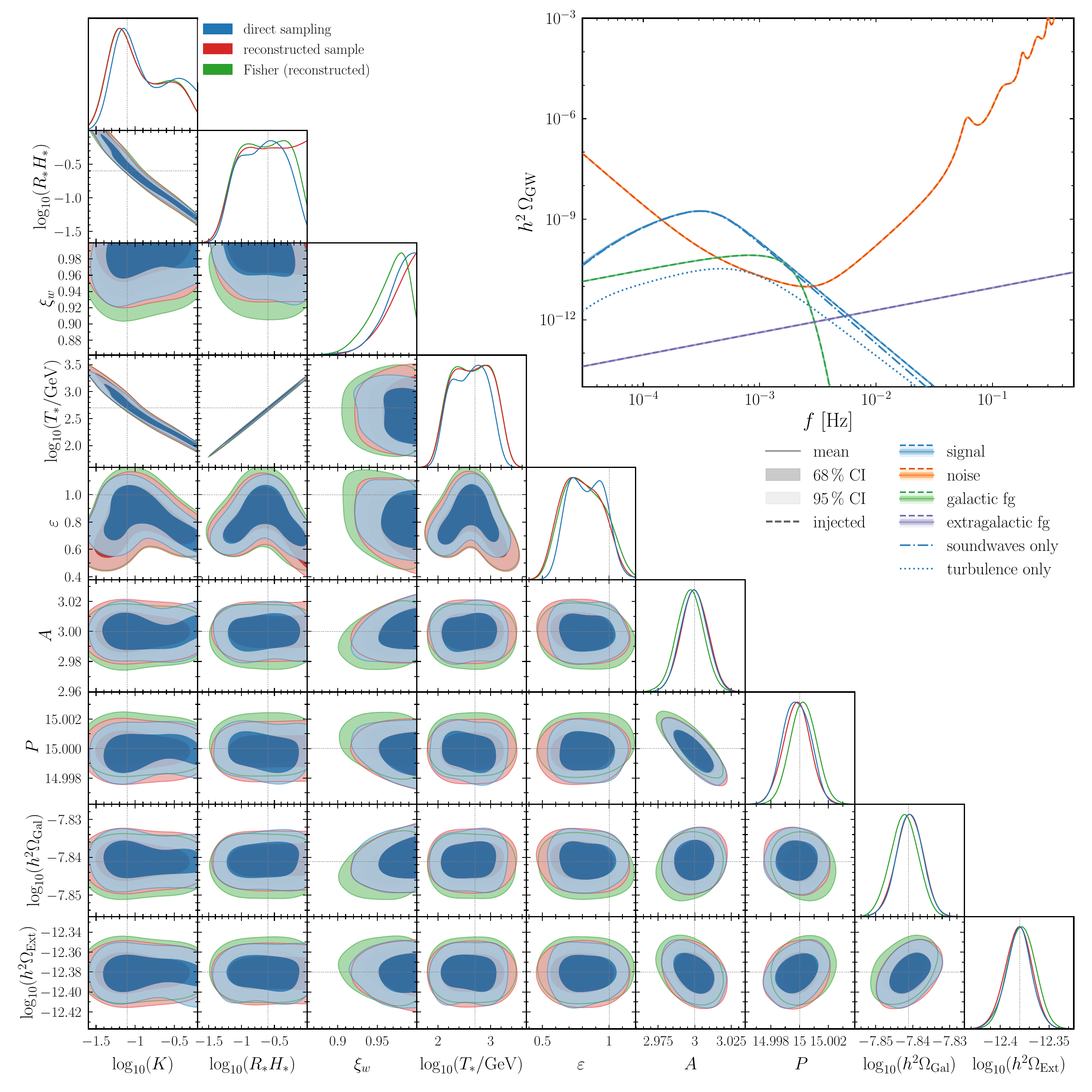}
    \caption{%
        Reconstruction of the sum of the sound waves and turbulence \gls{SGWB} spectra in terms of the thermodynamic parameters. 
        The blue contours are sampled directly using \Polychord in terms of the thermodynamic parameters, whereas the red contours are reconstructed from the sample in terms of the geometric parameters. 
        The green contours are obtained from a Fisher analysis performed in terms of the geometric parameters. 
    }
    \label{fig:reco_sw+turb}
\end{figure}

\Cref{fig:reco_sw+turb} depicts an example of parameter reconstruction in terms of the thermodynamic parameters of the \gls{PT} featuring both the sound waves and the turbulence contributions.
The blue contours correspond to directly sampling the thermodynamic parameters in \Polychord. For the red contours, instead, we first sample over the geometric parameters and then convert the chains in terms of the thermodynamic parameters with the procedure previously described. 
Parameter points corresponding to unphysical values ($\xi_w$, $K$, or $R_*H_* > 1$) were discarded.
The priors of the signal parameters in both cases are taken as uniform, with the corresponding ranges and the input values listed in \cref{tab:priors_sw+turb}.
In addition, the green contours depict the result of the Fisher approach calculating the Fisher matrix in terms of the geometric parameters, creating samples using a multivariate Gaussian distribution according to the Fisher matrix, and then converting these to the thermodynamic parameters.
While these approaches in principle differ in terms of the priors as discussed above, we again obtain consistent results.

Although in theory, the reconstruction of the sum of the two spectra should be unique, we can still see degeneracies between the kinetic energy fraction~$K$, the bubble size~$R_* H_*$, and the temperature~$T_*$.
These degeneracies arise since the first frequency break of the turbulence \gls{SGWB} spectrum is hidden below the sound waves signal and the Galactic foreground, such that it cannot be reconstructed with enough precision. 
The corresponding loss of information inhibits the full reconstruction of the two spectra, leading to degeneracies, as one now needs to constrain five thermodynamic parameters from only four reconstructed geometric parameters.
Furthermore,
the two possible solutions for $K$, corresponding to the case where the \gls{SGWB} generated from sound waves is cut off by their lifetime or at the Hubble time, lead to a kink in the 2D posteriors involving $K$ or $\varepsilon$.

\begin{figure}
    \includegraphics[width=\textwidth]{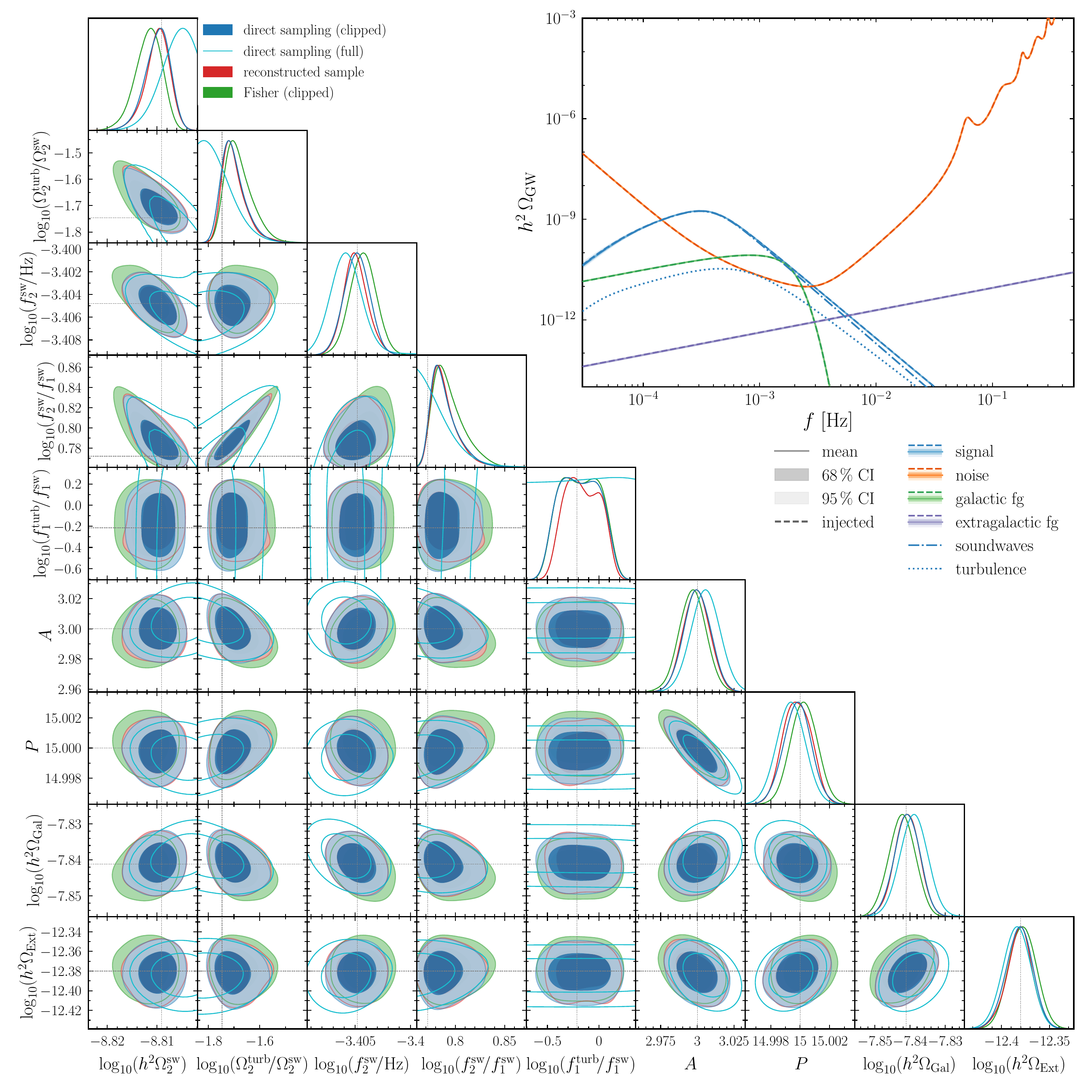}
    \caption{
        Reconstruction of the sum of two \glspl{DBPL} in terms of the respective amplitudes and frequency breaks, assuming spectral slopes corresponding to the sum of the sound wave and turbulence spectra.
        The blue contours are sampled directly using \Polychord, whereas the red contours are calculated from the sample in terms of the thermodynamic parameters. 
        The green contours are obtained from Fisher analysis.  
        Parameter points leading to unphysical thermodynamic parameters, i.e.~$\xi_w$, $K$, or $H_*R_* > 1$ are discarded.
        These points are retained in the cyan contours.
    }
    \label{fig:reco_2bpl}
\end{figure}

As a cross-check, \cref{fig:reco_2bpl} shows the reconstruction of the geometric parameters, i.e.\ the amplitudes and break frequencies of the two \glspl{DBPL}, for the same input signal.
The blue region now corresponds to directly sampling over the geometric parameters, whereas the red region shows the geometric parameters calculated from the sampling in terms of the thermodynamic parameters.
The green region depicts the result from Fisher analysis. Note that we clipped the samples of the geometric parameters by removing parameter points that would yield unphysical values of $\xi_w$, $K$, or $R_* H_*$ when converted to thermodynamic parameters.
While this would ideally be done directly in the priors or using other constraints, it turns out to be challenging to implement, due to the rather involved relations between the two sets of parameters (see, e.g., Ref.~\cite{Gowling:2022pzb} for a possible implementation).
For reference, we further include the unclipped results as cyan contours, where the failure of the reconstruction of $f_1^\text{turb}$, which leads to the degeneracies in the thermodynamic parameters reconstruction, becomes apparent.

\begin{figure}
    \centering
    \includegraphics[width=.75\textwidth]{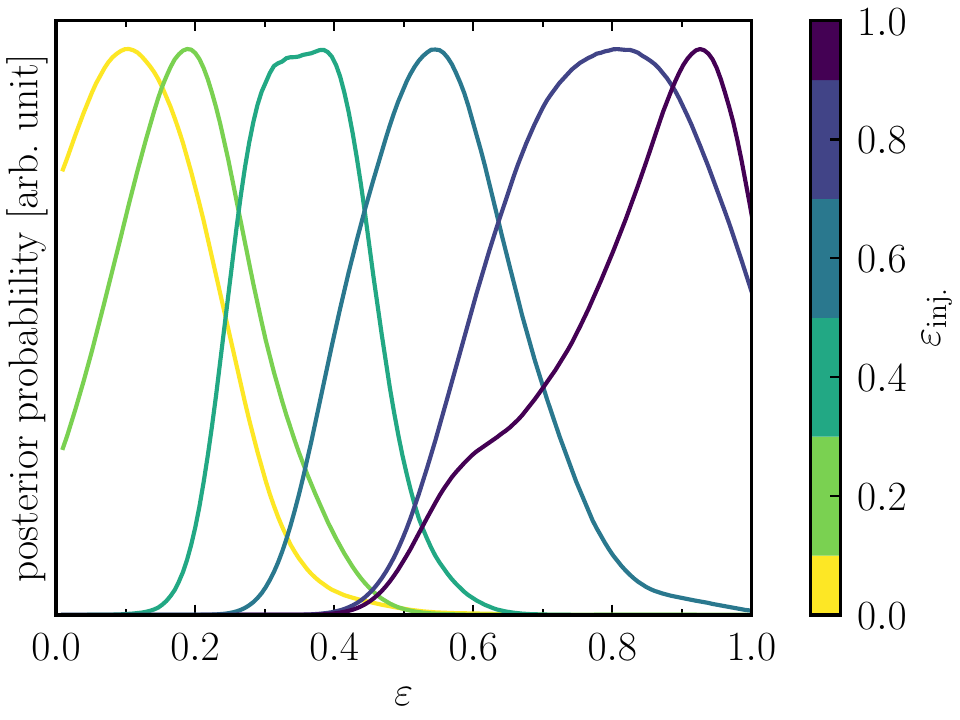}
    \caption{Posterior probability distributions for the reconstruction of $\varepsilon$ using \Polychord assuming $K=0.08$, $H_* R_* = 0.25$, $\xi_w=1$, and $T_* = \SI{500}{\GeV}$. The injected value of $\varepsilon$ is varied between 0 and 1 in steps of $0.2$.}
    \label{fig:epsilon_reco}
\end{figure}

Above we have treated the fraction $\varepsilon$ of bulk kinetic energy converted to turbulent motion as an additional independent parameter in our set. However, in principle 
we expect $\varepsilon$ to depend on the \gls{PT} parameters and on the properties of the primordial plasma. 
Therefore, if determined independently from an observed \gls{GW} signal from a \gls{PT} in \gls{LISA}, it would improve our understanding of the \gls{SGWB} source. In~\cref{fig:epsilon_reco}, we hence depict the posteriors for the reconstruction of~$\varepsilon$, where the colors indicate the injected value of~$\varepsilon$.
The remaining parameters are set to the ones listed in \cref{tab:priors_sw+turb}.
While a precise determination of~$\varepsilon$ appears to be unrealistic, a strong \gls{SGWB} signal from a \gls{PT} can be used to at least roughly constrain the range of~$\varepsilon$.

\section{Interpretation in terms of fundamental particle physics models}
\label{sec:interpretation_models}

In this section, we provide two examples of how the \gls{PT} \gls{SGWB} reconstruction can be recast as compelling bounds on the parameter space of particle physics models. 
Performing parameter inference starting from the \gls{LISA} \gls{SGWB} data directly on the parameters of a \gls{BSM} Lagrangian is currently prohibitive. 
Indeed, computing the thermodynamic quantities governing the \gls{PT} from the set of fundamental parameters within a given model requires numerical tools~\cite{Wainwright:2011kj, Basler:2018cwe, Athron:2020sbe} that, at the moment, are not ready for intensive, fast, automatised applications. For instance, numerical instabilities and the computation time can represent obstacles for likelihood samplers. 

For our reconstruction, we use a simple approach that circumvents these issues. We begin by scanning the parameter space of the models on a regular grid, and for each point, we determine the thermodynamic parameters. 
Next, we interpolate the data to find the parameters as smooth functions of the fundamental parameters of the model. 
These are then ready to be used in the reconstruction of the \gls{SGWB} signal at \gls{LISA}. We then map our reconstruction back onto the model parameter space. We do so for a handful of benchmarks of the \gls{SGWB} spectrum, appropriately chosen so that the geometrical parameters of the \gls{SGWB} can be well reconstructed.

We now proceed to explain how the \gls{PT} parameters can be calculated from the models we consider.
Below the critical temperature $T_c$ the new minimum in the potential of the theory becomes deeper than the initial one. From that moment the probability that the field tunnels into the new minimum due to thermal fluctuations can be approximated by~\cite{Linde:1980tt,Linde:1981zj}
\begin{equation}
\Gamma(T) = \left(\frac{S_3(T)}{2\pi T}\right)^{\frac{3}{2}} T^4 \textrm{e}^{-S_3(T)},
\end{equation}
where $S_3(T)$ is the action of the field configuration driving the decay. The field configuration is obtained by solving the equation of motion of the field in a given potential assuming $O(3)$-symmetry as appropriate for thermal systems. The probability of a point to remain in the initial vacuum is given by~\cite{Guth:1982pn} 
\begin{equation}
P(t) = \exp{\left[-\frac{4 \pi}{3} \int_{t_c}^{t} \!\dd t' \,\Gamma(t') \, a^3(t') \, r^3(t,t') \right]} \;,
\end{equation}
where the comoving radius of a bubble reads
\begin{equation}
\label{comoving_radius}
r(t,t') = \int_{t'}^{t}  \frac{\xi_w(\tilde{t})}{a(\tilde{t})} \dd \tilde{t} \, . 
\end{equation}
for a bubble with wall velocity $\xi_w(\tilde{t})$ growing up to time $t$ and nucleated at $t'$.
We assume the appropriate time to compute the thermodynamic quantities governing the \gls{GW} production is percolation time, when bubbles collide to fill the volume (we assume instantaneous reheating after the transition). This moment can be approximated by $P (t_*) \approx 0.7$, which allows us to find the corresponding percolation temperature $T_*$. 
The strength of the \gls{PT} is approximated as~\cite{Hindmarsh:2017gnf, Caprini:2019egz}
\begin{equation}\label{eq:alpha_def}
\alpha \equiv  \left. \frac{1}{\rho_r}\left( \Delta V(T) - \frac{T}{4}  \frac{\partial \Delta V(T)}{\partial T} \right) \right|_{T=T_*} \, ,
\end{equation}
where $\Delta V(T)$ is the energy difference between the two coexisting minima, and $\rho_r$ denotes the radiation energy density of the Universe.
The inverse duration of the transition can be computed using the action of the solution
\begin{equation} \label{eq:betaH}
\frac{\beta}{H_*} \equiv T_* \frac{d}{dT}  \left. \left(  \frac{S_3(T)}{T} \right) \right|_{T=T_*}\, ,
\end{equation}
and related to the average bubble size via \cref{eq:Rstar}. Using the methods described in Refs.~\cite{Laurent:2020gpg,Cline:2021iff,Lewicki:2021pgr,Laurent:2022jrs,Ellis:2022lft} we verify all the points of interest feature highly relativistic walls $\xi_w\approx 1$. 
In this case the efficiency factor reads~\cite{Espinosa:2010hh} 
\begin{equation}
\kappa=\frac{\alpha}{0.73 + 0.083\sqrt{\alpha}+\alpha}\, ,
\end{equation}
and with it, we have all the thermodynamic parameters needed to compute the \gls{SGWB} spectrum from bubble collisions and highly relativistic fluid shells (see \cref{sec:bubble_template}) as well as from sound waves (see \cref{sec:sw_template}).

We apply this procedure to two illustrative models, namely the \gls{SM} supplemented with a neutral singlet, and the classically conformal U$(1)_{B-L}$ extension of the \gls{SM}. The former model is arguably the simplest \gls{SM} extension allowing a strong first-order electroweak transition, while the latter is one of the minimal extensions featuring a wide parameter region with a very supercooled \gls{PT}. Both extensions have few free parameters, making the scan of their whole parameter space feasible.

\subsection{Gauge singlet extension with \texorpdfstring{\boldmath$\mathbb{Z}_2$}{Z2} symmetry}

One of the simplest extensions of the \gls{SM} giving rise to \glspl{PT} is that of an extra scalar singlet under the \gls{SM} gauge group endowed with a $\mathbb{Z}_2$ symmetry~\cite{McDonald:1993ey, Espinosa:1993bs, Espinosa:2007qk, Profumo:2007wc, Espinosa:2011ax, Barger:2011vm, Cline:2012hg, Alanne:2014bra, Curtin:2014jma, Vaskonen:2016yiu, Kurup:2017dzf, Beniwal:2017eik, Niemi:2021qvp, Lewicki:2021pgr, Ellis:2022lft}. The tree-level potential in this model is written as 
\begin{equation} \label{V0}
    V_{\text{tree}}(\Phi, s) = -\mu_h^2 \Phi^{\dagger} \Phi +\lambda ( \Phi^{\dagger} \Phi)^2 + \mu_s^2 \frac{s^2}{2} +\frac{ \lambda_{s}}{4} s^4 +\frac{ \lambda_{hs}}{2} s^2 \Phi^{\dagger} \Phi,
\end{equation}
where  $\Phi = (G^+ , (h + i G^0)/\sqrt{2})^\text{T}$ is the \gls{SM} Higgs doublet while $s$ is the singlet field. We consider a two-step transition where the singlet gets a non-zero vacuum expectation value at high temperature before the electroweak transition, and require that the global $T=0$ minimum of the potential lies at the electroweak vacuum $(h,s)=(v,0)$ with $v=\SI{246}{\GeV}$. As usual, the minimisation conditions allow us to relate $\lambda = \mu_h^2/v^2$,  $\mu_h^2 = m_h^2/2$, and $\mu_s^2 = m_s^2  - \lambda_{hs}v^2/2$, where $m_h$ and $m_s$ are the masses of the Higgs boson and the scalar singlet. The model is therefore described by the singlet mass $m_s$ and the quartic couplings $\lambda_s$ and $\lambda_{hs}$.

For a given value of singlet quartic coupling $\lambda_s$,  the parameter space consistent with a first-order \gls{PT}
has a \textit{banana} shape in the plane of $m_s$--$\lambda_{hs}$, such that larger $\lambda_{hs}$ values are required for larger values of $m_s$. This shape arises from the requirements that the electroweak vacuum is the global minimum of the potential at $T=0$, which gives an upper bound on $\lambda_{hs}$, and that the singlet gets a non-zero vacuum expectation value before the electroweak transition, which gives a lower bound on $\lambda_{hs}$~\cite{Vaskonen:2016yiu}. The strength of the transition is controlled only by the Higgs-portal coupling and larger $\lambda_{hs}$ values yield stronger \glspl{PT}. As the singlet mass increases and due to the \textit{banana} shape of the parameter space, very large $\lambda_{hs}$ values are required to produce sizable effects. Thus for sufficiently large $m_s$ ($\approx \SI{600}{\GeV}$) one reaches the perturbativity limit of the model. 

The prospects of detection at colliders have been explored in  
Refs.~\cite{Curtin:2014jma, Beniwal:2017eik}. For scalar singlet masses in the range $\SI{65}{\GeV} < m_s < \SI{300}{\GeV}$ the testability of this model by direct and indirect collider signatures is very limited and for this reason, it has been dubbed the \textit{nightmare scenario}~\cite{Curtin:2014jma}. We restrict our attention to the parameter range $\SI{65}{\GeV}<m_s<\SI{125}{\GeV}$ for which $\lambda_{hs}$ remains well within the perturbativity limit.
The simplified $\mathbb{Z}_2$ version of the model we consider also faces theoretical issues. Firstly, the entire parameter space capable of supporting a first-order \gls{PT} predicts an amount of scalar \gls{DM} which would already have been seen by direct detection experiments~\cite{Beniwal:2017eik}. Secondly, this version predicts the presence of scalar domain walls before the electroweak transition which would influence the nature of the transition~\cite{Blasi:2022woz}. While this is a challenge when building self-contained models we can simply assume minimal changes that would not affect the transition we want to model in a significant way. We can easily add new particles coupled to the singlet which would destabilise the scalar \gls{DM} and hide it from direct detection experiments while providing a \gls{DM} candidate with a large enough abundance~\cite{Beniwal:2018hyi}. Adding even a very small $\mathbb{Z}_2$ breaking term would cause the domain walls to annihilate before the electroweak transition~\cite{Azatov:2022tii}.
Thus the problems of the model can be fixed with minimal modifications that would not have a significant impact on the \gls{PT} dynamics and our parameter reconstruction forecast.

\begin{figure}
    \centering
    \includegraphics[height=\textwidth]{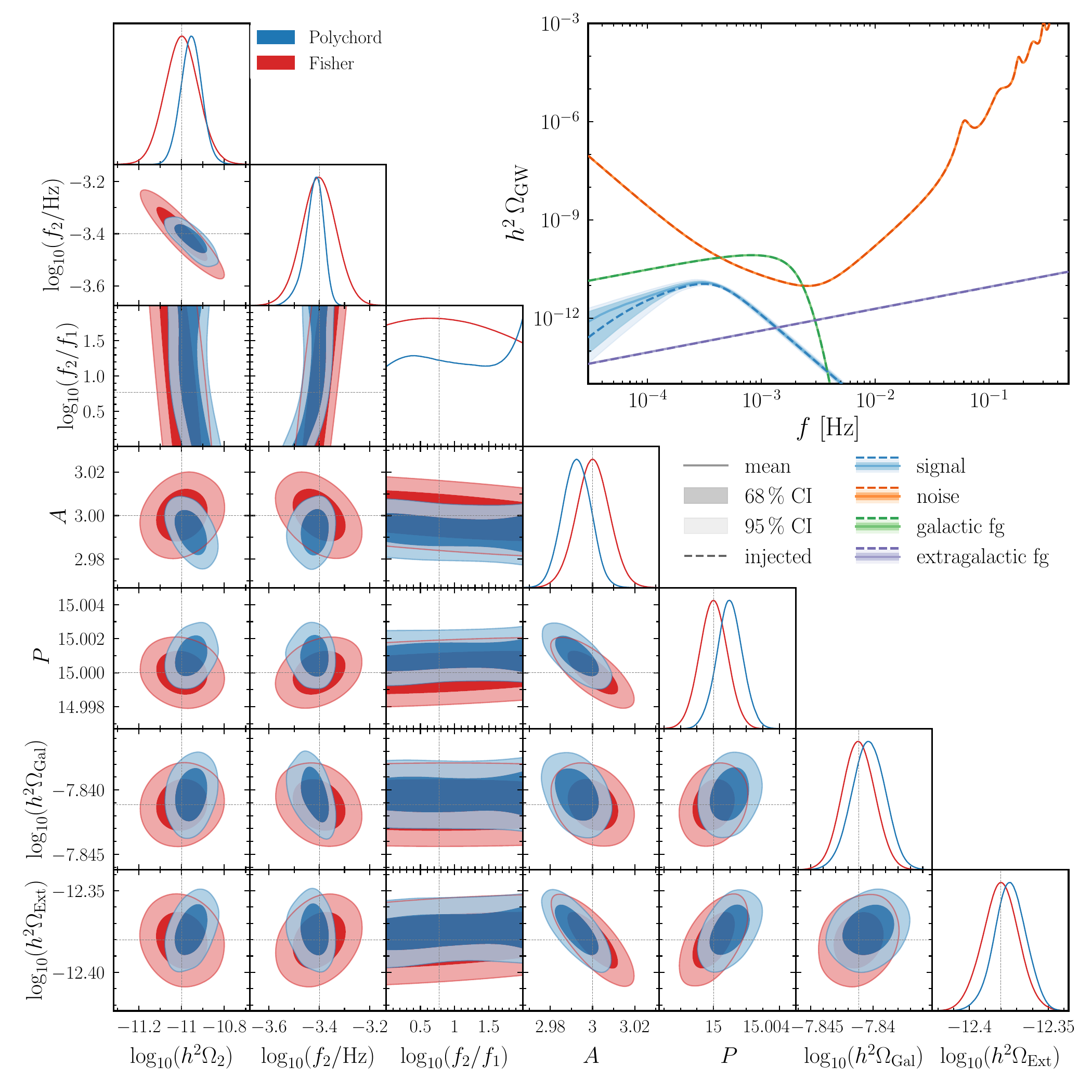}
        \caption{Reconstruction of the sound waves \gls{SGWB} spectrum with $h^2\Omega_2 = \num{e-11}$ and $f_2=\SI{0.4}{\milli\Hz}$,  where the injected separation between the breaks $f_2/f_1 \approx 5.9$ is fixed by $\xi_w \approx 1$ (see \cref{eq:sw_shape}).}
    \label{fig:triangle_singlet}
\end{figure}

\begin{figure}
\centering
\includegraphics[width=0.87\textwidth]{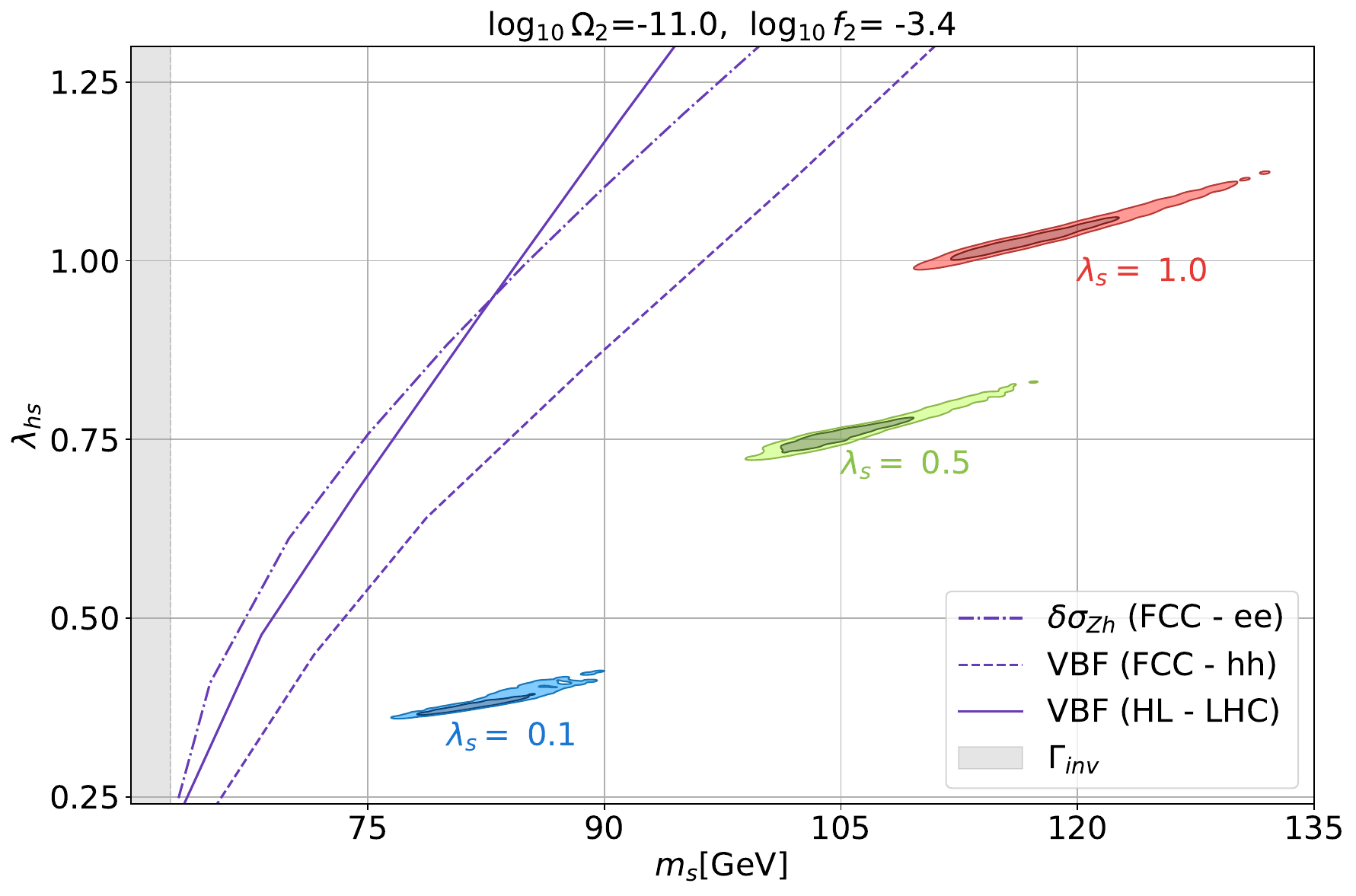}
\caption{\Glspl{CL} for different values of singlet quartic couplings. The violet curves are experimental upper limits from the \gls{FCC} and the high luminosity \gls{LHC}. The gray shaded region is excluded by the Higgs invisible decay.}
\label{fig:scalar_reconstruction}
\end{figure}

\begin{figure}
\centering
\includegraphics[width=0.48\textwidth]{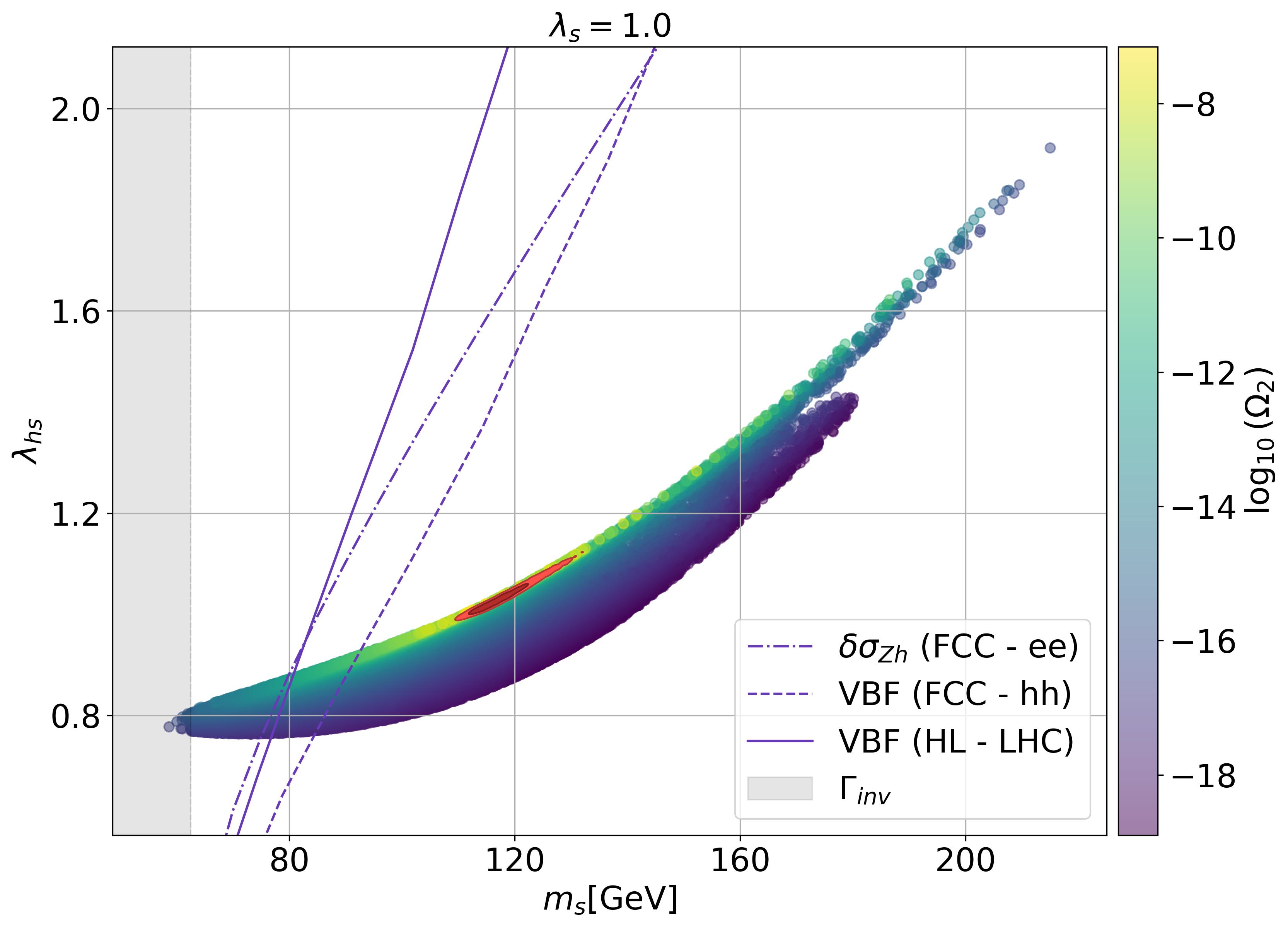}
\includegraphics[width=0.48\textwidth]{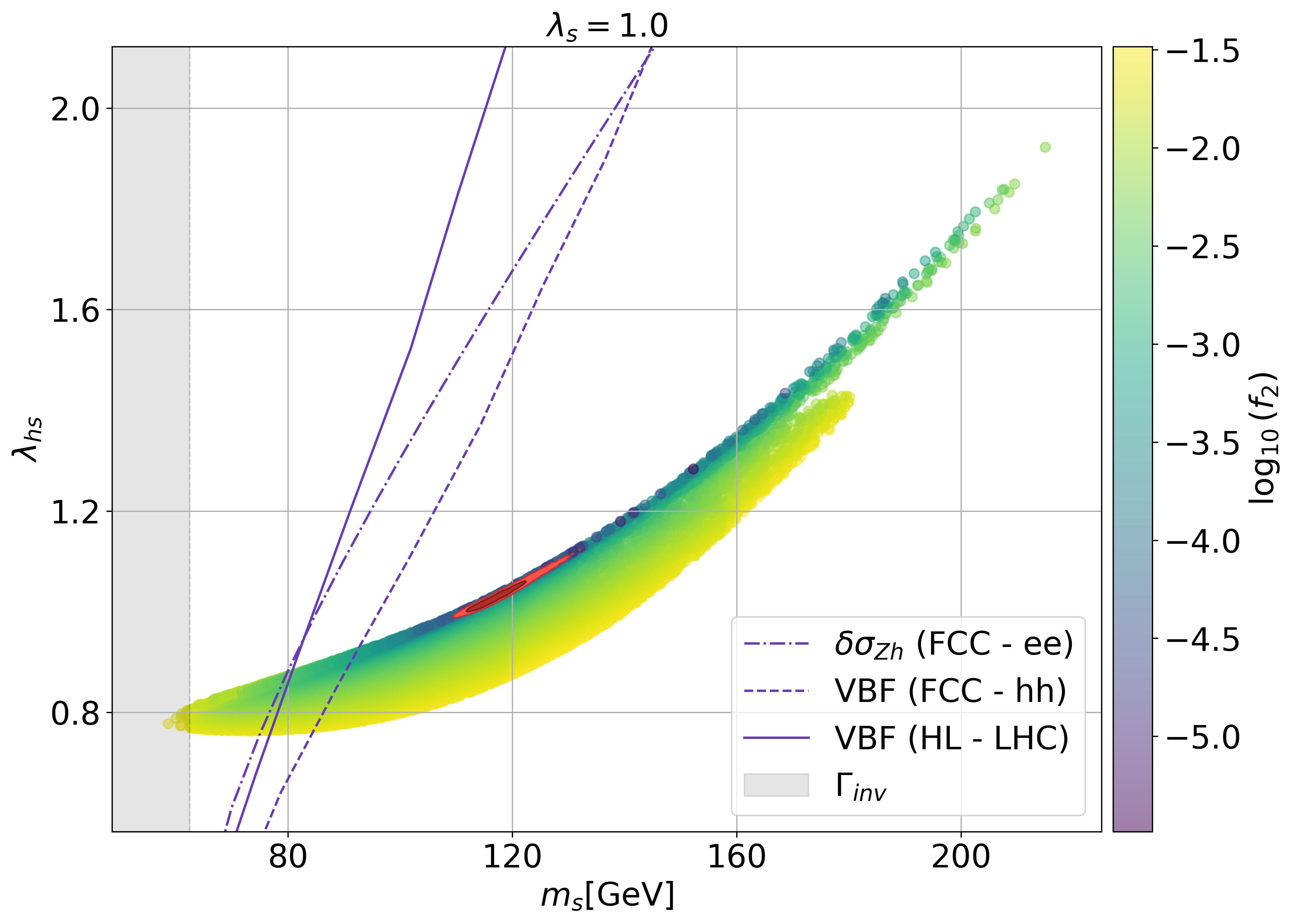}
\caption{Parameter space of the singlet model. The colour bar displays the magnitude of the \gls{SGWB} amplitude on the left panel and frequency on the right panel. The red region on top of the banana plots is the reconstructed ellipses. }
\label{fig:scalar_reconstruction_lseq1}
\end{figure}

\begin{figure}
\centering
\includegraphics[scale=.3]{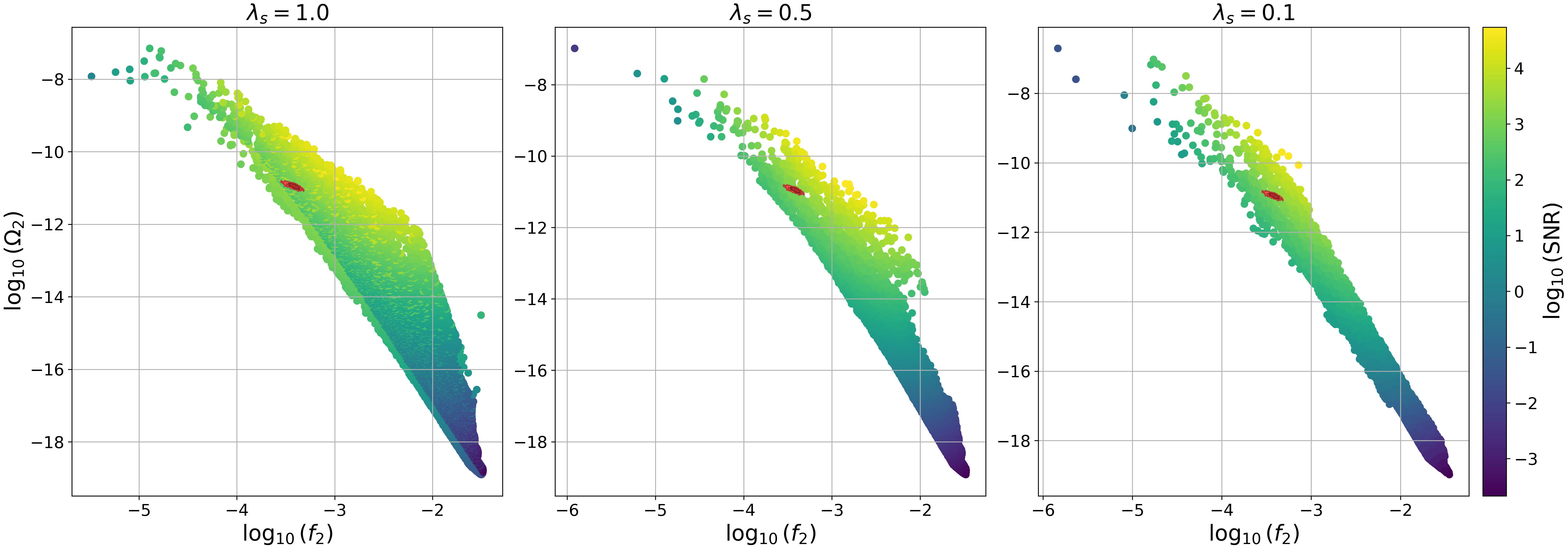}
\caption{  Geometric parameter space generated by the gauge singlet extension for different values of the singlet quartic coupling, displayed on top of each figure. The colour bar displays the \gls{SNR} computed from the sound waves \gls{SGWB} template. The small red ellipses on top of each plot are the \glspl{CL}. }
\label{fig:scalar_reconstruction_geom}
\end{figure}

We use the \gls{SGWB} spectrum produced by the sound waves source (see \cref{sec:sw_template}) as our benchmark for this model, further choosing $h^2\Omega_2 = \num{e-11}$ and $f_2=\SI{0.4}{\milli\Hz}$. We do not include the signal from  \gls{MHD} turbulence since no analysis so far has been able to predict the value of $\varepsilon$.
We implement a modified version of the \texttt{CosmoTransitions} code~\cite{Wainwright:2011kj} to scan the parameter space of this model. 
The reconstruction of the \gls{SGWB} spectrum with \gls{LISA} is shown in \cref{fig:triangle_singlet}. 
The corresponding area translated into the parameter space of the model, using the \texttt{ChainConsumer} package~\cite{Hinton2016}, is shown in \cref{fig:scalar_reconstruction,fig:scalar_reconstruction_lseq1,fig:scalar_reconstruction_geom}. The first one indicates the parameter space in terms of the scalar mass $m_s$ and portal coupling $\lambda_{hs}$ while the dependence on the scalar quartic is shown using three benchmark values  $\lambda_s=0.1, 0.5$, and $1$ to indicate its impact. \Cref{fig:scalar_reconstruction_lseq1} shows the entire parameter space populated by first-order \glspl{PT} by the singlet model for a fixed value of the quartic $\lambda_s=1$ with the red ellipsis indicating the reconstruction of our \gls{SGWB} benchmark point.
The future experimental sensitivities are shown as violet lines and these correspond to Higgs associated production $\sigma_{Zh}$, 
\gls{VBF} at a \gls{FCC} and the high luminosity \gls{LHC}~\cite{Craig:2014lda,Ellis:2018mja}.  \Cref{fig:scalar_reconstruction_geom} again shows the entire parameter space populated by first-order \glspl{PT} by the singlet model in terms of the geometric parameters of the spectrum $\Omega_2$ and $f_2$ (see \cref{sec:sw_template}) with the three panels using each of our three benchmark values $\lambda_s=0.1, 0.5$, and $1$ with the red ellipsis indicating the reconstruction of our benchmark point in each panel while the colour of points indicates the \gls{SNR} of the corresponding spectrum.\footnote{Note that our formula for the \gls{SNR} assumes a weak signal and very large values should be considered as qualitative indicators~\cite{Allen:1997ad}.}

\subsection{Classically conformal U\texorpdfstring{\boldmath$(1)_{B-L}$}{(1)\_{B-L}} model}

\begin{figure}
    \centering
    \includegraphics[width=0.9\textwidth]{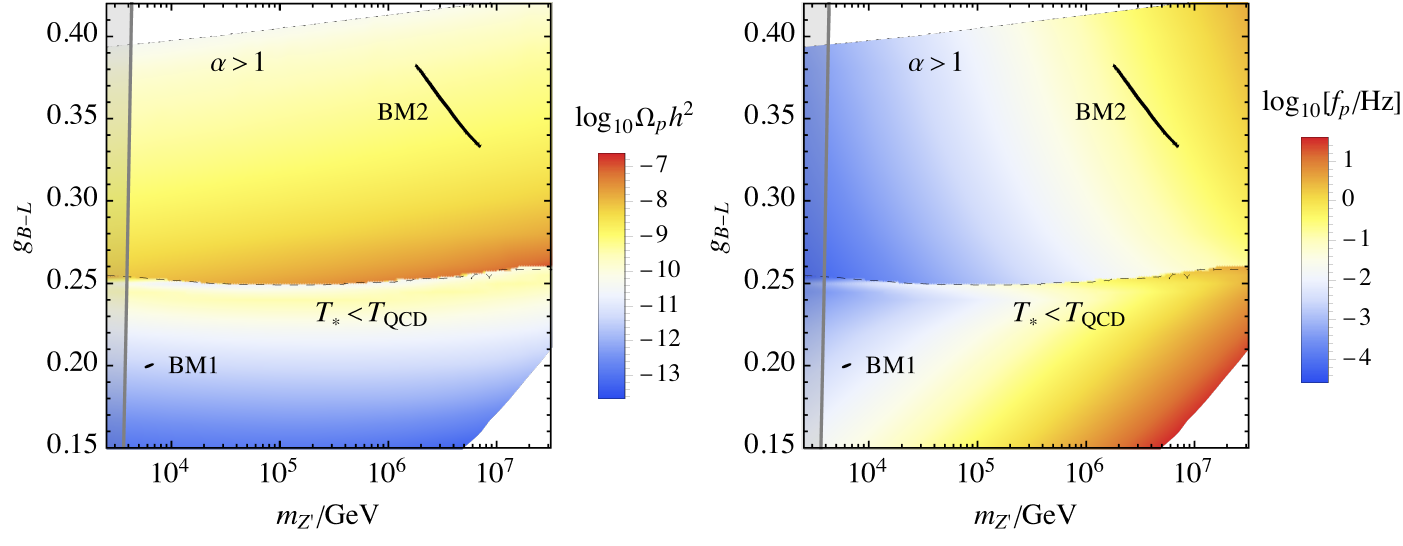}
    \caption{The colour coding shows in the left panel the peak amplitude and in the right panel the peak frequency of the \gls{SGWB} spectrum in the U$(1)_{B-L}$ model. In the white region at the top, $\alpha<1$. Below the dashed curve, the transition is triggered by the confinement of \gls{QCD} and the gray region on the left is excluded by collider searches of $Z'$.}
    \label{fig:U1BL2}
\end{figure}

The classically conformal U$(1)_{B-L}$ model~\cite{Iso:2009ss, Iso:2009nw, Jinno:2016knw, Iso:2017uuu, Marzo:2018nov, Ellis:2019oqb, Sagunski:2023ynd} is characterised by the gauge symmetry ${\rm SU}(3)_c \times {\rm SU}(2)_L \times {\rm U}(1)_{Y} \times {\rm U}(1)_{B-L}$ and the scalar potential that at tree level is scale-invariant,
\begin{equation}
    V = \lambda_H (H^\dagger H)^2 + \lambda_\phi (\phi^\dagger \phi)^2 - \lambda_p (H^\dagger H) (\phi^\dagger \phi) \,,
\end{equation}
where $H$ is the Higgs ${\rm SU}(2)_L$ doublet and $\phi$ is a complex scalar charged under ${\rm U}(1)_{B-L}$. In addition to the \gls{SM} parameters, this model includes three parameters: the ${\rm U}(1)_{B-L}$ gauge coupling $g_{B-L}$, the ${\rm U}(1)_{B-L}$ gauge boson mass $m_{Z'}$, and the gauge mixing parameter between ${\rm U}(1)_{Y}$ and ${\rm U}(1)_{B-L}$, $\tilde g$ (all the quantities are computed at the renormalisation scale corresponding to the vacuum expectation value of $\phi$). For $g_{\rm B-L} \gtrsim 0.1$, which is the region relevant for the \gls{GW} searches (see \cref{fig:U1BL2}), collider searches exclude $m_{Z'} \lesssim \SI{4}{\TeV}$~\cite{Escudero:2018fwn}. The \gls{PT} dynamics is not very sensitive to the mixing parameter and we fix it to $\tilde g = -0.5$ that ensures perturbativity and vacuum stability of the model up to the Planck scale~\cite{Marzo:2018nov}. 

The \gls{PT} dynamics and the resulting \gls{GW} signal in this model have been studied, e.g.\ in Refs.~\cite{Jaeckel:2016jlh, Jinno:2016knw, Iso:2017uuu, Marzo:2018nov, Ellis:2019oqb, Ellis:2020nnr, Dasgupta:2022isg, Sagunski:2023ynd}. As typical for classically conformal models, the transition can be very strongly supercooled and, consequently, the \gls{GW} signal arises from the bubble collisions and highly relativistic fluid shells.

\begin{figure}
    \centering
    \includegraphics[height=.49\textwidth]{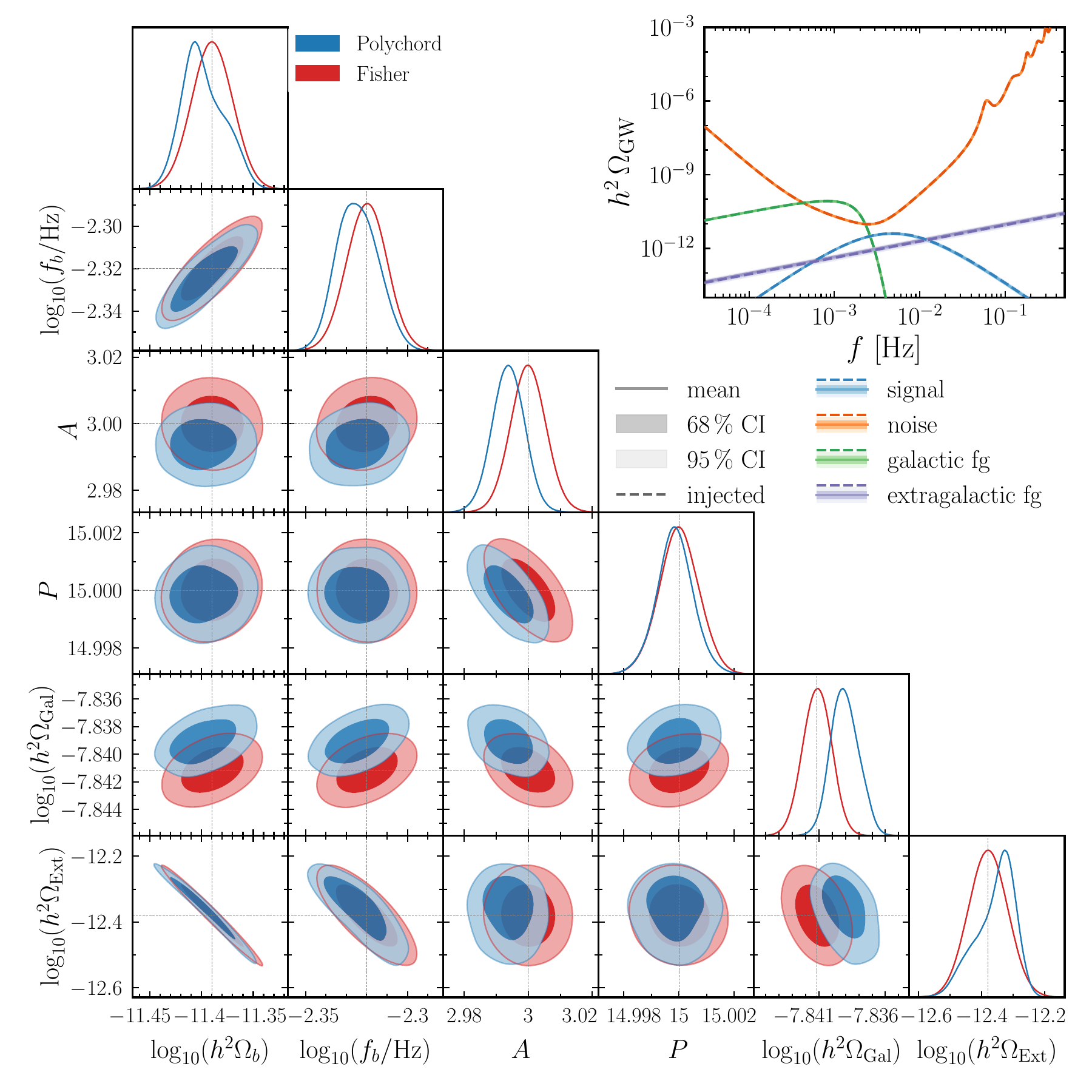}
    \includegraphics[height=.49\textwidth]{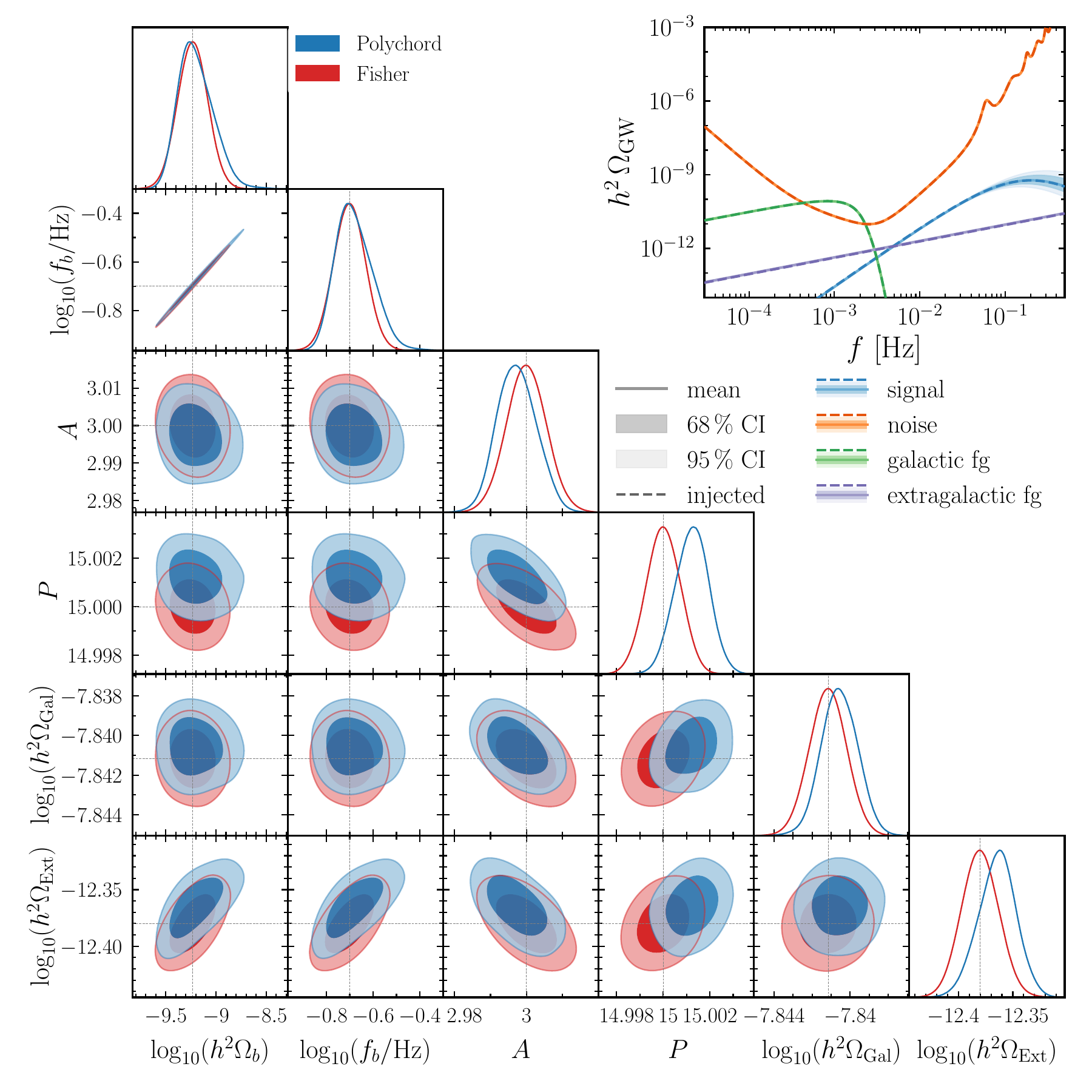}
    \caption{Reconstruction of the PT SGWB spectrum from bubble collisions and highly relativistic fluid shells with $h^2\Omega_b = \num{4e-12}$ and $f_b=\SI{0.5}{\milli\Hz}$ (left), and $h^2\Omega_b = \num{6e-10}$ and $f_b=\SI{0.2}{\Hz}$ (right).}
    \label{fig:triangle_U1B-L}
\end{figure}

\begin{figure}
    \centering
    \includegraphics[height=0.4\textwidth]{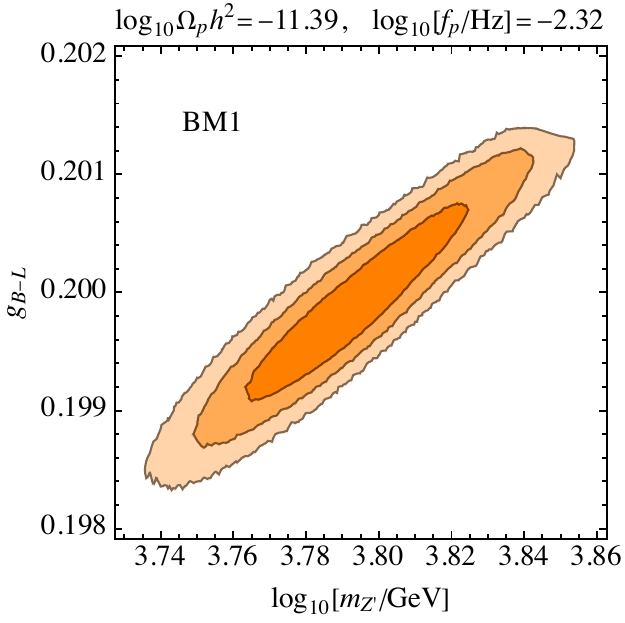} \hspace{2mm}
    \includegraphics[height=0.4\textwidth]{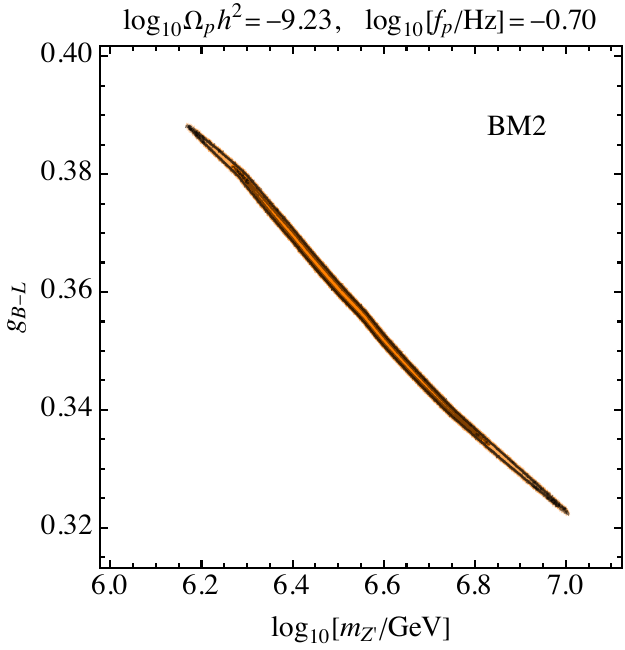}
    \caption{The shaded regions show the \SI{68}{\%}, \SI{95}{\%}, and \SI{99.7}{\%} \gls{CL} regions of \gls{LISA} in the U$(1)_{B-L}$ model parameter space for the bubble collisions \gls{SGWB} signal (gauged U$(1)$ case) with the peak amplitude and frequency given above the plots.}
    \label{fig:U1BL}
\end{figure}

We show the posteriors of the reconstruction in terms of the geometric parameters for two benchmark points with $h^2\Omega_b = \num{4e-12}$ and $f_b=\SI{0.5}{\milli\Hz}$, and $h^2\Omega_b = \num{6e-10}$ and $f_b=\SI{0.2}{\Hz}$ in~\cref{fig:triangle_U1B-L}, and the corresponding posterior probability distributions in the model parameter space in \cref{fig:U1BL}. The first benchmark point, shown in the left panels, corresponds to a case where the peak of the \gls{SGWB} spectrum is close to the best sensitivity of \gls{LISA} and, consequently, the correlation in the reconstruction of $\Omega_b$ and $f_b$ is small. This benchmark case is potentially within the reach of future collider experiments since the $Z'$ gauge boson is not very heavy. For example, Ref.~\cite{Marzo:2018nov} estimated that the \gls{FCC} could probe $Z'$ masses up to \SI{45}{\TeV} \cite{FCC:2018vvp}.

The second benchmark point, shown in the right panel of \cref{fig:triangle_U1B-L,fig:U1BL}, instead is at a much higher $Z'$ mass and cannot be probed in colliders. In this case, the \gls{SGWB} spectrum peak is in the high-frequency tail of the \gls{LISA} sensitivity. Therefore, the reconstructed $\Omega_b$ and $f_b$ are highly correlated. The \gls{SGWB} spectrum in this case is, however, also in the reach of \gls{ET} that will probe the opposite side of the \gls{GW} spectrum than \gls{LISA}, and combining these results will break the degeneracy in the reconstruction of $\Omega_b$ and $f_b$, consequently shrinking the posterior regions in the model parameter space. In \cref{fig:U1BL}, we show the peak amplitude and frequency of the \gls{SGWB} spectrum and benchmark points on a broader range of the model parameters.

\section{ Conclusions}
\label{sec:conclusions}

After the presence of an \gls{SGWB} has been identified in the \gls{LISA} data stream, an adequate interpretation of the \gls{SGWB} origin must be based on templates, linking the frequency shape of the \gls{SGWB} signal to the parameters of the sources.
In the present and companion papers, focusing on cosmic strings \cite{Blanco-Pillado:2024aca} and phenomena linked to inflation
\cite{Braglia:2024kpo}, we have selected a set of \gls{SGWB} templates suitable for this purpose, and
prototyped the \gls{LISA} template-based data analysis pipeline in \texttt{SGWBinner}, a code which has been used and tested for agnostic \gls{SGWB} reconstructions in previous \gls{LISA}-related analyses~\cite{Seoane:2021kkk, Colpi:2024xhw}. 
We forecast the detection and reconstruction of these sources. 
Although developed under simplifying assumptions, this program allows us to identify and quantify the bottlenecks that the \gls{LISA} community will have to tackle in the next years. 

In the present paper, we have reported the \gls{PT}-related part of the above program. As a first step, we have designed and implemented the templates for the main first-order \gls{PT} mechanisms sourcing an \gls{SGWB}, namely bubble collisions, sound waves and \gls{MHD} turbulence.
The former typically dominates the \gls{SGWB} signal in \gls{BSM} scenarios where the first-order \gls{PT} is extremely strong, while the others emerge in the case of mid-strength transition scenarios (weak transitions rarely lead to an observable signal at current \gls{GW} detectors). 
We have then adapted the \texttt{SGWBinner} to simulate data of the \gls{PT} \gls{SGWB} together with the astrophysical foregrounds and instrumental noises, and then forecast their parameter estimation. Crucially, the models used for the injection perfectly match the templates of the noise, astrophysical foregrounds, and \glspl{SGWB}  employed in the reconstruction. The forecasts then quantify the \gls{SGWB} parameter estimation that \gls{LISA} will perform if, at the time \gls{LISA} flies, the community will have shrunk the theoretical and experimental uncertainties below the statistical errors reconstructed in our analysis. 
In this sense, all our templates, although representing the current state of the art combining simplicity and accuracy, must be understood as placeholders of the precise templates that the community is urged to build in the next decade. On the other hand, our qualitative results are not expected to change if, as expected, the future templates will be similar to the current ones. 

We have performed an analysis of detection prospects as a function of the \gls{SGWB} template parameters, showing that for signals peaking around \SI{3}{\mHz}, the relative errors (\SI{68}{\%} \gls{CL}) in the parameter reconstruction are around or below \SI{10}{\%}  if the signal's peak has an amplitude of $\Omega_p\gtrsim \num{e-12}$. 
This result, already remarkable on its own, does not leverage the yearly modulation of the Galactic foreground; if one optimistically assumes that this feature allows complete removal of this astrophysical foreground, the above-mentioned lower bound decreases to $\Omega_p\gtrsim \num{e-13}$.  
These two bounds have been obtained with the \texttt{SGWBinner} code accurately sampling the full likelihood. 
We have verified that the (fast and simple) Fisher approximation would lead to similar values for these two bounds. The observed tendency is that, in the reconstruction of the typical first-order \gls{PT} signals, the values of the relative errors obtained with the Fisher approximation and with the full likelihood are in reasonably good agreement whenever they are about or below ten percent, while they quickly depart at higher values.

We have followed two approaches in the implementation of the first-order \gls{PT} templates. In one case we have parameterised the templates in terms of parameters that describe the geometrical features of the \gls{SGWB} signal frequency shape, i.e. the amplitude, frequency of the breaks, and the tilts of the spectrum.
This parameterisation reduces the parameter degeneracies and, in turn, simplifies the sampling of the likelihood. In the second case, we have expressed the templates as a function of the thermodynamic parameters, describing the physics behind the signal such as the temperature of the transition, its duration, and the kinetic energy fraction. This implies parameter degeneracies but allows a more direct physics interpretation of the signal.
For some benchmark signals with a satisfactory parameter reconstruction, we have verified that using the thermodynamic-parameter template leads to posteriors that are similar to those obtained by first using the geometric-parameter template and then converting the sampler chains into thermodynamic parameters. 
This is, however, only true provided one takes care to discard the regions corresponding to values of the geometric parameters leading to unphysical values for the thermodynamic ones.

The physics interpretation of the detection of (or upper bound on) the first-order \gls{PT} \gls{SGWB} cannot be limited to the thermodynamic parameters. Here we have proven that the \gls{SGWB} reconstruction can lead to compelling bounds on particle physics models. Currently, performing parameter inference directly between a model Lagrangian and the \gls{LISA} \gls{SGWB} data is prohibitive, as calculating the \gls{PT} (thermodynamic) parameters from a Lagrangian requires numerical tools that are not sufficiently automatised, fast or stable. We have shown how to circumvent the problem and applied the method to some concrete examples of BSM scenarios, namely the $\mathbb{Z}_2$-singlet extension and the scale-invariant $U(1)_{\rm B-L}$ extension. In the former model, with the detection of a first-order \gls{PT} \gls{SGWB} of amplitude $h^2\Omega_p \sim \num{e-11}$, \gls{LISA} can constrain the singlet mass and the Higgs-singlet coupling with an accuracy of about ten percent (for a known singlet quartic coupling). In the latter model, with a similar first-order \gls{PT} \gls{SGWB} detection, \gls{LISA} could bind the new gauge coupling with a relative error of one percent. The synergy with current and future particle physics experiments is manifest.

The synergy with other \gls{GW} detectors is instead less clear. 
We have found that the joint detection of the same first-order \gls{PT} signal both at \gls{LISA} and \gls{SKA} is unlikely. 
The prospects are more optimistic with \gls{ET}.
The observation in both \gls{LISA} and \gls{ET} would require first-order \gls{PT} signals with a peak frequency at $f_p\sim \SI{0.1}{\Hz}$, 
and peak amplitude $\Omega_p\gtrsim  \num{e-10}$ for a vacuum-dominated transition and $\Omega_p\gtrsim  \num{e-9}$ for sound waves and turbulence. This signal would indicate particle physics scenarios with a symmetry scale quite above the electroweak one. 
However, our synergy estimates with \gls{SKA} and \gls{ET} are likely optimistic, as they derive from \gls{SNR} computations that do not account for technical limitations such as the presence of foregrounds in \gls{SKA} and \gls{ET} frequency bands. They can be considered as indicative of a possible synergistic detection of the first-order \gls{PT} \gls{SGWB} at \gls{SKA} and \gls{ET}, however, reaching robust conclusions would require dedicated studies.   

In conclusion, the detection of the \gls{SGWB} by \gls{LISA} holds the potential for a historical, groundbreaking discovery. The precise reconstruction of this signal would further amplify the scientific impact of the observation. Our forecasts and their potential interpretations have highlighted this impact on scenarios featuring a first-order \gls{PT}. 

However, these results hinge on our simplifying assumptions concerning several theoretical and experimental issues. 
We assume that the data stream can be precisely cleaned from transient noise and signals. 
For the noise, foregrounds and \gls{SGWB}, we use the same templates for the injection and reconstruction, meaning that we neglect the systematic errors due to the current theoretical and experimental uncertainties on the noise and signals.
Thus, our results
constitute an approximate upper bound for such systematic error: above this bound, the \gls{PT} \gls{SGWB} reconstruction will not be limited by the instrument itself but by the insufficient knowledge of its actual sensitivity, the theoretical understanding of the \gls{SGWB} sources, and the data analysis capabilities.
Ultimately, we intend this paper to serve as a call to action for the worldwide scientific community to address these challenges and unleash the full potential of \gls{LISA}.

\acknowledgments
We acknowledge the LISA Cosmology Group members for seminal discussions. We especially thank the authors of Refs.~\cite{Blanco-Pillado:2024aca, Braglia:2024kpo} for the collaborative developments of the \texttt{SGWBinner} code shared among this and those works.
CC is supported by the Swiss National Science Foundation (SNSF Project Funding grant \href{https://data.snf.ch/grants/grant/212125}{212125}).
RJ is supported by JSPS KAKENHI Grant Numbers 23K17687 and 23K19048.
The work of ML is supported by the Polish National Agency for Academic Exchange within the Polish Returns Programme under agreement PPN/PPO/2020/1/00013/U/00001 and the Polish National Science Center grant 2018/31/D/ST2/02048.
EM acknowledges support from the Minerva foundation during the initial stage of the project.
MM received support from the Swedish Research Council (Vetenskapsr{\aa}det) through contract no.\ 2017-03934.
GN is partly supported by the grant Project.~No.~302640 funded by the
Research Council of Norway. MP acknowledges the hospitality of Imperial College London, which provided office space during some parts of this project.
ARP is supported by the Swiss National Science Foundation (SNSF Ambizione grant no.~\href{https://data.snf.ch/grants/grant/208807}{182044}).
VV is supported by the European Union's Horizon Europe research and innovation program under the Marie Sk\l{}odowska-Curie grant agreement No.~101065736, and by the Estonian Research Council grants PRG803, RVTT3 and RVTT7 and the Center of Excellence program TK202.

\appendix
\section{Response functions}
\label{app:response}

Here we report on the LISA response functions that \texttt{SGWBinner} adopts. Further details can be found in Ref.~\cite{Flauger:2020qyi}.

For our choice of TDI variables, the response functions ${R}_{ii}$ appearing in \cref{sec:data_in_each_channel} read
\begin{equation}
{R}_{ii}(f) = 16 \sin^2\left(\frac{2 \pi f L}{c}\right) \left(\frac{2 \pi f L}{c} \right)^{2} \tilde R_{ii}(f) \;  
\end{equation}
with $i=$A,E,T. The factors $\tilde{R}_{ii}(f)$ enclose the geometry of the detector. Their full expression, which are implemented in $\texttt{SGWBinner}$~\cite{Flauger:2020qyi}, can be well approximated as~\cite{Robson:2018ifk}
\begin{eqnarray}
	\tilde{R}_{ \text{AA} }(f) &=& \tilde{R}_{ \text{EE} }(f) \simeq \frac{9}{20} \left[ 
           1 +0.7 \left(\frac{2 \pi f L}{c}\right)^2 \right]^{-1}\,,\\
    \tilde{R}_{ \text{TT} }(f) &\simeq& \frac{9}{20}\left(\frac{2 \pi f L}{c}\right)^6 \left[\num{1.8e3} +0.7 \left(\frac{2 \pi f L}{c}\right)^8\right]^{-1} \,. 
\end{eqnarray}

Analogously, in the noise model of \cref{sec:data_in_each_channel}, the power spectra are given by 
\begin{equation}
\begin{aligned}
    P_{N, \text{AA} }(f, A, P)   = \;& P_{N, \text{EE} }(f, A, P) =   
   8 \sin^2\left(\frac{2 \pi f L}{c}\right) 
\tilde{R}_{ \text{AA} }(f)
\\
&\left\{ 4 \left[1 +\cos \left(\frac{2 \pi f L}{c} \right) + \cos^2 \left(\frac{2 \pi f L}{c} \right)\right] S^{\rm TM}_{\rm AA}(f, A) + \right.
\\  
& +  \left.\left[ 2 +\cos \left(\frac{2 \pi f L}{c} \right) \right] S^{\rm OMS}_{\rm AA}(f, P)  \right\}  \; , \\
\end{aligned}
\end{equation}
and
\begin{equation}
\begin{aligned}
    P_{N, \text{TT} }(f, A, P) & =  16 \sin^2\left(\frac{2 \pi f L}{c}\right) \tilde{R}_{ \text{TT} }(f) \left\{ 2 \left[ 1 - \cos \left(\frac{2 \pi f L}{c} \right) \right]^2  S^{\rm TM}_{\rm TT}(f, A) \;  + \right. \\ 
    & \hspace{3cm} \left. + \left[ 1 - \cos \left(\frac{2 \pi f L}{c} \right) \right] S^{\rm OMS}_{\rm TT}(f, P) \right\} \; ,
\end{aligned}
\label{eq:psdTT}
\end{equation}
where $S^{\rm OMS}_{ii}$ and $S^{\rm TM}_{ii}$ are defined in \cref{eq:Acc_noise_def,eq:OMS_noise_def} and are the same for any $i=$A,E,T. From \cref{eq:P_Sii}, one can see that $T^{\rm TM}_{ii}$ and $T^{\rm OMS}_{ii}$ are the factors in front of the corresponding $S^{\rm OMS}_{ii}$ and $S^{\rm TM}_{ii}$.

\printnoidxglossaries

\bibliographystyle{JHEP}
\bibliography{references.bib}

\end{document}